\documentclass[a4paper,11pt]{article}
\usepackage{jheppub} 
\usepackage{lineno}
\usepackage{bm}
\usepackage{subfigure}

\title{Non-extensive Hard Thermal Loop Resummation and Its Applications: Analysis in Zero and Finite Magnetic Fields}
\author[a]{He-Xia Zhang,}
\author[b]{Yu-Xin Xiao \footnote{Corresponding author}}

\affiliation[a]{Key Laboratory of Information Functional Material for Fujian Higher Education, College of Physics and Information Engineering,  Quanzhou Normal University,
\\ Quanzhou, 362000, China}

\affiliation[b]{
Key Laboratory of Quark \& Lepton Physics (MOE) and Institute of 
Particle Physics, Central China Normal University,\\ Wuhan, 430079, China
}

\emailAdd{hxzhang@qztc.edu.cn}
\emailAdd{xiaoyx@mails.ccnu.edu.cn}

\abstract{
The impact of non-extensive statistics on the hard thermal loop (HTL) resummation technique is investigated, in the absence and presence of a magnetic field. By utilizing the non-extensive bare propagators in the real-time formalism of finite temperature field theory, we determine the non-extensive deformations of both HTL gluon self-energies and resummed gluon propagators at the one-loop order. We observe that the introduction of non-extensivity results in distinct shifts in the Debye masses for the retarded/advanced and symmetric gluon self-energies. Applying the non-extensive modified resummed gluon propagators to obtain the dielectric permittivity of a quark-gluon plasma (QGP), we thereby derive the static heavy quark potential, which incorporates both short-range Yukawa and long-range string-like interactions between heavy quarks and the QGP medium. The real part of the potential exhibits increased screening as the non-extensive parameter $q$ ($q \geq  1$) increases, reducing the binding energies of heavy quarkonia. Furthermore, including non-extensivity enhances the magnitude of the imaginary part of the potential, causing a broadening in the decay widths of heavy quarkonia. Based on these observations, we estimate the melting temperatures of heavy quarkonia. Our results indicate that non-extensivity lowers the melting temperatures of heavy quarkonia, thus facilitating their dissociation, whereas the presence of a magnetic field inhibits this dissociation.
}


\begin{document}
\maketitle

\flushbottom

\section{Introduction}

Studying the properties of strongly interacting matter at high temperatures and/or high densities is a frontier topic in nuclear physics. Strongly interacting matter under such extreme conditions is expected to exhibit a deconfined state of quarks and gluons called quark-gluon plasma (QGP).
The wealth of data harvested in the Relativistic Heavy Ion Collider (RHIC) at BNL and the Large Hadron Collider (LHC) at CERN provide compelling evidence for the existence of QGP and characterize its properties~\cite{STAR:2005gfr, PHENIX:2004vcz}. Furthermore, a strong magnetic field is also expected to be generated in the early stage of the non-central heavy-ion collisions. 
Theoretical estimations showed that the maximum magnetic field strength can reach $eB\sim 5 m_{\pi}^2$ in Au+Au collisions at the top RHIC energy and $eB\sim 70~m_{\pi}^2$ in Pb+Pb collisions at the LHC energies~\cite{Deng:2012pc, Skokov:2009qp, Voronyuk:2011jd} where $m_\pi$ is pion mass, and $e$ represents the charge of the proton.

At sufficiently high temperatures, the QGP behaves as a weak coupling plasma ($T\gg \Lambda_{\rm QCD}$, $g\ll 1$, $g\equiv\sqrt{4\pi \alpha_s}$ is the quantum chromodynamics (QCD) gauge coupling), the perturbative QCD at finite temperature  and density is an essential theoretical framework
for the description of QGP properties, both in the absence and presence of a magnetic field~\cite{ Toimela:1984xy, Weldon:1982aq, Frenkel:1989br, Braaten:1991gm, Thoma:2000dc, Elmfors:1996fx, Kapusta:1989tk, Bellac:2011kqa, Andersen:2011sf, Haque:2014rua}. 
However, in the naive perturbative theory calculation, several results for the gauge fields become gauge-dependent and/or suffer infrared divergences using the bare propagators and vertices.
To resolve these issues,  Braatten and Pisarski developed the hard thermal loop (HTL) resummation technique~\cite{Pisarski:1990ds, Braaten:1991gm}.
Under the HTL approximation, the typical momenta of thermal quarks and gluons, comprising the internal lines in the loop expansion diagrams, are hard scales ($\sim T$ for gluon loop, $\sim T, \sqrt{eB}$ for quark loop), while the momentum transfers via gluon exchanges, comprising the external lines in the loop expansion diagrams, are soft scales ($\sim g eB$ and $g T^2$). Using the HTL resummation technique, the effective propagators and vertices which take into account the summation of higher-order expansion terms take over the bare propagator and vertices. 
The HTL resummation perturbative theory allows for systematic computations of various observables of the QGP, such as parton self-energies~\cite{Mrowczynski:2000ed, Thoma:1995ju}, in-medium complex heavy quark potential~\cite{Burnier:2015nsa, Laine:2007gj, Laine:2006ns, Beraudo:2007ky, Zhang:2023ked}, heavy quark diffusion coefficients~\cite{Caron-Huot:2007rwy, Moore:2004tg, Fukushima:2015wck, Zhang:2023ked}, dilepton production rate~\cite{Burnier:2007qm, Braaten:1990wp, Wong:1991be}, and the energy loss of partons~\cite{Thoma:1990fm},  etc. 
Recently, the use of the HTL resummation technique has been extended to some specific nonequilibrium scenarios in the absence of a magnetic field. 
Specifically, the momentum anisotropy due to the rapid longitudinal expansion in the early time of heavy-ion collision and the bulk viscous effects of the medium are embedded in real-time bare propagators by deforming distribution function of thermal particles~\cite{Romatschke:2003ms, Romatschke:2004jh, Kasmaei:2018yrr}, to influence the HTL parton self-energies and resummed propagators~\cite{Dumitru:2009fy, Du:2016wdx}, subsequently impact the in-medium properties of heavy quarkonia~\cite{Dong:2022mbo, Dumitru:2009ni, Thakur:2020ifi, Thakur:2021vbo, Thakur:2012eb}.

Over the last decades, there has been a widespread consensus that strong dynamical correlations, long-range color interactions, and microscopic memory effects can exist in high-energy collisions.
Non-extensive statistical mechanism, first proposed by C. Tsallis~\cite{Tsallis:1987eu}, provides an essential theoretical framework to deal with these physical phenomena. 
In non-extensive statistics, a (real) non-extensive parameter $q$ is introduced to incorporate the intrinsic fluctuating ambiance and measure the degree of non-extensivity in the system~\cite{Wilk:1999dr}.
The successful high-energy physics applications of non-extensive statistics are the accurate fitting of transverse momentum spectra of final state particles in the high-energy particle collision experiments~\cite{Shao:2009mu, Wong:2013sca, Che:2020fbz, Tang:2008ud, Wilk:1999dr, Alberico:1999nh, Huovinen:2012is, ALICE:2010syw, PHENIX:2010qqf, Chen:2020zuw, Su:2021bfm, Sharma:2018utk}. 
There have also been a considerable variety of issues to explore the possible non-extensive effects on the hydrodynamics~\cite{Alqahtani:2022xvo, Biro:2011bq, Osada:2008sw}, thermodynamics~\cite{Gervino:2012xh, Kyan:2022eqp, Bhattacharyya:2016lrk}, transport coefficients~\cite{Tiwari:2017aon, Rath:2019nne}, and the structure of phase transition~\cite{Rozynek:2009zh, Lavagno:2010xu}, in the high energy physics.

In this article, we extend the HTL  resummation technique to the non-extensive situation, starting from the non-extensive version of the bare propagators for quarks and gluons within real-time quantum field theory. We compute the retarded, advanced, and symmetric HTL gluon self-energies and corresponding resummed propagators for both zero and finite magnetic fields in the presence of non-extensivity to the leading order of $(q-1)$. As an application of these results, we utilize the non-extensive modified resummed gluon propagators to derive the dielectric permittivity of the non-extensive QGP, and then obtain the complex heavy quark potential incorporating both the perturbative Yukawa and the non-perturbative string-like interactions between heavy quarks and the medium. The responses of the heavy quark potential to the non-extensive statistics effects are studied. Furthermore, we use the real part of the non-extensive modified potential to solve the Schr$\ddot{\rm o}$dinger equation to obtain the binding energies of heavy quarkonia, while the imaginary part of the modified potential folding with radial quarkonium wave function is used to determine the decay widths of heavy quarkonia, in both the absence and presence of a magnetic field. Subsequently,  the melting temperatures of heavy quarkonia are estimated for different values of the non-extensive parameter.

The article is organized as follows.
In section~\ref{sec:Formalism}, we provide the basic formalism,  including the non-extensive distribution functions of quarks and gluons, the non-extensive real-time bare propagators of quarks and gluons, and the general expressions of HTL resummed gluon propagators to the leading order of $(q-1)$ in  Keldysh presentation. In section~\ref{sec:energy&propagator}, we compute the non-extensive correction to the retarded, advanced, and symmetry (time-order) HTL gluon self-energies as well as the corresponding resummed gluon propagators, both in the absence and presence of a magnetic field.  We also discuss the impacts of non-extensivity and magnetic field on Debye masses. 
In section~\ref{sec:potential}, we apply the obtained non-extensive modified resummed gluon propagators to derive the dielectric permittivity of QGP, then compute the static complex heavy quark potential, analyzing the effect of non-extensivity on its real and imaginary parts. Section~\ref{sec:E&Decay} presents calculations of binding energies, decay widths, and melting temperatures for heavy quarkonia, discussing their responses to non-extensive correction and the magnetic field. In the Appendix,  we present some derivations of the HTL gluon self-energy, both in the absence and presence of a magnetic field, within non-extensive statistics, using real-time formalism. 

\section{Formalism}\label{sec:Formalism}

\subsection{Non-extensive  distribution functions of  (anti)quarks and gluons}\label{sec:q_distribution}
 Following Refs.~\cite{PhysRevE.80.011126, Rahaman:2019dhp}, the non-extensive versions of single-particle distribution functions for (anti)quarks and gluons are  respectively  given as  
\begin{eqnarray}
f_{q,F}^{\pm}(E^f)=\frac{1}{[\exp_q(\beta(E^f\mp \mu_f))]^q+1},~~f_{q,B}(k)=\frac{1}{[\exp_q(\beta E)]^q- 1},\label{eq:q_distribution}
\end{eqnarray}
 where the left equation relates to quarks (superscript ``$+$") and antiquarks (subscript ``$-$"), and the right equation relates to gluons. $\mu_f$ stands for the chemical potential of $f$-th flavor quark, in this work we takes $\mu_u=\mu_d=\mu_s=\mu$. $\beta=1/T$ is the inverse temperature of the system.
In the absence of the magnetic field, the dispersion relation of $f$-th flavor (anti)quarks is given by  
 $E^f=\sqrt{k^2+m_f^2}$  with
 $k\equiv |\bm{k}|$, wherein $m_f$ is current mass of $f$-th quark. 
In the presence of the magnetic field oriented along the $z$-axis,  we take the scale hierarchy $T^2\sim eB$,  the dispersion relation of light (anti)quarks  is converted to the Landau quantized ones, which is given as 
$E_{k_z,n}^f=\sqrt{k_z^2+2n|e_f B|+m_f^2}$ with quantum number of Landau level $n$, where $e_f$ is the electric charge of $f$-flavor quark. From now on, we will take the massless quark limit ($m_f=0$), then arrive at $E^f=k$ as well as $E_{k_z,n}^f=\sqrt{k_z^2+2n|e_f B|}$. In eq.~(\ref{eq:q_distribution}), $\exp_{q}(x)$ is the non-extensive exponential. For $x\leq 0$ and $q>1$,  $\exp_{q}(x)$  is given as 
 \begin{equation}
     \exp_q(x)=
 \left[1+(q-1)x\right]^{  1/(q-1)}.
 \end{equation}
From a phenomenological perspective, the non-extensive parameter $q$ can be considered as a free parameter, and its values are generally greater than 1 in the realm of high energy collisions (see, for instance, Refs.~\cite{Alberico:1999nh, Alberico:2009gj, Lavagno:2002hv, Biyajima:2004ub, Biyajima:2006mv, Wilk:1999dr}). Notice that in the limit  $q\to 1$, $\exp_q(x)=\exp(x)$, eq.~(\ref{eq:q_distribution}) recovers to the standard Fermi-Dirac and Bose-Einstein distributions, which are respectively presented as 
 \begin{align}
   f_{F}^{0\pm}(k)=\frac{1}{\exp(\beta(k\mp \mu))+1},~~f_{B}^0(k)=\frac{1}{\exp(\beta k)- 1}. 
 \end{align}

In the present work,  we will focus our study on small deviations from the standard statistics and assume that $q$ differs from 1 by a small amount,  
eq.~(\ref{eq:q_distribution}) can be expanded in a Taylor series of powers of $(q-1)$, yielding the following result,
\begin{equation}
f_{q,F}^{\pm}(k)=f_{F}^{0\pm}(k)+\delta f_{q,F,(1)}^{\pm}(k).
\end{equation}
The correction term $f_{q,F,(1)}^{\pm}$ is a measure of the degree of non-extensivity of the system, and
its specific form is given as
\begin{eqnarray}\label{eq:eq:delta f_F}
 f_{q,F,(1)}^{\pm}(k)= \frac{[(k\mp \mu)^2-2(k\mp \mu)T](q-1)}{2T^2}f^{0\pm}_{F}(k)(1- f_{F}^{0\pm}(k)).
\end{eqnarray}
It is worth noting that the linear expansion holds when $k/T$ is not too large. The HTL approximation, based on the assumption that soft momenta of the order $k\sim gT$ and hard ones of the order $k\sim T$ can be distinguished in the weak coupling limit $g\ll 1$, satisfies this condition. In the presence of a magnetic field, the non-extensive correction term of distribution functions for (anti)quarks can be obtained by replacing  $k$ in eq.~(\ref{eq:eq:delta f_F}) with $E_{k_z,n}^f$.
For gluons, their non-extensive correction term of the distribution function is given as 
\begin{eqnarray}
 f_{q,B,(1)}(k)= \frac{{(k^2-2kT)}(q-1)}{2T^2}f_{q,B}^0(k)(1+ f_{q,B}^0(k)).
\end{eqnarray}

\subsection{Real-time bare propagators with the scope of non-extensive statistics} 

Following Ref.~\cite{Rahaman:2019dhp, Rahaman:2021xqv},  the real-time bare propagator for massless quarks  within the non-extensive statistic at a finite chemical potential is a  $2\times 2$ matrix, which takes the following form, 
\begin{align}\label{eq:Matrix_S}
iS(K)=& \displaystyle{\not}  K\left (\begin{array}{cc}
\frac{i}{K^2+i\epsilon} & 0\\
0 & \frac{-i}{K^2-i\epsilon} \\
       \end{array} \right )
       + \displaystyle{\not}  K 2 \pi \delta(K^2)
\left (\begin{array}{cc}
N(k_0) &  N(k_0)-\Theta(-k_0)\\
N(k_0)- \Theta(k_0) &  N(k_0)  \\
\end{array} \right ) \, ,
        \end{align}
 with four-momentum $K=(k_0,\bm{k})$ and   $N(k_0)=\Theta(k_0)f_{q,F}^{+}(k_0)+\Theta(-k_0)f_{q,F}^{-}(-k_0)$. $\Theta(x)$ is the Heaviside step function. In the limit $q\to 1$, we return to the standard real-time bare quark propagator.
 In the Landau level presentation, the real-time bare propagator for massless quarks of the $f$-th flavor in a magnetic field within the framework of non-extensive statistics is given by
\begin{align}
i S^{n,f}(K)=& e^{-\frac{\bm{k}_{\perp}^2}{|e_f B|}}\sum_{n=0}^{\infty}(-1)^nD_{n}^f(K)\bigg[\left (\begin{array}{cc}
\frac{i}{K_\|^2-2n|e_fB|+i\,\epsilon} & 0\\
0 & \frac{-i}{K_\|^2-2n|e_fB|-i\,\epsilon} \\
       \end{array} \right )\nonumber\\
        + &2 \pi \delta(K_{\|}^2-2n|e_fB|)
\left (\begin{array}{cc}
N(k_0) &  N(k_0)-\Theta(-k_0)\\
N(k_0)- \Theta(k_0) &  N(k_0)  \\
\end{array} \right ) \,\bigg] , 
\end{align}
where $K_{\|}^2=k^2_0-k_z^2$ and $\bm{k}_{\perp}^2=k_x^2+k_y^2$, the transverse function is given as \begin{align}
D_n^f(K)=&2\displaystyle{\not} K_{\|}\left[\mathcal{P}^f_+L^0_n\bigg(\frac{2\bm{k}_{\perp}^2}{|e_f B|}\bigg)-\mathcal{P}^f_-L^0_{n-1}\bigg(\frac{2\bm{k}_{\perp}^2}{|e_f B|}\bigg)\right]+4\displaystyle{\not} \bm{k}_{\perp}L_{n-1}^{1}\bigg(\frac{2\bm{k}_{\perp}^2}{|e_f B|}\bigg),
\end{align}
with $\mathcal{P}_{\pm}^f=[1\pm i\gamma^x\gamma^y\mathrm{sgn}(e_fB)]/2$ being the spin projectors. $L_{n}^\alpha(x)$ are the generalized Laguerre polynomials, and $L_{-1}^\alpha=0$ by definition.
For the real-time bare propagator for gluons in the non-extensive statistics, its $2\times 2$ matrix can be formulated as 
\begin{align}
i G(K)=& \left (\begin{array}{cc}
\frac{i}{K^2+i\,\epsilon} & 0\\
0 & \frac{-i}{K^2-i\,\epsilon} \\
       \end{array} \right )
       + 2 \pi \delta(K^2)
\left (\begin{array}{cc}
f_{q,B}(k_0) &  f_{q,B}(k_0)+\Theta(-k_0)\\
f_{q,B}(k_0)+ \Theta(k_0) &  f_{q,B}(k_0)  \\
\end{array} \right ) \, .
      \end{align}

The four components of the real-time bare propagator are not independent, which satisfy $D_{11}-D_{12}-D_{21}+D_{22}=0$, where $D_{ij}$ stands for $S_{ij} (S_{ij}^{n,f})$ or $G_{ij}$. It is more useful to write the bare propagators in terms of three independent components, i.e., the retarded ($R$), advanced ($A$), and symmetric ($F$) components, in Keldysh representation~\cite{Chou:1984es,Keldysh:1964ud}. Accordingly, one gets 
\begin{align}
 S_{R/A}(K)&=S_{11}(K)-S_{12/21}(K)
 =\frac{ \displaystyle{\not} K}{K^2\pm i\,\mathrm{sgn}(k_0)\epsilon}\label{eq:SRA},\\
    S_{F}(K)&=S_{11}(K)+ S_{22}(K)=
   - 2\pi i\displaystyle{\not} K [1-2N(k_0)]\delta(K^2),\label{eq:SF}
\end{align}
where ``$\pm$" in eq.~(\ref{eq:SRA}) represent retarded propagator and advanced propagator, respectively.
Only the symmetric component of the bare propagator depends on the temperature, chemical potential, as well as non-extensive parameter.
In the finite magnetic field,  three independent components of the bare quark propagator are given by
\begin{align}
    S^{n,f}_{R/A}(K)=S^{n,f}_{11}(K)-S^{n,f}_{12/21}(K)
    =& \frac{e^{-\frac{\bm{k}_{\perp}^2}{|e_f B|}}\sum_{n=0}^{\infty}(-1)^n D_n^f(K)}{K_{\|}^2-2n|e_fB|\pm i\,\mathrm{sgn}(k_0)\epsilon},\\
     S^{n,f}_{F}(K)=S^{n,f}_{11}(K)+S^{n,f}_{22}(K)
     =&-2\pi i D_n^f(K) [1-2N(k_0)] \delta(K_{\|}^2-2n|e_fB|).
\end{align}
For the bare gluon propagator, three independent components in the Keldysh representation take the following forms:
\begin{align}
 G_{R/A}(K)&=G_{11}(K)-G_{12/21}(K)
 =\frac{1}{K^2\pm i\,\mathrm{sgn}(k_0)\epsilon}\label{eq:SRA_gluon},\\
    G_{F}(K)&=G_{11}(K)+ G_{22}(K)=
    -2\pi i [1+2f_{q,B}(k_0)]\delta(K^2).\label{eq:SF_gluon}
\end{align}
Using real-time formalism, the self-energy also becomes a $2\times 2$ matrix and  fulfills the relation $\Pi_{11}(K)+\Pi_{12}(K)+\Pi_{21}(K)+\Pi_{22}(K)=0$.  The three components of gluon self-energy in Keldysh representation are defined as~\cite{Carrington:1996rx} 
\begin{align}    \Pi_R(K)&=\Pi_{11}(K)+\Pi_{12}(K),\label{eq:Pirelation1}\\
\Pi_A(K)&=\Pi_{11}(K)+\Pi_{21}(K),\label{eq:Pirelation2}\\
\Pi_F(K)&=\Pi_{11}(K)+\Pi_{22}(K).\label{eq:Pirelation3}
\end{align}

\subsection{Resummed gluon propagators in the presence of  non-extensivity}
\label{sec:self-energies} 
Having the bare propagators and gluon self-energies, one can compute the resummed gluon propagator, which describes the propagation of a collective plasma mode. Here we restrict ourselves to the system in Coulomb gauge\footnote{Throughout this paper, we use the Coulomb gauge, which is convenient for later applications. Since the final results for physical quantities are gauge-independent using the HTL resummed technique, we can choose any gauge.}, where only the temporal components of the self-energies and bare or resummed propagators, such as $\Pi_R^{00}$ and $G_R^{00}$, are considered. In the following, we will omit the superscript ``00"  of temporal components for simplification unless otherwise specified. Similar to the case in thermal field theory with the scope of extensive quantum  statistics, the resummed retarded/advanced gluon propagator within non-extensive statistics in the Coulomb gauge can also be determined from the following Dyson-Schwinger equation,
 \begin{equation}\label{eq:{G}_{R/A}^*}
 {G}_{R/A}^* (K)= G_{R/A}(K) + G_{R/A}(K) \Pi_{R/A}(K)  G_{R/A}^*(K), 
 \end{equation}
 where $G_{R/A}(K)= \frac{1}{\bm k^2}$ is the temporal component of bare retarded/advanced propagator.
 We use the superscript ``$*$" here to label a resummed propagator. The resummed symmetric gluon propagator satisfies the following Dyson-Schwinger equation 
\begin{equation}
G_F^*(K)
=
G_F(K)
+ G_R(K) \Pi_R(K)  G^*_F (K)
+ G_F (K)\Pi_A(K)  G^*_A (K)
+ G_R(K)\Pi_F(K)  G^*_A(K).
\label{eq:ds-s}
\end{equation} 
Using the identity for the bare symmetric gluon propagators in non-extensive statistics, $G_{F}(K) = (1 + 2f_{q,B}(k_0))\mathrm{sgn}(k_0)(G_{R}(K) - G_{A}(K))$, the solution to eq.~(\ref{eq:ds-s}) takes the following form:
\begin{align}\label{eq:GF}
{G}^{*}_F(K)&=(1+2f_{q,B}(k_0))\mathrm{sgn}(k_0)({G}^{*}_R(K)-{G}^{*}_A(K))\nonumber\\
& +{G}^{*}_R(K)\left[{\Pi}^{}_F(K)-\left(1+2 f_{q,B}(k_0)\right)\mathrm{sgn}(k_0)\left({\Pi}^{}_R(K)-{\Pi}^{}_A(K)\right)\right]{G}^{*}_A.
\end{align}

In the presence of small non-extensivity,  we only consider the contributions at leading order in $(q-1)$.  Consequently, the temporal components of the resummed gluon propagator (either retarded, advanced, or symmetry) and  gluon self energies can be expanded as:
\begin{align}
G_{R/A/F}^*(K)&\approx G_{R/A/F,(0)}^*(K)+
    G_{R/A/F,(1)}^*(K),\label{eq:GRAF*}\\
    \Pi_{R/A/F}(K)&\approx \Pi_{R/A/F,(0)}(K)+
    \Pi_{R/A/F,(1)}(K).
\end{align}
The temporal component of resummed retarded/advanced/symmetric propagator to the order of  $(q-1)^0$, denoted as $G^*_{R/A/F,(0)}(K)$, which satisfies the relation, i.e., $G_{R/A,(0)}^*(K)=G_{R/A}(K)+G_{R/A}(K)\Pi_{R/A,(0)}(K)G_{R/A,(0)}^*(K)$. For the linear term of order $(q-1)$ in eq.~(\ref{eq:GRAF*}), the expression is presented as, 
  \begin{equation}
      G_{R/A,(1)}^*(K)=G_{R/A}(K)\Pi_{R/A,(1)}(K)G_{R/A,(0)}^*(K)+G_{R/A}(K)\Pi_{R/A,(0)}(K)G_{R/A,(
      1)}^*(K).
  \end{equation}
Finally, we can get
  \begin{align}
     G^*_{R/A,(0)}(K)&=\frac{1}{G_{R/A}^{-1}(K)-\Pi_{R/A,(0)}(K)},\label{eq:G^*_R(0)}\\
      G^*_{R/A,(1)}(K)&=\frac{\Pi_{R/A,(1)}}{\left(G_{R/A}^{-1}(K)-\Pi_{R/A,(0)}(K)\right)^2}.\label{eq:G^*_R(1)}
  \end{align}
In the absence of non-extensivity, the term in the square brackets of eq.~(\ref{eq:GF}) vanishes as a consequence of the Kubo-Martin-Schwinger boundary condition:
  \begin{align}\label{eq:GF_p1}
      G_{F,(0)}^*(K)&=(1+2f^0_{B}(k_0))\mathrm{sgn}(k_0)[G_{R,(0)}^*(K)-G_{A,(0)}^*(K)],
  \end{align}
which is free of possible pinch problems and reflects the dissipation-fluctuation theorem.
The non-extensive correction term of the temporal component of the resummed symmetric gluon propagator at the leading order in $(q-1)$ is given by
  \begin{align}\label{eq:GF_p2}
      G_{F,(1)}^*(K)&=(1+2f^0_{B}(k_0))\mathrm{sgn}(k_0)\left[G_{R,(1)}^*(K)-G_{A,(1)}^*(K)\right]\nonumber\\
      &+2f_{q,B,(1)}(k_0)\mathrm{sgn}(k_0)\left[G_{R,(0)}^*(K)-G_{A,(0)}^*(K)\right]\nonumber\\
      &+G_{R,(0)}^*(K)\bigg\{\Pi_{F,(1)}(K)-(1+2f_{B}^0(k_0))\mathrm{sgn}(k_0)\left[\Pi_{R,(1)}(K)-\Pi_{A,(1)}(K)\right]\nonumber\\
     &-2f_{q,B,(1)}(k_0)\mathrm{sgn}(k_0)\left[\Pi_{R,(0)}(K)-\Pi_{A,(0)}(K)\right]\bigg\}G_{A,(0)}^*(K).
  \end{align}
All expressions in this subsection are general for zero and finite magnetic field backgrounds. To obtain the definite expressions of these resummed gluon propagators, we need to compute the HTL gluon self-energy first,  which will be addressed in detail in the following chapter.

\section{Non-extensive correction to  retarded, advanced, and symmetric HTL gluon self-energies and resummed gluon propagators}\label{sec:energy&propagator}
In the HTL resummed technique, gluon self-energy at the one-loop order comprises two distinct contributions: quark contributions, and gluon contributions (including the gluon-loop, ghost-loop, and tadpole contributions). Furthermore, due to the Landau level quantization of light quark motions in the presence of a magnetic field, the one-loop contribution from quarks to the gluon self-energy will be strikingly different from that in a zero magnetic field. Therefore, we will divide the computation of non-extensive modified gluon self-energies into two subsections: a case of the zero magnetic field and a case of the finite magnetic field.

\subsection{Case of the  zero magnetic field}
In the HTL approximation, the one-loop contributions from $N_f=3$ quarks and $2N_c=6$ gluons to the temporal component of the retarded gluon self-energy, denoted as $ \Pi_{R}^{\rm quark}$ and $ \Pi_{R}^{\rm gluon}$ within the framework of non-extensive statistics are respectively given as (for a detailed derivation, see Appendix \ref{sec:appendixA}),
\begin{align}
\Pi^{\rm quark}_{R}(Q)&=\frac{  N_{f}g^2}{(2\pi)^3}\sum_{b=\pm}\int
kdk   f_{q,F}^{b} (k)\frac{d\Omega_k}{2}\bigg[  \frac{1-x^2}{[x+(\omega+i\epsilon)/\widetilde{q}]^2}+\frac{1-x^2}{[-x+(\omega+i\epsilon)/\widetilde{q}]^2}\bigg],
\label{eq:PiR_quark}\\
\Pi^{\rm gluon}_{R}(Q)&=\frac{ 2 N_{c}g^2}{(2\pi)^3}\int_{}^{} kdk   f_{q,B} (k)\frac{d\Omega_k}{2}\bigg[  \frac{1-x^2}{[x+(\omega+i\epsilon)/\widetilde{q}]^2}+\frac{1-x^2}{[-x+(\omega+i\epsilon)/\widetilde{q}]^2}\bigg],\label{eq:PiR_gluon}
\end{align}
where $Q = (\omega, \widetilde{\bm{q}})$ denotes the external four-momentum in the one-loop diagram and is a soft scale. The differential solid angle is given by $d\Omega_k=\sin\theta d \theta d\phi=dxd\phi$, where  $x=\bm{k}\cdot\widetilde{\bm{q}}/(k\widetilde{q})$ with $\widetilde{q}\equiv |\widetilde{\bm{q}}|$. The variable $x$ ranges from $-1$ to 1. As $q$ approaches 1,  $f_{q,F}^{\pm}(k)$ and $f_{q,B}(k)$  recover to  $f_{F}^{0\pm}(k)$ and $f_{B}^0(k)$, respectively.
If the distributions are angular-independent, the square bracket term in eqs.~(\ref{eq:PiR_quark}) and (\ref{eq:PiR_gluon}) after the integration over $\Omega_k$ arrives at 
\begin{eqnarray}
\int_{}^{}d\Omega_k\bigg[  \frac{1-x^2}{[x+(\omega+i\epsilon)/\widetilde{q}]^2}+\frac{1-x^2}{[-x+(\omega+i\epsilon)/\widetilde{q}]^2}\bigg]=16\pi\left(\frac{\omega}{2\widetilde{q}}\ln\frac{\omega+\widetilde{q}+i\epsilon}{\omega-\widetilde{q}+ i\epsilon}-1\right).
\end{eqnarray}
Subsequently, to the order of $(q-1)^0$, eq.~(\ref{eq:PiR_quark}) and  eq.~(\ref{eq:PiR_gluon}) are respectively computed as 
\begin{align}
\Pi^{\rm quark}_{R,(0)}(Q) &= (m^{\rm quark
}_{D})^2\left(\frac{\omega}{2\widetilde{q}}\ln\frac{\omega+\widetilde{q}+ i\epsilon}{\omega-\widetilde{q}+ i\epsilon}-1\right) , 
\label{eq:pir_quark_eq}\\
\Pi^{\rm gluon}_{R,(0)}(Q) &= (m^{\rm gluon
}_{D})^2\left(\frac{\omega}{2\widetilde{q}}\ln\frac{\omega+\widetilde{q}+ i\epsilon}{\omega-\widetilde{q}+ i\epsilon}-1\right). 
\label{eq:pir_gluon_eq}
\end{align}
Here, $m^{\rm quark/gluon}_{D}$  denotes the one-loop contribution from quarks/gluons to the Debye mass in the standard quantum statistics.
By summing eq.~(\ref{eq:pir_quark_eq}) and eq.~(\ref{eq:pir_gluon_eq}),  
the total retarded gluon self-energy  to the  order of  $(q-1)^0$
is obtained as 
\begin{equation}
\Pi^{}_{R,(0)}(Q) = m^{2}_{D,R}\left(\frac{\omega}{2\widetilde{q}}\ln\frac{\omega+\widetilde{q}+ i\epsilon}{\omega-\widetilde{q}+ i\epsilon}-1\right) , 
\label{eq:pir-eq}
\end{equation}
which is retarded gluon self-energy in the equilibrium QGP within the standard quantum statistics. In eq.~(\ref{eq:pir-eq}), $m_{D,R}$ presents the usual retarded Debye mass, and is given by 
\begin{eqnarray}
m^{2}_{D,R}=m^{2}_{D}=(m^{\rm quark}_{D})^2+(m_{D}^{\rm gluon})^2=\frac{g^{2}T^{2}}{6}\left[N_{f}\left(1+\frac{3 \alpha^{2}}{\pi^{2}}\right)+2N_{c}\right],
\label{eq:debye-eq}
\end{eqnarray} 
where $\alpha=\mu/T$.
In the space-like region where  $\omega^2<\widetilde{q}^2$, eq.~(\ref{eq:pir-eq}) has an imaginary part, and the bracket term has the following structure
\begin{eqnarray}
\left(\frac{\omega}{2\widetilde{q}}\ln\frac{\omega+\widetilde{q}+ i\epsilon}{\omega-\widetilde{q}+ i\epsilon}-1\right)=\frac{\omega}{2\widetilde{q}}\bigg[\ln \bigg|\frac{\omega+\widetilde{q}}{\omega-\widetilde{q}}\bigg|- i\pi \Theta(\widetilde{q}^2-\omega^2)\bigg]-1.
\end{eqnarray} 

 In the presence of non-extensivity, specifically the small value of $(q-1)$, the temporal component of retarded gluon self-energy is   modified as
$\Pi_{R}(Q)  = \Pi^{}_{R,(0)}(Q)   +  \Pi_{R,(1)}(Q) $, where $\Pi_{R,(1)}(Q)= \Pi_{R,(1)}^{\rm quark}(Q)+\Pi_{R,(1)}^{\rm gluon}(Q)$ represents the non-extensive correction to $\Pi_{R}(Q)$ in the leading order of $(q-1)$. Here, one-loop contributions from quarks and  gluons to  $\Pi_{R,(1)}(Q)$,  
denoted as $\Pi^{\rm quark}_{R,(1)}(Q)$ and $\Pi^{\rm gluon}_{R,(1)}(Q)$, respectively, are computed as follows:
\begin{align}
\Pi^{\rm quark}_{R,(1)}(Q)
&=
\frac{  N_{f}g^2}{\pi^2}\sum_{b=\pm}\int kdk f_{q,F,(1)}^{b}(k)\left(\frac{\omega}{2\widetilde{q}}\ln\frac{\omega+\widetilde{q}+i\epsilon}{\omega-\widetilde{q}+ i\epsilon}-1\right) 
\\
&=\left( m_{D,R,(1)}^{\rm quark}\right)^2\left(\frac{\omega}{2\widetilde{q}}\ln\frac{\omega+\widetilde{q}+ i\epsilon}{\omega-\widetilde{q}+ i\epsilon}-1\right),\label{eq:Debye_R_quark}\\
\Pi^{\rm gluon}_{R,(1)}(Q)
&=
\frac{2N_{c}g^2}{\pi^2}\int k dk   f_{q,B,(1)}^{}(k) 
\left(\frac{\omega}{2\widetilde{q}}\ln\frac{\omega+\widetilde{q}+ i\epsilon}{\omega-\widetilde{q}+ i\epsilon}-1\right)\\
&=\left( m_{D,R,(1)}^{\rm gluon}\right)^2\left(\frac{\omega}{2\widetilde{q}}\ln\frac{\omega+\widetilde{q}+ i\epsilon}{\omega-\widetilde{q}+i\epsilon}-1\right).\label{eq:Debye_R_gluon}
\end{align}
Here, $m_{D,R,(1)}^{\rm quark/gluon}$ denotes the non-extensive correction term of the quark/gluonic contribution to the retarded Debye mass.
By combining eq.~(\ref{eq:pir-eq}), eq.~(\ref{eq:Debye_R_quark}), and eq.~(\ref{eq:Debye_R_gluon}), the temporal component of total retarded/advanced gluon self-energy, including non-extensive correction, can be expressed as
\begin{equation}
\Pi_{R}(Q)
=\widetilde {m}_{D,R}^2
\left(\frac{\omega}{2\widetilde{q}}\ln\frac{\omega+\widetilde{q}+ i\epsilon}{\omega-\widetilde{q}+ i\epsilon}-1\right).
\label{eq:pir-nonextensive}
\end{equation}
Here,
$\widetilde {m}_{D,R}^2 = m^{2}_{D}+m^{2}_{D,R,(1)}$
represents the non-extensive modified  retarded Debye mass, 
and  the total correction term $m^{2}_{D,R,(1)}$ is written as 
\begin{align}
 m^{2}_{D,R,(1)}
&= ( m_{D,R,(1)}^{\rm gluon })^2+( m_{D,R,(1)}^{\rm quark })^2\\
&=\frac{q-1}{2}\left[(m_{D}^{\rm gluon})^2a^{\rm gluon}_{R}+(m_{D}^{\rm quark})^2 a^{\rm quark}_{R}\right],
\label{eq:mDR_NES}
\end{align}
where the dimensionless quantities
$ a^{\rm quark}_{R} $ and $ a^{\rm gluon}_{R}$ are respectively defined as 
\begin{align}
a^{\rm quark}_{R} = \frac{2}{(q-1)} \frac{\sum_{b=\pm}\int k dk  f_{q,F,(1)}^{b}(k)}{\sum\limits_{b=\pm}\int k dk f_F^{0b}(k)},~ a^{\rm gluon}_{R} = \frac{2}{(q-1)} \frac{\int k dk f_{q,B,(1)}(k)}{\int k dk f^0_{B}(k)}, 
\end{align}
and their explicit forms are respectively given as  
\begin{align}
a^{\rm quark}_{R} =&
-\frac{36}{\pi^2+3 \alpha^2 }
\left[ {\rm Li}_{3}(-e^{\alpha})
+ {\rm Li}_{3}(-e^{-\alpha})\right]
+\frac{24}{\pi^2+3 \alpha^2 }
\left[\mathrm{Li}_{2}(-e^{\alpha})+\mathrm{Li}_{2}(-e^{-\alpha})\right]\nonumber\\
+&\frac{24}{\pi^2+3 \alpha^2 }\alpha 
\left[\mathrm{Li}_{2}(-e^{\alpha})-\mathrm{Li}_{2}(-e^{-\alpha})\right]+\frac{6}{\pi^2+3 \alpha^2 }\alpha^2
\left[ \ln (1+e^{\alpha})+\ln (1+e^{-\alpha})\right]\nonumber\\
+&\frac{12}{\pi^2+3 \alpha^2 }\alpha
\left[ \ln (1+e^{\alpha})-\ln (1+e^{-\alpha})\right],\\
a^{\rm gluon}_{R/A} =&
\frac{36}{\pi^2}\zeta(3)-4,
\end{align}
with ${\rm Li}_{n} (x)$ being the polylogarithm functions. Finally, the non-extensive modified  retarded/advanced Debye mass is given as
\begin{equation}
\widetilde{m}_{D,R}^2=
\frac{g^2{T^2}}{6}\left[2N_{c}
\left(1+a^{\rm gluon}_{R}\frac{q-1}{2}\right)
+N_{f}\left(1+\frac{3  \alpha^{2}}{\pi^{2}}\right)
\left(1+a^{\rm quark}_{R}\frac{q-1}{2}\right)\right].
\end{equation}
When the chemical potential is  equal to zero, $a^{\rm quark}_{R}$ can be simplified to:
$a^{\rm quark}_{R}= \frac{54}{\pi^2}\zeta(3)-4$.
Consequently, the non-extensive statistic effect leads to a shift in the retarded/advanced  Debye mass, yielding,
\begin{align}
\frac{\widetilde m^2_{D,R}}{m^2_D}
&=\left[1+ \frac{(72N_c+54N_f)\zeta(3)-4\pi^2(2N_c+N_f)}{\pi^2(2N_c+N_f)}\frac{(q-1)}{2}\right]
\\
&=\left[1+\left(\frac{21}{\pi^2} \zeta(3)-2
\right)(q-1)\right].\label{eq:ratio_mDR} 
\end{align}

In the HTL approximation, 
the one-loop contributions from $N_f$ quarks and $2N_c$ gluons  to the temporal component of the  symmetric  gluon self-energy denotes as $ \Pi_{F}^{\rm quark}$ and $ \Pi_{F}^{\rm gluon}$,  in non-extensive statistics  are respectively computed as (for a detailed derivation, see Appendix \ref{sec:appendixA}),
 \begin{eqnarray}
 \Pi^{\rm quark}_F(Q)&=&-iN_{f}g^2\sum_{b=\pm}\int \frac{dk k^{2}}{2\pi}f_{q,F}^{b} (k)(1-f^{b}_{q,F} (k))\frac{2}{\widetilde{q}}\Theta(\widetilde{q}^{2}-\omega^{2}), 
 \label{eq:pif-quark}\\
\Pi^{\rm gluon}_{F}(Q)&=&-i2N_{c}g^2\int \frac{dk k^{2}}{2\pi}f_{q,B} (k)(1+f_{q,B} (k))\frac{2}{\widetilde{q}}\Theta(\widetilde{q}^{2}-\omega^{2}). 
\label{eq:pif-gluon}
\end{eqnarray}
Upon substitution of the distributions in eq.~(\ref{eq:pif-quark}) and eq.~(\ref{eq:pif-gluon}) with $f_F^{0\pm }(k)$ and $f_{B}^0(k)$, eq.~(\ref{eq:pif-quark}) and eq.~(\ref{eq:pif-gluon}) to order $(q-1)^0$ are respectively expressed as 
 \begin{align}
\Pi^{\rm quark}_{F,(0)}(Q)&=-2\pi i (m_{D}^{\rm quark})^2 \frac{T}{\widetilde{q}}\Theta(\widetilde{q}^{2}-\omega^{2}),
\label{eq:pif_quark_eq}\\
\Pi^{\rm gluon}_{F,(0)}(Q)&=-2\pi i (m_{D}^{\rm gluon})^2 \frac{T}{\widetilde{q}}\Theta(\widetilde{q}^{2}-\omega^{2}).
\label{eq:pif_gluon_eq}
\end{align}
Accordingly, the temporal component of total symmetric gluon self-energy in the equilibrium QGP within the standard quantum statistics can be written as
\begin{equation}
\Pi^{\rm}_{F,(0)}(Q)
= -2\pi i m_{D,F}^2 \frac{T}{\widetilde{q}}\Theta(\widetilde{q}^{2}-\omega^{2}),
\label{eq:pif-eq}
\end{equation}
where $m_{D,F}=m_{D}$ is the usual symmetric Debye mass in the standard quantum statistics.
Similar to the retarded gluon self-energy, considering the small non-extensivity, the temporal component of total symmetric gluon  self-energy is expressed as
$\Pi_{F}(Q)= 
\Pi^{\rm }_{F,(0)}(Q)
+  \Pi_{F,(1)}(Q)$, 
and $\Pi_{F,(1)}(Q)= \Pi_{F,(1)}^{\rm quark}(Q)+\Pi_{F,(1)}^{\rm gluon}(Q)$ represents the non-extensive correction to $\Pi_{F}(Q)$ in the leading order of $(q-1)$. 
The one-loop contribution from quarks and gluons to $\Pi_{F,(1)}(Q)$, denoted as $\Pi^{\rm quark}_{F,(1)}(Q)$ and $\Pi^{\rm gluon}_{F,(1)}(Q)$, respectively, are computed as follows:
\begin{eqnarray}
 \Pi^{\rm quark}_{F,(1)}(Q)&=&-iN_{f}g^2\sum_{b=\pm}\int \frac{dk k^{2}}{2\pi} f^b_{q,F,(1)}(k) (1 - 2 f^{0b}_{F} (k) )\frac{2}{\widetilde{q}}\Theta(\widetilde{q}^{2}-\omega^{2})\\
 &=&-2\pi i\big(m_{D,F,(1)}^{\rm quark}\big)^2\frac{T}{\widetilde{q}}\Theta(\widetilde{q}^{2}-\omega^{2}), 
 \label{eq:pif-quark_NES}\\
 \Pi^{\rm gluon}_{F,(1)}(Q)&=&-i2N_{c}g^2\int \frac{dk k^{2}}{2\pi}f_{q,B} (k)(1+2f_{q,B} (k))\frac{2}{\widetilde{q}}\Theta(\widetilde{q}^{2}-\omega^{2})\\
 &=&-2\pi i\big(m_{D,F,(1)}^{\rm gluon}\big)^2\frac{T}{\widetilde{q}}\Theta(\widetilde{q}^{2}-\omega^{2}). 
 \label{eq:pif-gluon_NES}
 \end{eqnarray}
Here, $m_{D,F,(1)}^{\rm quark/gluon}$ denotes the non-extensive correction term for the quark/gluonic contributions to the symmetric Debye mass.
Finally,  the temporal component of total  symmetric gluon self-energy including the effect of non-extensivity is written as
\begin{equation}
\Pi_{F}(Q)
=-2\pi i\widetilde {m}_{D,F}^2
\frac{T}{\widetilde{q}}\Theta(\widetilde{q}^{2}-\omega^{2}), 
\label{eq:pif-nonextensive}
\end{equation}
where 
$\widetilde {m}_{D,F}^2 = m^{2}_{D}+m^{2}_{D,F,(1)}$
represents the non-extensive modified symmetric Debye mass and the associated
 non-extensive correction term to order $(q-1)^1$ is given 
 by 
\begin{align}
 m^2_{D,F,(1)}&=(m^{\rm gluon}_{D,F,(1)})^2+(m^{\rm quark}_{D,F,(1)})^2\\
&=
\frac{q-1}{2}\left[(m_{D}^{\rm gluon})^2a^{\rm gluon}_{F}+(m_{D}^{\rm quark})^2 a^{\rm quark}_{F}\right].
\end{align}
Here, the dimensionless quantities 
$ a^{\rm quark}_{F} $ and $ a^{\rm gluon}_{F}$ are respectively defined as
\begin{eqnarray}
a^{\rm quark}_{F}
&=& \frac{2}{q-1} 
\frac{ \sum_{b=\pm}
\int  dk \ k^2  f^b_{q,F,(1)}(k) (1 - 2 f^{0b}_{F} (k) )}
{ \sum_{b=\pm}\int dk\ k^2 
f_F^{0b} (k) ( 1 -f_F^{0b} (k) )}, \\
a^{\rm gluon}_{F}
&=& \frac{2}{q-1} 
\frac{\int  dk \ k^2  f_{q,B,(1)}(k) (1  +2 f^0_B (k) )
}{ \int  dk\ k^2 
f^0_B (k) ( 1 +f^0_B (k) )}, 
\end{eqnarray}
and their explicit forms respectively are 
\begin{align}
a^{\rm quark}_{F}=&-
\frac{72}{ \pi^{2}+3\alpha^{2} }
\left[{\rm Li}_{3}(-e^{-\alpha})+{\rm Li}_{3}(-e^{\alpha})\right]+\frac{36}{ \pi^{2}+3\alpha^{2} }
\left[{\rm Li}_{2}(-e^{-\alpha})+{\rm Li}_{2}(-e^{\alpha})\right]\nonumber\\
-&\frac{36\alpha}{ \pi^{2}+3\alpha^{2} }\left[{\rm Li}_{2}(-e^{-\alpha})-{\rm Li}_{2}(-e^{\alpha})\right]+\frac{6\alpha^2}{ \pi^{2}+3\alpha^{2} }\left[\ln (1+e^{\alpha})+\ln (1+e^{-\alpha})\right]\nonumber\\
+&\frac{12\alpha}{ \pi^{2}+3\alpha^{2} }\left[\ln (1+e^{\alpha})-\ln (1+e^{-\alpha})\right],\\
a^{\rm gluon}_{F}=&\frac{72
}{\pi^{2}}\zeta(3)-6. 
\end{align}
Finally, the symmetric Debye mass including the small non-extensive correction is expressed  as
\begin{equation}
\widetilde{m}_{D,F}^2=
\frac{g^2{T^2}}{6}\left[2N_{c}
\left(1+a^{\rm gluon}_{F}\frac{q-1}{2}\right)
+N_{f}\left(1+\frac{3  \alpha^{2}}{\pi^{2}}\right)
\left(1+a^{\rm quark}_{F}\frac{q-1}{2}\right)\right].
\end{equation}
At zero chemical potential, the expression of $a^{\rm quark}_F$  reduces to $a^{\rm quark}_{F} = \frac{108}{\pi^2}\zeta(3)-6$, one consequence of the non-extensive statistic effect is a shift in the symmetric Debye mass, as follows:
 \begin{align}
 \frac{\widetilde m^2_{D, F}}{m^2_D}
 &=\bigg[1+ \frac{(144N_c+108N_f)\zeta(3)-6\pi^2(2N_c+N_f)}{\pi^2(2N_c+N_f)}\frac{(q-1)}{2}\bigg]\\
 &=\bigg[1+\left(\frac{42}{\pi^2}\zeta(3)-3\right)(q-1)\bigg].\label{eq:ratio_mDF}
 \end{align}
Comparing eq.~(\ref{eq:ratio_mDF}) with eq.~(\ref{eq:ratio_mDR}), when $q>1$,  we find that $\widetilde{m}_{D,R}^2\neq \widetilde{m}_{D,F}^2$ and $\frac{\widetilde{m}_{D,F}^2}{\widetilde{m}_{D,R}^2}=\frac{\pi^2+(42\zeta(3)-3\pi^2)(q-1)}{\pi^2+(21\zeta(3)-2\pi^2)(q-1)}>1$,  indicating that the presence of non-extensivity alters the equivalence between the symmetric Debye mass and the retarded Debye mass.

\subsection{Case of the finite magnetic field}
Following the procedures outlined in the above subsection, we determine the non-extensive correction to retarded, advanced, and symmetric gluon self-energies, as well as the corresponding resummed propagators, in a magnetic field background. The Landau quantization of motions of light quarks significantly alters the computation of the quark-loop contribution to gluon self-energy in a magnetic field, making it strikingly different from the case with the zero magnetic field. 
Furthermore, unlike the HTL approximation in $(3+1)$ dimensions, the quark-loop contribution to the vacuum part of the temporal component of retarded gluon self-energy in a magnetic field, denoted as  $\Pi_{R,B,\rm vac}^{\rm quark}$, persists even at high temperature and/or density, which is presented in eq.~(\ref{eq:Pi_quark_vac}). 
To attain a clearer comprehension, we divide the computations of the real and imaginary components of gluon self-energy. In the HTL approximation, characterized by the hierarchy of scale $T^2\sim eB\gg g^2T^2$~\cite{Elmfors:1996fx, Zhang:2023ked, Fukushima:2015wck}, we compute the one-loop contribution from quarks to the medium part of temporal component of  retarded gluon self-energy in a magnetic field, denoted as  $\Pi_{R,B,\rm med}^{\rm quark}$ (see detailed derivation in Appendix \ref{sec:appendixB}). Its real part  under the static limit ($\omega\to0$) is written as 
\begin{align}\label{eq:RePi_quark}
  \underset{\omega\to 0}{\mathrm{lim}} \mathrm{Re}\, \Pi_{R,B,\rm med}^{\rm quark}(Q)
     =&-\sum_f\sum_{n=1}^{\infty}\sum_{b=\pm }\frac{g^2|e_fB|}{4\pi}
     \int\frac{dk_z }{(2\pi)}\bigg\{-\frac{2(E^f_{k_z,n})^2+\widetilde{q}_zk_z}{E_{k_z,n}^f[(E_{k_z,n}^f)^2-(E_{p_z,n}^f)^2]}\nonumber\\
     \times& f^{b}_{q,F}(E_{k_z,n}^f) +\frac{(E^f_{p_z,n})^2+(E^f_{k_z,n})^2+\widetilde{q}_zk_z}{E_{p_z,n}^f[(E_{k_z,n}^f)^2-(E_{p_z,n}^f)^2]}f^{b}_{q,F}(E_{p_z,n}^f)\bigg\},
\end{align}
where $E_{p_z,n}^f=\sqrt{(k_z+\widetilde{q}_z)^2+2n|e_fB|}$ with $p_z=k_z+\widetilde{q}_z$.  
In the presence of non-extensivity,  when takes $\widetilde{q}_z\to 0$,  the curly bracket in eq.~(\ref{eq:RePi_quark}), to the leading order of $(q-1)$, can be expanded as follows:
\begin{align}
    &\bigg\{\dots\bigg\}\approx\frac{H_{b}^f(E^f_{k_z,n})+ f_{q,F,(1)}^{b}(E_{k_z,n}^f)(1-2f_F^{0b }(E_{k_z,n}^f))}{T}-\frac{1}{T^2}(q-1)(E_{k_z,n}^f-b \mu)H_{b }^f(E^f_{k_z,n})\nonumber\\
  & +\frac{1}{T}(q-1)H_{b}^f(E^f_{k_z,n}).\label{eq:expansion}
\end{align}
Here, $H_{b}^f(E^f_{k_z,n})=f_F^{0b}(E^f_{k_z,n})(1-f_F^{0b}(E^f_{k_z,n}))$.
Adding the real part of $\Pi_{R, B,\rm vac}^{\rm quark}$ in eq.~(\ref{eq:Pi_quark_vac}) and eq.~(\ref{eq:RePi_quark}),  the total real part of the one-loop contribution from quarks to the temporal component of retarded gluon self-energy in a magnetic field,  denoted as $\mathrm{Re}\,\Pi_{R,B}^{\rm quark}$, is obtained.
To order $(q-1)^0$, it in the static limit ($\omega\to0, \widetilde{q}_z\to 0$) is written as 
\begin{eqnarray}\label{eq:RePiq}
  \underset{\substack{\omega\to 0 \\ \widetilde{q}_z\to 0}}{\mathrm{lim}} \mathrm{Re}\,\Pi_{R,B,(0)}^{\rm quark}(Q)
&=&-\sum\limits_{f}\sum\limits_{n=0}\limits^{\infty}\sum_{b=\pm }\frac{g^2\alpha_{0n}|e_fB|}{4\pi T}
\int\frac{dk_z}{2\pi}H_{b}^f(E^f_{k_z,n}),
\end{eqnarray}
with $\alpha_{0n}=(2-\delta_{0n})$ being the Landau level-dependent spin degeneracy.
The above equation is just the standard Debye mass from quark contribution to the gluon self-energy in a magnetic field, i.e., $(m_{D,B}^{\rm 
quark})^2=-\mathrm{Re}\,\Pi_{R,B,(0)}^{\rm quark}(\omega\to 0,\bm{\widetilde{q}}_{}\to 0)$.
Since thermal gluons are not directly affected by the magnetic field, the computation of gluonic contribution to gluon self-energy in the presence of a magnetic field is identical to that in the absence of a magnetic field. Therefore, the total magnetic field-dependent Debye mass from retarded/advanced gluon self-energy in the equilibrium within the standard quantum  statistics is given as 
$m_{D,B}^2=(m_{D,B}^{\rm quark})^2+(m_{D}^{\rm gluon})^2$, which is also consistent with the result using the semi-classical transport theory in the magnetic field~\cite{Kurian:2019nna}.

Inserting eq.~(\ref{eq:expansion}) into eq.~(\ref{eq:RePi_quark}), the non-extensive correction term of $\mathrm{Re}\,\Pi_{R,B}^{\rm quark}$ to  order $(q-1)^1$, denoted as $\mathrm{Re}\, \Pi_{R,B,(1)}^{\rm quark}$,  within the static limit ($\omega\to 0, \widetilde{q}_z\to 0$) is expressed as 
\begin{equation}\label{eq:Repiq1}
\underset{\substack{\omega\to 0 \\ \widetilde{q}_z\to 0}}{\mathrm{lim}}\mathrm{Re}\ \Pi_{R,B,(1)}^{\rm quark} (Q)
=-\sum\limits_{f}\sum\limits_{n=1}\limits^{\infty}\sum_{b=\pm }\frac{g^2|e_fB|}{4\pi T}\int\frac{dk_z}{2\pi} \frac{(q-1)}{2}M_{b}^f(E^f_{k_z,n}),
\end{equation}
where  the function $M_b^f(E_{k_z,n}^f)$ is defined as 
\begin{align}
M^{f}_b(E_{k_z,n}^f)&=  H_{b}^f(E_{k_z,n}^f)\frac{(E_{k_z,n}^f-b \mu)}{T}\bigg[\frac{(E_{k_z,n}^f-b \mu-2T)}{T}\tanh\left(\frac{E_{k_z,n}^f-b \mu}{2T}\right)\nonumber\\
&-2+\frac{2T}{(E_{k_z,n}^f-b \mu)}\bigg].
\end{align}
 Accordingly, we also get the non-extensive correction term of the retarded Debye mass from quark contributions in a magnetic field, namely,
 \begin{align}
   (m_{D,R,B,(1)}^{\rm 
quark})^2=-\mathrm{Re}\,\Pi_{R,B,(1)}^{\rm quark}(\omega\to 0,\bm{\widetilde{q}}_{}\to 0).
 \end{align}
Finally, the total non-extensive modified retarded Debye mass in a magnetic field is given as 
\begin{align}
\widetilde{m}_{D,R,B}^2 =(m_{D,B}^{\rm quark})^2\left(1+\frac{q-1}{2}a_{R,B}^{\rm quark}\right)+\frac{g^2{T^2}}{6}2N_{c}
\left(1+\frac{q-1}{2}a^{\rm gluon}_{F}\right),
\end{align}
with dimensionless quantity $a_{R,B}^{\rm quark}=2(m_{D,R,B,(1)}^{\rm 
quark})^2/((q-1)(m_{D,B}^{\rm quark})^2)$.

Next, we explore the quark contributions to the temporal component of the imaginary part of retarded gluon self-energy in a magnetic field, denoted as $\mathrm{Im}\,\Pi_{R,B}^{\rm quark}$. 
After some calculations listed in Appendix~\ref{sec:appendixB}, the medium part of  $\mathrm{Im}\, \Pi_{R,B}^{\rm quark}$ in a non-extensive QGP within the static limit is given as 
 \begin{align}\label{eq:ImPi_HLL}
    \underset{\omega\to 0}{\mathrm{lim}} \frac{ \mathrm{Im}\, \Pi^{\rm quark}_{R,B,\rm med}(Q)}{\omega}
   =&-\sum_{f}\sum_{n=1}^{\infty}\sum_{b=\pm }\frac{g^2}{4\pi}\frac{2n|e_fB|^2}{E_{\widetilde{q}_z/2,n}^f|\widetilde{q}_z|}\frac{1  }{2T}\nonumber\\
\times&\bigg[q\exp_q\bigg(\frac{E_{\widetilde{q}_z/2,n}^f- b\mu}{T}\bigg)\left(f_{q,F}^{b}(E_{\widetilde{q}_z/2,n}^f)\right)^2\bigg].
 \end{align} 
In the presence of  small non-extensivity, the square bracket term in eq.~(\ref{eq:ImPi_HLL}) is expanded to the leading order of $(q-1)$, yielding 
\begin{align}\label{eq:expansion3}
 \bigg[\dots \bigg]\approx  
H_b^f(E_{\widetilde{q}_z/2,n}^f)+\frac{q-1}{2}M_b^f(E_{\widetilde{q}_z/2,n}^f).
\end{align}
Putting eq.~(\ref{eq:expansion3}) into eq.~(\ref{eq:ImPi_HLL}) and summing  the vacuum part of   $\mathrm{Im}\, \Pi_{R,B}^{\rm quark}$ given in eq.~(\ref{eq:Pi_quark_vac}), the expression of $\mathrm{Im}\, \Pi_{R,B}^{\rm quark}$  in the order of $(q-1)^0$ within the static limit ($\omega\to 0$)  is written as 
\begin{align}\label{eq:Impiq}
  \underset{\omega\to 0}{\mathrm{lim}}\frac{\mathrm{Im}\, \Pi^{ \rm quark }_{R, B,(0)} (Q)}{\omega}&=-
\sum_{f}\sum_{n=1}^{\infty}\sum_{b=\pm}\frac{g^2}{4\pi}
\frac{ 2n|e_fB|^2}{TE_{\widetilde{q}_z/2,n}^f|\widetilde{q}_z|}H^f_{b}(E^f_{\widetilde{q}_z/2,n})-\sum_f\frac{g^2}{4\pi}|e_fB|\delta(\widetilde{q}_z).
    \end{align} 
To the leading order in $(q-1)$, we get
    \begin{align}\label{eq:Impiq1}
  \underset{\omega\to 0}{\mathrm{lim}}\frac{\mathrm{Im}\,\Pi^{ \rm quark }_{R, B,(1) }(Q)}{\omega}&=-
\sum_{f}\sum_{n=1}^{\infty}\sum_{b=\pm}\frac{g^2}{4\pi}
\frac{ 2n|e_fB|^2}{TE_{\widetilde{q}_z/2,n}^f|\widetilde{q}_z|}\frac{q-1}{2}M^f_{b}(E^f_{\widetilde{q}_z/2,n}).
\end{align} 
By summing the set of equations ( (\ref{eq:pir_gluon_eq}), (\ref{eq:Debye_R_gluon}), (\ref{eq:RePiq}), (\ref{eq:Repiq1}), (\ref{eq:Impiq}), and (\ref{eq:Impiq1})), the temporal component of total retarded gluon self-energy in the presence of a magnetic field, incorporating the non-extensive correction, is expressed as
\begin{align}\label{eq:Pir_B}
  \Pi_{R,B}(Q)&=\mathrm{Re}\,\Pi^{ \rm quark }_{R, B,(0)} (Q)+i\,\mathrm{Im }\,\Pi^{ \rm quark }_{R, B,(0)} (Q)+\mathrm{Re}\,\Pi^{ \rm quark }_{R, B,(1)} (Q)+i\,\mathrm{Im }\,\Pi^{ \rm quark }_{R, B,(1)} (Q)\nonumber\\
  &+ \Pi_{R,(0)}^{\rm gluon}(Q)+\Pi_{R,(1)}^{\rm gluon}(Q).
\end{align}

The one-loop contribution from quarks to the symmetric gluon self-energy in a magnetic field depends purely on the medium and only has an imaginary part. Using eqs.~(\ref{eq:Pimunu_F_quark_B})-(\ref{eq:appendix_Pi_quark_F3}) from Appendix~\ref{sec:appendixB}, we compute its temporal component in the static limit within non-extensive statistics, which is written as
\begin{align}\label{eq:PiFquark_B}
  \underset{\omega\to 0}{\mathrm{lim}} \Pi^{\rm quark}_{F,B}(Q)
=&-i\sum_f\sum_{n=1}^{\infty}\sum_{b=\pm}\frac{g^2}{4\pi}\frac{4n|e_fB|^2}{E_{\widetilde{q}_z/2,n}^f|\widetilde{q}_z|} \bigg[f_{q,F}^{b}(E_{\widetilde{q}_z/2,n}^f)\left(1-f_{q,F}^{b}(E_{\widetilde{q}_z/2,n}^f)\right)\bigg].
 \end{align}
If we take into account the small non-extensive correction, the square bracket in the above equation can be expanded up to the leading order of $(q-1)$, yielding,  
\begin{align}
\big[\dots\big]
\approx
H_{b}^f(E_{\widetilde{q}_z/2,n}^f)+f_{ q,F,(1)}^{b}(E_{\widetilde{q}_z/2,n}^f)\left[1-2f_F^{0b}(E_{\widetilde{q}_z/2,n}^f)\right].  \end{align}
Finally, eq.~(\ref{eq:PiFquark_B}) to the order of $(q-1)^0$ is obtained as follows:
\begin{align}\label{eq:IMPi_quark_F}
  \underset{\omega\to 0}{\mathrm{lim}}\Pi^{\rm  quark }_{F,B,(0)} (Q)
=&-i\sum_{f}\sum_{n=1}^{\infty}\sum_{b=\pm}\frac{g^2}{4\pi}
\frac{8n|e_fB|^2 }{E_{\widetilde{q}_z/2,n}^f|\widetilde{q}_z|}H^f_{b}
(E_{\widetilde{q}_z/2,n}^f).
\end{align}
The associated non-extensive correction term to order $(q-1)^1$ is expressed as
 \begin{align}\label{eq:IMPi_quark_F1}
  \underset{\omega\to 0}{\mathrm{lim}}\Pi^{\rm quark}_{F,B,(1)}(Q) =-i\sum_{f}\sum_{n=1}^{\infty}\sum_{b=\pm}\frac{g^2}{4\pi}\frac{8n|e_fB|^2 }{E_{\widetilde{q}_z/2,n}^f|\widetilde{q}_z|}\frac{q-1}{2}W^f_{b}
(E_{\widetilde{q}_z/2,n}^f),
 \end{align}
where the function $  W_b^f(E_{\widetilde{q}_z/2,n}^f)$ is defined as
 \begin{equation}
W_b^f(E_{\widetilde{q}_z/2,n}^f)
=\frac{(E_{\widetilde{q}_z/2,n}^f-b \mu)(E_{\widetilde{q}_z/2,n}^f-b \mu-2T)}{T^2}\tanh\bigg(\frac{E_{\widetilde{q}_z/2,n}^f-b \mu}{2T}\bigg)H_b^f(E_{\widetilde{q}_z/2,n}^f).
 \end{equation}

Due to the electrically neutral nature of gluons, the one-loop contribution from gluons to the symmetric gluon self-energy in the magnetic field is the same as that in the zero magnetic field. Thus, the temporal component of  HTL symmetric gluon self-energy, including a small non-extensive correction, in the magnetic field is expressed as 
\begin{align}\label{eq:Pif_B}
   \Pi_{F,B}(Q)=\Pi^{\rm  quark }_{F,B,(0)}(Q)+\Pi^{\rm  quark }_{F,B,(1)}(Q)+\Pi^{\rm  gluon }_{F,(0)}(Q)+\Pi^{\rm  gluon }_{F,(1)}(Q) .
\end{align}

\begin{figure}[tb]
\centering
\subfigure{\hspace{-0mm}\includegraphics[width=0.5\textwidth]{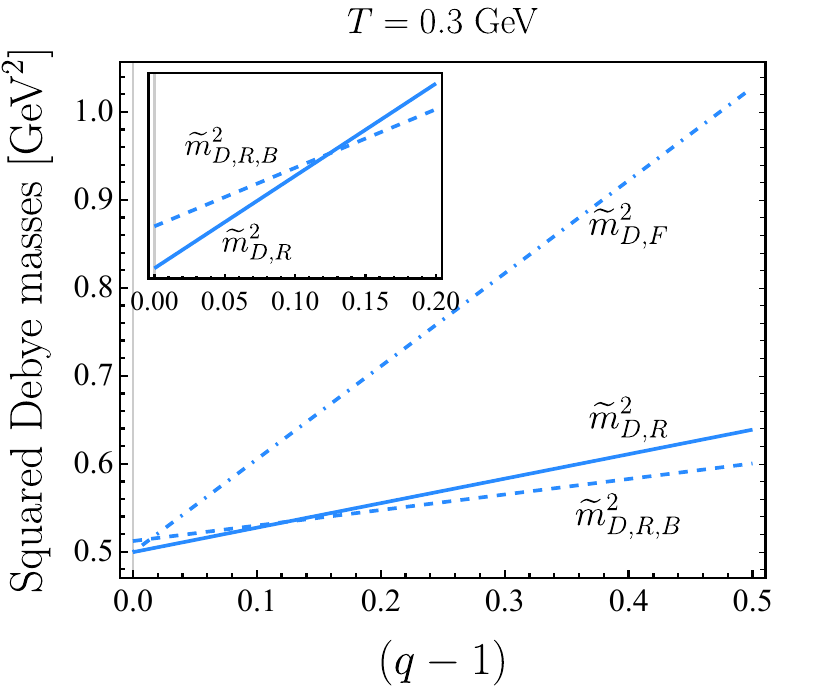}} 
\caption{ The non-extensive modified Debye masses as a function of the non-extensive parameter $(q-1)$. The solid line and dot-dashed line are the results of retarded Debye mass ($\widetilde{m}_{D,R}^2$) and symmetric Debye mass ($\widetilde{m}_{D,F}^2$) for a vanishing magnetic field ($eB=0$), respectively. The dashed line is the result of the retarded Debye mass for a finite magnetic field $eB=15~m_{\pi}^2$ ($\widetilde{m}_{D,R,B}^2$). In the illustration we compare $\widetilde{m}_{D,R}^2$ and  $\widetilde{m}_{D,R,B}^2$ in the small $(q-1)$ region.  All numerical results are performed at a fixed temperature of $T=0.3~$GeV and zero chemical potential.}
\label{Plot_Debyemass}
\end{figure}
In figure~\ref{Plot_Debyemass}, we present the non-extensive modified retarded Debye mass $\widetilde{m}_{D,R}^2$ and symmetric Debye mass $\widetilde{m}_{D,F}^2$ in the zero magnetic field, alongside with the non-extensive modified retarded Debye mass under a finite magnetic field $eB=15~m_{\pi}^2$, i.e., $\widetilde{m}_{D,R,B}^2$. We observe that all Debye masses monotonically increase with increasing $q$, indicating that the screening effect of the medium is strengthened by non-extensivity. Specially, $\widetilde{m}_{D,R}^2$ is smaller than $\widetilde{m}_{D,F}^2$, and this difference can be more pronounced in the larger $(q-1)$ region. In the absence of non-extensivity, $\widetilde{m}_{D,R,B}^2$ is larger than $\widetilde {m}_{D,R}^2$, suggesting the screening effect of the medium is increased by the presence of a magnetic field. 
From the illustration in figure~\ref{Plot_Debyemass}, we note that as $(q-1)$ increases the differece between $\widetilde{m}_{D,R,B}^2$ and $\widetilde{m}_{D,R}$ reduces.
The steeper slope of $\widetilde{m}_{D,R} ^2(q-1)$ compared to that of  $\widetilde{m}_{D,R,B}^2 (q-1)$ results in  a situation where $\widetilde{m}_{D,R}^2$ can overtake the dominance of $\widetilde{m}^2_{D,R,B}$ at larger values of $(q-1)$ for $eB=15~m_{\pi}^2$.

Inserting eq.~(\ref{eq:pir-eq}) into eq.~(\ref{eq:G^*_R(0)}), the temporal component of HTL resummed retarded gluon propagator to the order $(q-1)^0$, $G_{R,(0)}^*$, in the zero magnetic field can be determined. 
It is in the static limit ($\omega\to 0$) written as
  \begin{align}
   \underset{\omega\to 0}{\lim }G_{R,(0)}^* (Q)=
\frac{1}{\widetilde{ q}^2+m_{D}^2}-i\frac{m_D^2}{2\widetilde{q}}\frac{\pi\omega}{(\widetilde{ q}^2+m_{D}^2)^2},\label{eq:GR*_p1}
 \end{align}
 which follows from 
$\underset{\omega\to 0}{\mathrm{lim}} \left[\ln\frac{\omega+\widetilde{q}+ i\epsilon}{\omega-\widetilde{q}+ i\epsilon}\right]=- i\pi$.
By inserting  eq.~(\ref{eq:pir-nonextensive}) to  eq.~(\ref{eq:G^*_R(1)}),  the non-extensive correction term of the temporal component of resummed retarded gluon propagator in the order of $(q-1)^1$, denoted as $G_{R,(1)}^*$   within the static limit, is obtained as 
 \begin{align}
     \underset{\omega\to 0}{\rm lim} G_{R,(1)}^* (Q)=
-\frac{m_{D,R,(1)}^2}{(\widetilde{ q}^2+m_{D}^2)^2}-i\frac{m_{D,R,(1)}^2}{2\widetilde{q}}\frac{(\widetilde{q}^2-m_D^2)\omega}{(\widetilde{q}^2+m_D^2)^3}.\label{eq:GR*_p2}
 \end{align}

We explicitly determine the temporal component of the resummed symmetric gluon propagator, $G_F^*$. To facilitate calculation, we will use the following identity:   
\begin{align}
     \underset{\omega\to 0}{\mathrm{lim}} (1+2f_{q,B,(0)}(\omega))\mathrm{sgn}(\omega)&=\frac{2T}{\omega}+\dots.\label{eq:limit1}
\end{align}
Utilizing the set of equations ( (\ref{eq:pir-eq}),~(\ref{eq:Debye_R_quark}), (\ref{eq:Debye_R_gluon}),  (\ref{eq:pif-quark_NES}), (\ref{eq:pif-gluon_NES}), and (\ref{eq:limit1})) into eqs.~(\ref{eq:GF_p1}) and (\ref{eq:GF_p2}), the  expressions of $G_F^*$ up to  order  $(q-1)^0$ and order  $(q-1)^1$,  in the static limit within a zero magnetic field, are respectively derived as 
\begin{align}
  \underset{\omega\to 0}{\mathrm{lim}}G_{F,(0)}^*(Q)
  &=-i\frac{2\pi  T  
	{m}_{D}^2
}{\widetilde{q}(\widetilde{ q}^{2}+
{m}_{D}^2
)^{2}},\label{eq:GF*_p1}  \\
  \underset{\omega\to 0}{\mathrm{lim}}G_{F,(1)}^*(Q)
  &=-i\frac{m_{D,R,(1)}^2}{2\widetilde{q}}\frac{4\pi T (\widetilde{q}^2-m_D^2)}{(\widetilde{ q}^2+m_D^2)^3}-i \frac{2\pi  T  
	[{m}_{D,F,(1)}^2-{m}_{D,R,(1)}^2]
}{\widetilde{q}(\widetilde{ q}^{2}+
{m}_{D}^2
)^{2}}\label{eq:GF*_p2}.
\end{align}

In the finite magnetic field, inserting eq.~(\ref{eq:pir_gluon_eq}) and eq.~(\ref{eq:RePiq}) into eq.~(\ref{eq:G^*_R(0)}),  the temporal component of resummed retarded gluon propagator to order $(q-1)^0$, denoted as   $G^*_{R,B,(0)}$, within the static limit  is obtained as 
\begin{align}
\underset{\omega\to 0}{\mathrm{lim}}G^*_{R,B,(0)}(Q)=&\frac{1}{{\widetilde{q}}^2+{m}_{D,B}^2}-i\frac{\omega\sum_{f} g^2|e_fB|)}{4\pi({\widetilde{q}}^2+{m}_{D,B}^2)^2}\delta(\widetilde{q}_z)-i\frac{\omega\pi ({m}_{D}^{\rm gluon})^2}{2\widetilde{q}({\widetilde{q}}^2+{m}_{D,B}^2)^2}\nonumber\\
-&i
\sum_{f}\sum_{n=1}^{\infty}\sum_{b=\pm}\frac{\omega g^2 2n|e_fB|^2H_b^f(E_{\widetilde{q}_z/2,n}^f)}{4\pi T({\widetilde{q}}^2+{m}_{D,B}^2)^2|\widetilde{q}_z|E^f_{\widetilde{q}_z/2,n}}.
\label{eq:GR_B_p1}
\end{align}
Inserting the set of equations ((\ref{eq:pir_gluon_eq}), (\ref{eq:Debye_R_gluon}),~(\ref{eq:RePiq}), and (\ref{eq:Repiq1})) into eq.~(\ref{eq:G^*_R(1)}), the correction term of  temporal component of resummed retarded gluon propagator to order  $(q-1)^1$, denoted as $G^*_{R,B,(1)}$, in the static limit arrives at
\begin{align}
\underset{\omega\to 0}{\mathrm{lim}}G^*_{R,B,(1)}(Q)=&-\frac{(m_{D,R,B,(1)}^{\rm quark})^2+(m_{D,R,(1)}^{\rm gluon})^2}{({\widetilde{q}}^2+{m}_{D,B}^2)^2}-i\frac{\omega\pi ({m}_{D,R,(1)}^{\rm gluon})^2(\widetilde{q}^2+m_{D,B}^2)}{2\widetilde{q}({\widetilde{q}}^2+{m}_{D,B}^2)^3}\nonumber\\
+&i\frac{2\omega\pi ({m}_{D}^{\rm gluon})^2m_{D,R,B,(1)}^2}{2\widetilde{q}({\widetilde{q}}^2+{m}_{D,B}^2)^3}+i\frac{2\omega\sum_{f} g^2|e_fB|m_{D,R,B,(1)}^2}{4\pi ({\widetilde{q}}^2+{m}_{D,B}^2)^3}\delta(\widetilde{q}_z)\nonumber\\
-&i
\sum_{f}\sum_{n=1}^{\infty}\sum_{b=\pm}\frac{\omega g^2 
2n|e_fB|^2M_b^f(E_{\widetilde{q}_z/2,n}^f)}{4\pi T({\widetilde{q}}^2+{m}_{D,B}^2)^2|\widetilde{q}_z|E^f_{\widetilde{q}_z/2,n}}\frac{q-1}{2}\nonumber\\
+&i
\sum_{f}\sum_{n=1}^{\infty}\sum_{b=\pm}\frac{2\omega g^ |e_fB|^2H_b^f(E_{\widetilde{q}_z/2,n}^f)m_{D,R,B,(1)}^2}{4\pi T({\widetilde{q}}^2+{m}_{D,B}^2)^3|\widetilde{q}_z|E^f_{\widetilde{q}_z/2,n}}.
\label{eq:GR_B_p2}
\end{align}

Similarly, by inserting eq.~(\ref{eq:GR_B_p1}) into eq.~(\ref{eq:GF_p1}), the temporal component of resummed symmetric gluon propagator to the order of $(q-1)^0$ in  the  magnetic field, denoted as $G_{F,B,(0)}^{*}$,  within the static limit ($\omega\to 0$)  is obtained as, 
\begin{align}
\underset{\omega\to 0}{\mathrm{lim}}G_{F,B,(0)}^{*}(Q)
&=-
i\sum_{f}\sum_{n=1}^{\infty}\sum_{b=\pm}\frac{ 
g^28n|e_fB|^2H_b^{f}(E^f_{\widetilde{q}_z/2,n})}{4\pi ({\widetilde{q}}^2+m_{D,B}^2)^2|\widetilde{q}_z|E^f_{\widetilde{q}_z/2,n}}-i\frac{2T\pi  ({m}_{D}^{\rm gluon})^2}{\widetilde{q}({\widetilde{q}}^2+m_{D,B}^2)^2}\label{eq:GF_B_p1}.
\end{align}

By inserting  the set of equations ((\ref{eq:pir_gluon_eq}), (\ref{eq:Debye_R_gluon}), (\ref{eq:pif_gluon_eq}), (\ref{eq:pif-gluon_NES}), (\ref{eq:Impiq}), (\ref{eq:Impiq1}),(\ref{eq:IMPi_quark_F}), (\ref{eq:IMPi_quark_F1}),  (\ref{eq:GR_B_p1}), and (\ref{eq:GF_B_p2})) into eq.~(\ref{eq:GF_p2}), 
one can finally get the non-extensive correction term of the temporal component of resummed symmetric gluon propagator to order $(q-1)^1$ in a magnetic field, denoted as $G_{F,B,(1)}^{*}$, which in the static limit ($\omega\to 0$) is expressed as 
\begin{align}
\underset{\omega\to 0}{\mathrm{lim}}G_{F,B,(1)}^{*}(Q)
&=-i\frac{2T\pi ({m}_{D,R,(1)}^{\rm gluon})^2}{\widetilde{q}({\widetilde{q}}^2+{m}_{D,B}^2)^2}- i\frac{2T\pi 
 \left[(m^{\rm gluon}_{D, F,(1)})^2-(m^{\rm gluon}_{D, R,(1)})^2\right]}{\widetilde{q}({\widetilde{q}}^2+{m}_{D,B}^2)^2}\nonumber\\
&+i\frac{4T\pi ({m}_{D}^{\rm gluon})^2m_{D,R,B,(1)}^2}{\widetilde{q}({\widetilde{q}}^2+{m}_{D,B}^2)^3}+i\frac{8T\sum_{f} g^2|e_fB|
m_{D,R,B,(1)}^2}{4\pi ({\widetilde{q}}^2+{m}_{D,B}^2)^3}\delta(\widetilde{q}_z)\nonumber\\
&+i\sum_{f}\sum_{n=1}^{\infty}\sum_{b=\pm}\frac{ g^2 16n|e_fB|^2H_b^f(E^f_{\widetilde{q}_z/2,n})m_{D,R,B,(1)}^2}{4\pi ({\widetilde{q}}^2+{m}_{D,B}^2)^3|\widetilde{q}_z|E^f_{\widetilde{q}_z/2,n}}\nonumber\\
&-i\sum_{f}\sum_{n=1}^{\infty}\sum_{b=\pm}
\frac{g^2 8n|e_fB|^2 W^f_{b}
(E_{\widetilde{q}_z/2,n}^f)}{4\pi ({\widetilde{q}}^2+{m}_{D,B}^2)^2|\widetilde{q}_z|E_{\widetilde{q}_z/2,n}^f}\frac{q-1}{2}.\label{eq:GF_B_p2}
\end{align}
Compared to eq.~(\ref{eq:GF_B_p1}), we note that the term related to the vacuum contribution from quark-loop to the symmetric gluon self-energy is present in the non-extensive correction term.

\section{Heavy quark potential in a non-extensive QGP for zero and finite magnetic fields}
\label{sec:potential}
Applying the non-extensive modified resummed gluon propagator in the static limit, we investigate the in-medium heavy quark potential, which describes the interactions between a quark and its antiquark within a QGP medium.
To facilitate the use of non-relativistic approaches to study the in-medium properties of heavy quarkonia such as binding energies, decay widths, and melting temperatures, we first calculate the heavy quark potential in a non-extensive QGP in both zero and finite magnetic fields.
In a vacuum, the heavy quark potential can be well parameterized by the Cornell potential~\cite{Eichten:1974af, Matsui:1986dk}:
\begin{align}\label{eq:CornellV}
    V_{\rm Cornell}(r)=-C_F\alpha_s/r+\sigma r,
\end{align}
where $r\equiv|\bm{r}|$ denotes the separation distance between heavy quark and antiquark, $\alpha_s$ represents the strong coupling constant,
$C_{F}=(N_c^2-1)/2N_c$, $ \sigma $ is the string tension which is determined to reproduce the vacuum quarkonium property~\cite{Jacobs:1986gv}. The first term corresponds to the Coulombic part, reflecting the asymptotic freedom at small separation distances, while the second term represents the string-like part, responsible for color confinement at large separation distances. In the QGP, the in-medium heavy quark potential can be obtained by modifying the vacuum potential with the dielectric permittivity~\cite{Thakur:2013nia, Agotiya:2008ie, Thakur:2016cki}. 

\subsection{Dielectric permittivity}
As described in refs.~\cite{Thakur:2013nia, Agotiya:2008ie, Thakur:2016cki}, the dielectric permittivity $\varepsilon$,  which encodes in-medium effects, such as finite temperature and non-extensive effects, is obtained by using the temporal component of the 11-part of the HTL resummed gluon propagator. 
It is expressed as 
\begin{equation}\label{eq:permittivity}
\varepsilon^{-1}(\widetilde{{q}})
=\lim_{\omega  \to 0} \widetilde{q}^2 G_{11}^*(Q)=\lim_{\omega  \to 0}  \widetilde{q}^2\left(  G_R^*(Q)  + G_A^* (Q)+ G_F^* (Q) \right)/2
\end{equation}
where $(G_R^*(Q)+G_A^*(Q))/2=\mathrm{Re}\, G_{R}^*(Q)$.
 By inserting eqs.~(\ref{eq:GR*_p1}), (\ref{eq:GR*_p2}), (\ref{eq:GF*_p1}), and (\ref{eq:GF*_p2}) into eq.~(\ref{eq:permittivity}),
the non-extensive modified dielectric permittivity in the presence of small non-extensivity for vanishing magnetic field can be determined as:
\begin{align}
\varepsilon^{-1}({\widetilde{q}}) 
&= 
\frac{\widetilde{q}^2}{{\widetilde{q}}^2 + {m}^2_{D}}-\frac{\widetilde{q}^2 m_{D,R,(1)}^2}{({\widetilde{q}}^2 + {m}^2_{D})^2}
-i\frac{\pi T \widetilde{q} 
	({m}_{D}^{\rm gluon})^2
}{(\widetilde{q}^{2}+
{m}_{D}^2
)^{2}} -i\frac{\pi T\widetilde{q} (\widetilde{q}^2-m_D^2)}{(\widetilde{q}^2+m_D^2)^3}(m_{D,R,(1)}^{\rm gluon})^2 \nonumber\\
&  - i\frac{\pi T \widetilde{q}
	\left[({m}_{D,F,(1)}^{\rm gluon})^2-({m}_{D,R,(1)}^{\rm gluon})^2\right]
}{(\widetilde{q}^{2}+
{m}_{D}^2
)^{2}}-i\frac{\pi T \widetilde{q} 
({m}_{D}^{\rm quark})^2}{(\widetilde{q}^{2}+
{m}_{D}^2
)^{2}} -i\frac{\pi T\widetilde{q} (\widetilde{q}^2-m_D^2)}{(\widetilde{q}^2+m_D^2)^3}(m_{D,R,(1)}^{\rm quark})^2
 \nonumber\\
& - i\frac{\pi  T  \widetilde{q}
	\left[({m}_{D,F,(1)}^{\rm quark})^2-({m}_{D,R,(1)}^{\rm quark})^2\right]
}{(\widetilde{q}^{2}+
{m}_{D}^2
)^{2}}.
\label{eq:epsilon}
\end{align}
As $q$ approaches 1,
${m}_{D,R,(1)}^{\rm quark/gluon}$ and ${m}_{D, F,(1)}^{\rm quark/gluon}$  vanish, then the standard equilibrium form of $\varepsilon^{-1}(\widetilde{q})$ is reproduced. 
Similarly, by putting eqs.~(\ref{eq:GR_B_p1}), (\ref{eq:GR_B_p2}), (\ref{eq:GF_B_p1}), and (\ref{eq:GF_B_p2}) into eq.~(\ref{eq:permittivity}), the non-extensive modified dielectric permittivity for the finite magnetic field is written as
\begin{align}
\varepsilon^{-1}_{B}({\widetilde{q}}) =&
\frac{\widetilde{q}^2}{{\widetilde{q}}^2+{m}_{D,B}^2}-\frac{\widetilde{q}^2m_{D,R,B,(1)}^2}{({\widetilde{q}}^2+{m}_{D,B}^2)^2}-i\frac{\pi T  \widetilde{q}({m}_{D}^{\rm gluon})^2}{({\widetilde{q}}^2+m_{D,B}^2)^2} -i\frac{\pi T  \widetilde{q}
 (m^{\rm gluon}_{D, F,(1)})^2
}{({\widetilde{q}}^2+{m}_{D,B}^2)^2}
\nonumber\\
+&i\frac{2\pi T \widetilde{q}({m}_{D}^{\rm gluon})^2m_{D,R,B,(1)}^2}{({\widetilde{q}}^2+{m}_{D,B}^2)^3} -i\sum_{f}\sum_{n=1}^{\infty}\sum_{b=\pm}\frac{ 
\alpha_s4n|e_fB|^2H_b^{f}(E^f_{\widetilde{q}_z/2,n}) \widetilde{q}^2}{({\widetilde{q}}^2+m_{D,B}^2)^2|\widetilde{q}_z|E^f_{\widetilde{q}_z/2,n}} 
\nonumber\\
+&i\frac{4T\sum_{f} \alpha_s|e_fB|
m_{D,R,B,(1)}^2\widetilde{q}^2}{({\widetilde{q}}^2+{m}_{D,B}^2)^3}\delta(\widetilde{q}_z)\nonumber\\
+&i
\sum_{f}\sum_{n=1}^{\infty}\sum_{b=\pm}\frac{ \alpha_s 
8n|e_fB|^2H_b^f(E^f_{\widetilde{q}_z/2,n})m_{D,R,B,(1)}^2\widetilde{q}^2}{({\widetilde{q}}^2+{m}_{D,B}^2)^3|\widetilde{q}_z|E^f_{\widetilde{q}_z/2,n}}\nonumber\\
-&i\sum_{f}\sum_{n=1}^{\infty}\sum_{b=\pm}
\frac{\alpha_s4n|e_fB|^2 W^f_{b}
(E_{\widetilde{q}_z/2,n}^f)\widetilde{q}^2}{({\widetilde{q}}^2+{m}_{D,B}^2)^2|\widetilde{q}_z|E_{\widetilde{q}_z/2,n}^f}\frac{q-1}{2}.
\label{eq:epsilon_B}
\end{align}

\subsection{Real part of in-medium heavy quark potential}
Following the approach proposed in \cite{Thakur:2013nia,Agotiya:2008ie, Thakur:2016cki}, the in-medium heavy quark potential in the non-extensive QGP can be determined through the convolution of the Cornell potential and the  non-extensive modified dielectric permittivity,
\begin{equation}\label{eq:heavy_potential_q}
V({\widetilde{q}}) = V_{\rm Cornell}  ({\widetilde{q}} )  \varepsilon^{-1}({\widetilde{q}} ) . 
\end{equation}
Here, the Fourier transform of the Cornell potential in the momentum space,
$V_{\rm Cornell} ({\widetilde{q}})$, is given as~\cite{Thakur:2013nia} 
\begin{equation}
V_{\rm Cornell} (\widetilde{q})=-\sqrt{(2/\pi)} \frac{C_{F} \alpha_{s}}{\widetilde{q}^2}-\frac{4\sigma}{\sqrt{2 \pi} \widetilde{ q}^4}.
\label{eq:vcornell}
\end{equation} 
Through the Fourier transform, eq.~(\ref{eq:heavy_potential_q}) is transformed into real coordinate space, which is expressed as
\begin{eqnarray}
\label{defn}
V(\bm{r})&=&\int \frac{d^3 \bm{\widetilde{q}}}{{(2\pi)}^{3/2}}
(e^{i\bm{\widetilde{\bm q}} \cdot \bm{r}}-1)V_{\rm Cornell} (\widetilde{q})\varepsilon^{-1}({\widetilde{q}}) .\label{eq:potential}
\end{eqnarray}
Without loss of generality,    $\widetilde{\bm{q}} 
$  and  $\bm{r}
$ are  chosen as $\bm{\widetilde{q}}=\widetilde{q}(\sin\theta\cos\phi,\sin \theta\sin\phi,\cos\theta)$ and $\bm{r}=r(\sin\chi, 0, \cos\chi)$, respectively.
Here, $\chi$ denotes the angle between $\bm{\widetilde{q}}$  and $\bm{r}$. The dot product $\bm{\widetilde{q}}\cdot\bm{r}$ is given by $\widetilde{q}r(\sin\chi\cos\phi\sin\theta+\cos\chi\cos\theta)$. Consequently, eq.~(\ref{defn}) can be rewritten as 
\begin{align}\label{eq:V_general}
V(r,\chi,q,T,eB)= \int \frac{ \widetilde{q}^2 \sin\theta d\theta d\widetilde{q} }{(2\pi)^{1/2}}\left[J_{0}(\widetilde{q}r\sin\theta\sin\chi)e^{i\widetilde{q}r\cos\theta\cos\chi}-1\right] V_{\rm Cornell}  ({\widetilde{q}} )  \varepsilon^{-1}({\widetilde{q}} ),
\end{align}
where $J_{m}(x)$ is the Bessel function of the first kind. 

In the present work, the string tension is chosen as $\sigma=0.18~\mathrm{GeV}^2$~\cite{Satz:2005hx}. The running  strong coupling  strength $g$ is obtained from~\cite{Ayala:2018wux}
\begin{eqnarray}
\alpha_s(\Lambda^2,eB)=\frac{g^2}{4\pi}=\frac{\alpha_s(\Lambda^2)}{1+\frac{11N_c-2N_f}{12\pi}\alpha_s(\Lambda^2)\ln (\frac{\Lambda^2}{\Lambda^2+eB})},
\end{eqnarray}
where the  one-loop QCD strong  coupling constant $\alpha_s$ for $eB=0$ is given as $\alpha_s(\Lambda^2)=\frac{12\pi}{(11N_c-2N_f)\ln\left(\Lambda^2/\Lambda_{\bar{MS}}^2\right)}$ at $\Lambda_{\bar{MS}}=176~\mathrm{MeV}$ for $N_f=N_c=3$~\cite{Bazavov:2012ka}. The  scale $\Lambda$ is taken as $2\pi\sqrt{T^2+\mu^2/\pi^2}$ for quarks and $2\pi T$ for gluons.

By inserting the real part of eq.~(\ref{eq:epsilon}) into eq.~(\ref{eq:V_general}), we compute the real part of the potential,  denoted as $\mathrm{Re}\,V$, in the absence of a magnetic field. The expression of $\mathrm{Re}\,V$ to order $(q-1)^0$ is written as follows:
\begin{align}
\mathrm{Re}\, V_{(0)}(r,T)
&=\mathrm{Re} \,V_{ \rm HTL,(0)}^{}(r,T)+\mathrm{Re}\, V_{\rm string,(0)}^{}(r,T) \label{eq:general_ReV(0)}\\
&= 
-C_F\alpha_s \, 
{m}_{D}
\left(\frac{e^{-{m}_{D}\, r}}{{m}_{D}\, r}+1\right) 
+ 
\frac{2\sigma}{
{m}_{D} 
}\left(\frac{e^{-{m}_{D}\,{r}}-1}{{m}_{D}\,r}+1\right).
\label{eq:ReV(0)}
\end{align}
The non-extensive correction term  of $\mathrm{Re}\,V$ to order $(q-1)^1 $ is obtained as 
\begin{align}
\mathrm{Re}\, V_{(1)}(r,q,T)
&=\mathrm{Re} \,V_{\rm HTL,(1)}^{}(r,q,T)+\mathrm{Re}\, V_{\rm string,(1)}^{}(r,q,T)\\
&= C_F\alpha_s\frac{m_{D,R,(1)}^2}{2m_D }\left( e^{-m_D r}-1\right)\nonumber\\
&+\frac{\sigma m_{D,R,(1)}^2}{m_D^3}\left(\frac{2-(2+m_Dr)e^{-m_Dr}}{m_D r}-1\right).
\label{eq:ReV(1)}
\end{align}
Similarly,  by inserting the real part of eq.~(\ref{eq:epsilon_B}) into eq.~(\ref{eq:V_general}), the $\mathrm{Re}V$ in the presence of a magnetic field  is also  computed, and  to order $(q-1)^0$ it can be written as 
\begin{align}
\mathrm{Re} \,V_{(0)}(r,T, eB)
&= \mathrm{Re}\, V_{\rm HTL,(0)}^{}(r,T, eB)+\mathrm{Re}\, V_{\rm string,(0)}^{}(r, T,eB)\\
&= 
-C_F\alpha_s \, {m}_{D,B}
\left(\frac{e^{-{m}_{D,B}\, r}}{{m}_{D,B}\, r}+1\right) 
+ \frac{2\sigma}{{m}_{D,B} 
}\left(\frac{e^{-{m}_{D,B}\,{r}}-1}{{m}_{D,B}\,r}+1\right).\label{eq:ReV(0)_B}
\end{align}
 To order  $(q-1)^1$, its non-extensive correction term is obtained as 
\begin{align}
\mathrm{Re}\, V_{(1)}(r,q, T,eB)
&= \mathrm{Re}\, V_{\rm HTL,(1)}^{}(r, q,T, eB)+\mathrm{Re} \,V_{\rm string,(1)}^{}(r,q, T,eB), \label{eq:general_ReV(1)_B}\\
&=C_F\alpha_s\frac{m_{D,R,B,(1)}^2}{2m_{D,B} }\left( e^{-m_{D,B} r}-1\right)\nonumber\\
&+\frac{\sigma m_{D,R,B,(1)}^2}{m_{D,B}^3}\left(\frac{2-(2+m_{D,B}r)e^{-m_{D,B}r}}{m_{D,B} r}-1\right) .\label{eq:ReV(1)_B}
\end{align}
The first and second terms on the right side of eqs.~(\ref{eq:general_ReV(0)})-(\ref{eq:ReV(1)_B}) are the HTL  and string-like parts of the potential, which are related to the short-range perturbative Yukawa and long-range non-perturbative color confining or string-like interactions between heavy quarks and the QGP medium, respectively. 

\begin{figure}[htbp]
\centering
\subfigure{\includegraphics[width=0.34\textwidth]{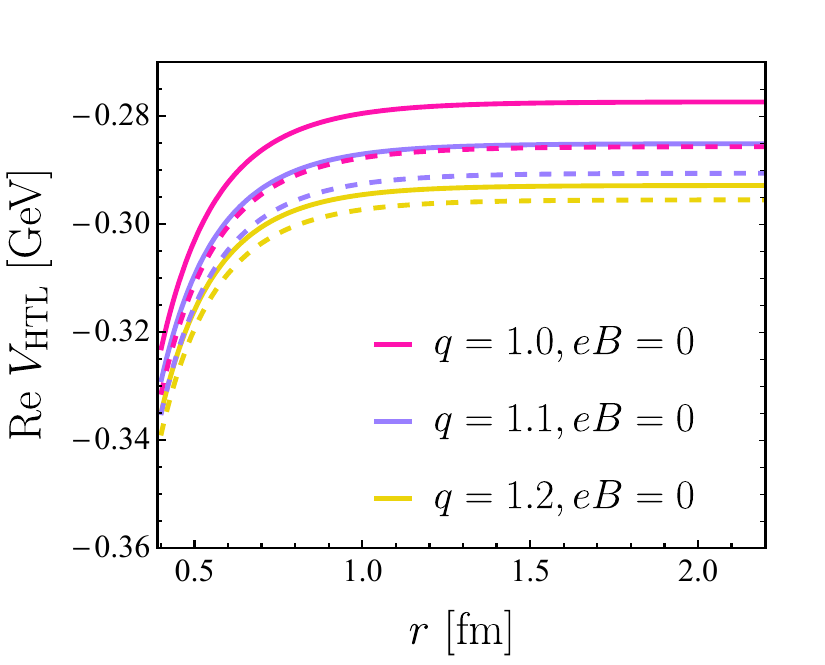}\includegraphics[width=0.32\textwidth]{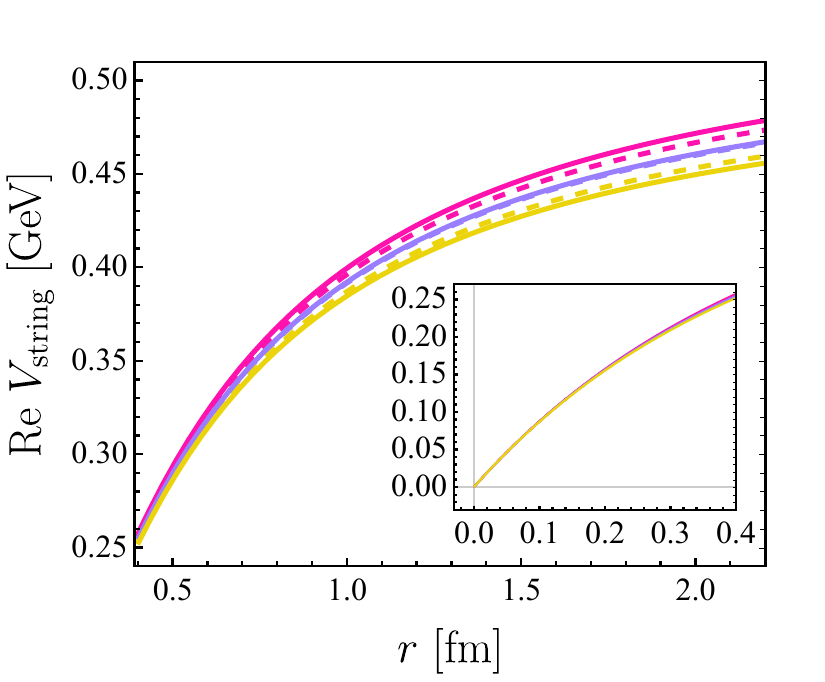}\includegraphics[width=0.33\textwidth]{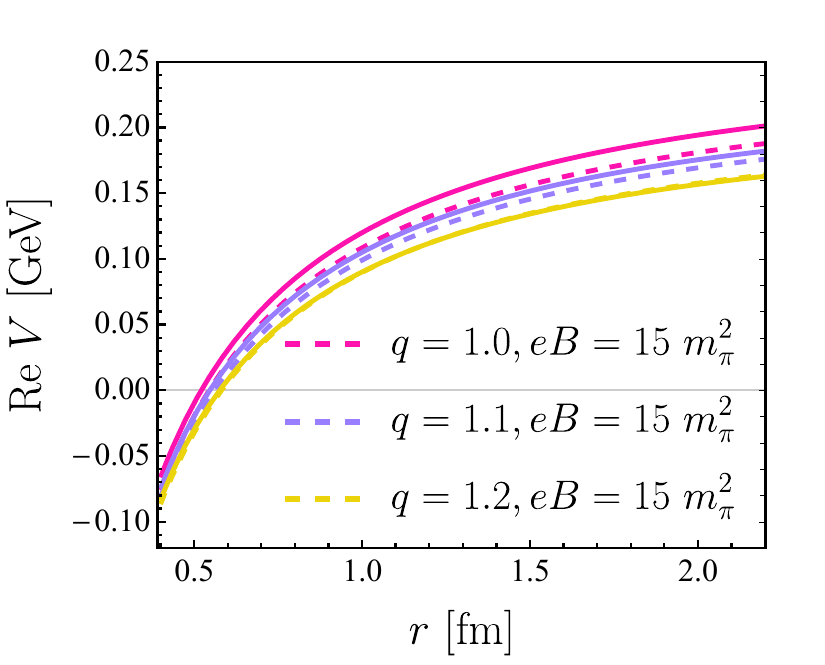}} 

\vspace{-5 mm}
\subfigure{\includegraphics[width=0.33\textwidth]{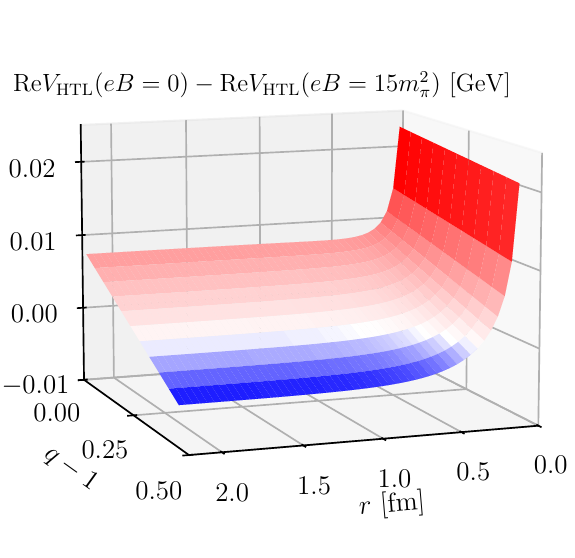}
\includegraphics[width=0.33\textwidth]{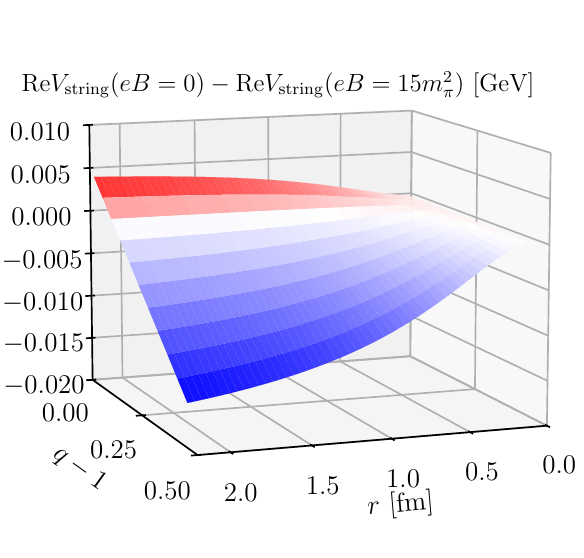}
\includegraphics[width=0.33\textwidth]{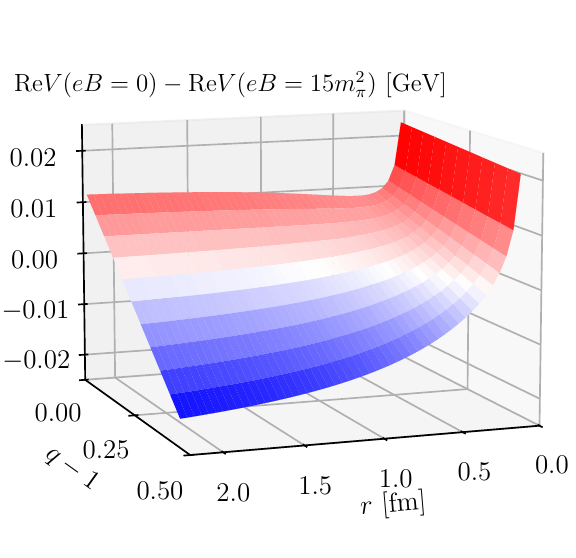}} 
\caption{
(Upper panel) The  HTL component (left),  string component (middle), and total component (right) of the real part of heavy quark potential, $\mathrm{Re}\,V$, as functions of quark-antiquark separation distance $r$ at different values of non-extensive parameter $q$. The solid lines and dashed lines are the results for a zero magnetic field ($eB=0$) and the results for a finite magnetic field ($eB=15~m_{\pi}^2$), respectively.  (Lower panel)  The 3D figures of the difference in the various components of $\mathrm{Re}\,V$ between zero and finite magnetic fields, as a function of both $r$ and $(q-1)$. All numerical results are obtained at a fixed temperature of $T=0.3~$GeV and zero chemical potential.
}\label{Plot_ReV}
\end{figure}

In figure~\ref{Plot_ReV}, 
we display the real part of the heavy quark potential, $\mathrm{Re}\,V$, 
for different values of non-extensive parameter $q$  with varying quark-antiquark separation distances $r$ in both $eB=0$ (solid lines) and $eB=15~m_{\pi}^2$ (dashed lines). 
The magnetic field influences $\mathrm{Re}\,V$ by affecting the quark contributions to the gluon self-energy and the QCD coupling constant. 
The non-extensive correction alters $\mathrm{Re}\, V$  by modifying the Debye masses, transforming $m_D$ and  $m_{D,B}$ into $\widetilde{m}_{D, R}$ and $\widetilde{m}_{D, R,B}$, respectively.
To gain a deeper understanding of the qualitative behavior of $\mathrm{Re}\,V$ under these conditions, we also construct three-dimensional plots to visualize the difference in $\mathrm{Re}\,V$ between scenarios with zero and finite magnetic fields. This difference is depicted as a function of the variables  $r$ and $(q-1)$.  In these 3D plots, regions exhibiting positive differences are colored red for clarity, while those with negative differences are colored blue. 
In the upper panel of figure~\ref{Plot_ReV}, we observe that both  HTL and string components of $\mathrm{Re}\,V$ ($\mathrm{Re}\,V_{\rm HTL}$  and $\mathrm{Re}\,V_{\rm string}$) increase rapidly at first and then gradually flattens out as $r$ increases.
In the absence of non-extensivity $(q=1)$, both $\mathrm{Re}\,V_{\rm HTL}$ and $\mathrm{Re}\,V_{\rm string}$ become flatter (more screened) in a nonzero magnetic field compared to that in a zero magnetic field. 
Furthermore, introducing non-extensivity also leads to a shorter Debye screening length (or equivalently, a larger Debye screening mass), which results in both $\mathrm{Re}\,V_{\rm HTL}$ and $\mathrm{Re}\,V_{\rm string}$ flattening with increasing $q$. 
Compared with the case when $q=1$, we note that the difference in $\mathrm{Re}\,V_{\rm HTL}$ and the difference in  $\mathrm{Re}\,V_{\rm string}$ between the zero magnetic field and finite magnetic field becomes smaller for $q=1.2$. 
As the $q$ further increases, this difference can diminish and change sign from negative to positive, as shown in the lower panel of figure~\ref{Plot_ReV}. 
Summing $\mathrm{Re}\,V_{\rm HTL}$ and $\mathrm{Re}\,V_{\rm string}$, we observe that the qualitative feature of $\mathrm{Re}V_{}$  in the small $r$ region is dominated primarily by $\mathrm{Re}\,V_{\rm HTL}$, while  in the large $r$ region is governed by  $\mathrm{Re}\,V_{\rm string}$.

\subsection{Imaginary  part of in-medium heavy quark potential}

Next, we study the imaginary part of in-medium heavy quark potential, denoted as  $\mathrm{Im }V$, which relates to the inelastic scattering of the light constituents of the medium with heavy quarkonium via exchanged gluons (Landau damping phenomenon) in HTL resummed perturbation theory~\cite{Brambilla:2011sg}.
A larger magnitude of $\mathrm{Im}\,V$ could trigger a broadening decay width for quarkonium states, which plays an important role in quarkonium dissociation. Since the gluonic contributions and quark contributions to the gluon self-energy, particularly in a finite magnetic field are strikingly different in form. Therefore, we will separately analyze their respective contributions to the $\mathrm{Im}\,V$, to better elucidate the qualitative features of the potential.
Accordingly, the $\mathrm{Im}\, V$  can be decomposed into $\mathrm{Im} \,V=\mathrm{Im}\, V^{\mathrm{gluon}}+\mathrm{Im} \,V^{\mathrm{quark}}$.

First, we look at the imaginary part of the potential associated with the gluonic contribution to gluon self-energy, denoted as $\mathrm{Im}\,V^{\rm gluon}$. 
By inserting the third term in the right-hand side of eq.~(\ref{eq:epsilon}) to eq.~(\ref{eq:V_general}),  $\mathrm{Im}\,V^{\rm gluon}$ to order $(q-1)^0$ in a zero magnetic field is computed as
  \begin{align}
    \mathrm{Im}\,V^{\rm gluon}_{(0)}(r,T)&=\mathrm{Im}\,V_{\rm HTL,(0)}^{\rm gluon}(r,T)+\mathrm{Im}\,V_{\rm string,(0)}^{\rm gluon}(r,T)\\
&=-C_F\alpha_sT\frac{({m}_{D}^{\rm gluon})^2}{{m}_{D}^2}\phi_2({m}_{D}r)-\frac{2\sigma T ({m}_{D}^{\rm gluon})^2}{{m}_{D}^4}\chi_2({m}_{D} r)\label{eq:ImV_gluon(0)}.
\end{align}  
Then, inserting the fourth and fifth terms in the right-hand side of eq.~(\ref{eq:epsilon})  to eq.~(\ref{eq:V_general}), the non-extensive correction term of  $\mathrm{Im}\,V^{\rm gluon}$ in a zero  magnetic field is obtained  as
 \begin{align}
    \mathrm{Im~V}^{\rm gluon}_{(1)}(r,q,T)&=\mathrm{Im}V_{\rm HTL,(1)}^{\rm gluon}(r,q,T)+\mathrm{Im}V_{\rm string,(1)}^{\rm gluon}(r,q,T),\\
&=-C_F\alpha_sT\bigg[\frac{(m_{D,R,(1)}^{\rm gluon})^2}{m_D^2}\bigg(\psi(m_D r)-\phi_2(m_D r)-\phi_3(m_D r)\bigg)\nonumber\\
&+\frac{(m_{D,F,(1)}^{\rm gluon})^2}{{m}_{D}^2}\phi_2({m}_{D}r)\bigg]-\frac{2\sigma T}{m_D^2}\bigg[\frac{(m_{D,F,(1)}^{\rm gluon})^2}{{m}_{D}^2}\chi_2({m}_{D} r)\nonumber\\
&+\frac{ ({m}_{D,R,(1)}^{\rm gluon})^2}{{m}_{D}^2}\bigg(\phi_3(m_D r)-\chi_2(m_D r)-\chi_3(m_D r)\bigg)\bigg]\label{eq:ImV_gluon(1)}.
\end{align}  
In the presence of a magnetic field, by inserting eq.~(\ref{eq:epsilon_B})  to eq.~(\ref{eq:V_general}), $\mathrm{Im}V_{(0)}^{\rm gluon}$ and  $\mathrm{Im}V_{(1)}^{\rm gluon}$ are computed as 
\begin{align}
\mathrm{Im}\,V_{(0)}^{\rm gluon}(r,T,eB)&=\mathrm{Im}\,V_{\rm HTL,(0)}^{\rm gluon}(r,T,eB)+\mathrm{Im}\,V_{\rm string,(0)}^{\rm gluon}(r,T,eB)\\
&=-C_F\alpha_sT\frac{({m}_{D}^{\rm gluon})^2
}{{m}_{D,B}^2}\phi_2({m}_{D,B}r)
-\frac{2\sigma T ({m}_{D}^{\rm gluon})^2}{{m}_{D,B}^4}\chi_2({m}_{D,B} r)\label{eq:ImV_gluon_B(0)},
\end{align}
and
\begin{align}
\mathrm{Im}\,V_{(1)}^{\rm gluon}(r,q,T, eB)
&=\mathrm{Im}\,V_{\rm HTL,(1)}^{\rm gluon}(r,q,T ,eB)+\mathrm{Im}\,V_{\rm string,(1)}^{\rm gluon}(r,q,T, eB)\\
&=-C_F\alpha_sT\bigg[\frac{(m_{D,F,(1)}^{\rm gluon})^2
}{{m}_{D,B}^2}\phi_2({m}_{D,B}r)\nonumber\\
&-\frac{2m_{D,R,B,(1)}^2(m_{D}^{\rm gluon})^2}{m_{D,B}^4}\phi_3(m_{D,B} r)\bigg]-\frac{2\sigma T}{m_{D,B}^2}\bigg[\frac{ (m_{D,F,(1)}^{\rm gluon})^2
}{{m}_{D,B}^2}\chi_2({m}_{D,B} r)\nonumber\\
&-\frac{2 {m}_{D,R,B,(1)}^2({m}_{D}^{\rm gluon})^2}{{m}_{D,B}^4}\chi_3(m_{D,B} r)\bigg]\label{eq:ImV_gluon_B(1)},
\end{align} 
respectively. It is clear that eq.~(\ref{eq:ImV_gluon_B(0)}) is formally consistent with eq.~(\ref{eq:ImV_gluon(0)}) except for the Debye mass. The functions $\phi_n(x)$ and $\chi(x)$ are defined by 
$\phi_n (x)
\equiv 
2 \int_{0}^{\infty} dz
\frac{z}{(z^{2}+1)^{n}}\left[1-\frac{\sin(x z)}{x z}\right] $,
$
\chi_n (x) 
\equiv
2 \int_{0}^{\infty}  \frac{dz}{z(z^{2}+1)^{n}}
\left[1-\frac{\sin(x z)}{x z}\right] $, respectively, as well as $\psi(x)=2\int_0^{\infty}\frac{z^3}{(z+1)^3}\left[1-\frac{\sin(x z)}{x z}\right]$. Comparing eq.~(\ref{eq:ImV_gluon(1)}) and eq.~(\ref{eq:ImV_gluon_B(1)}), we note that the non-extensive correction terms of $\mathrm{Im}\,V^{\rm gluon} $ in the absence and presence of a magnetic field are different.

\begin{figure}[tb]	\centering	\subfigure{\includegraphics[width=0.335\textwidth]{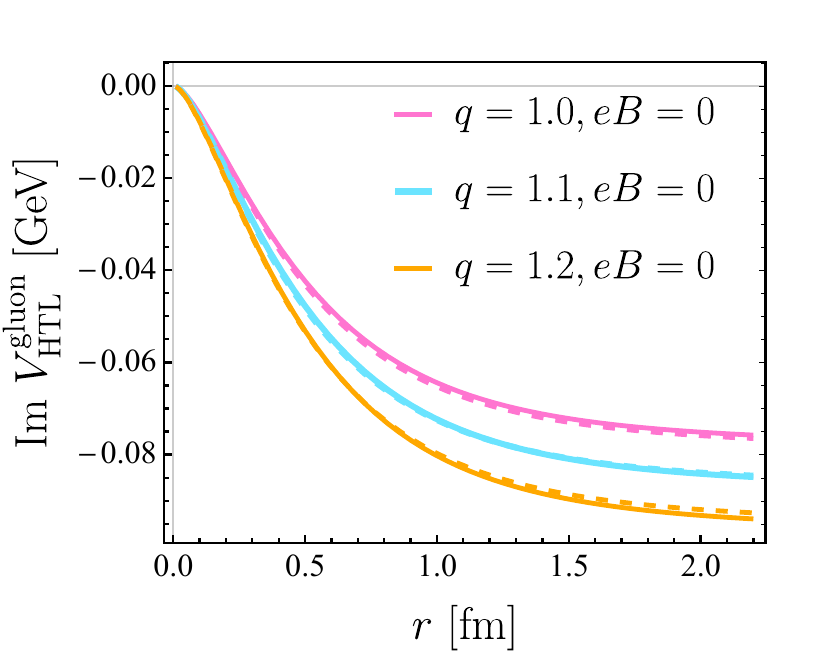}\includegraphics[width=0.3275\textwidth]{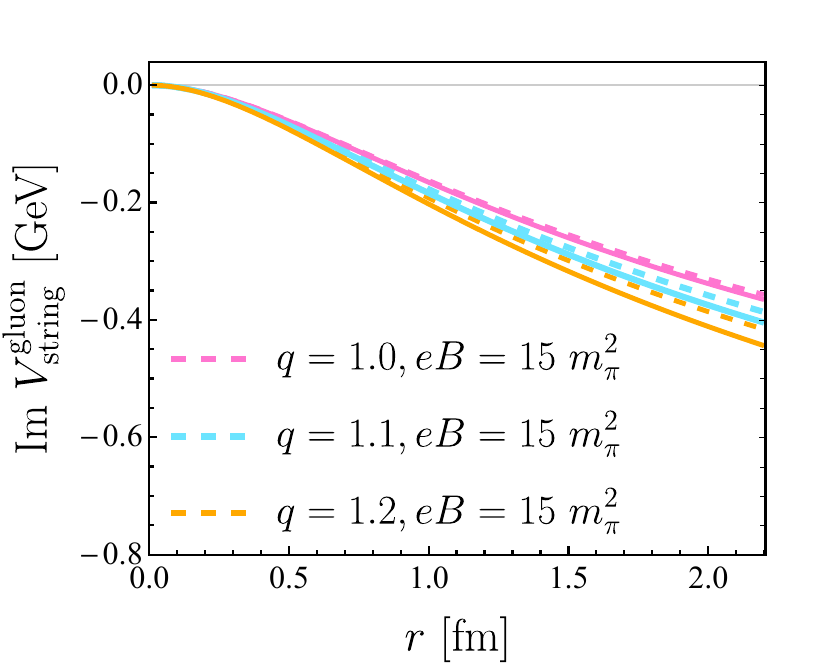}\includegraphics[width=0.3275\textwidth]{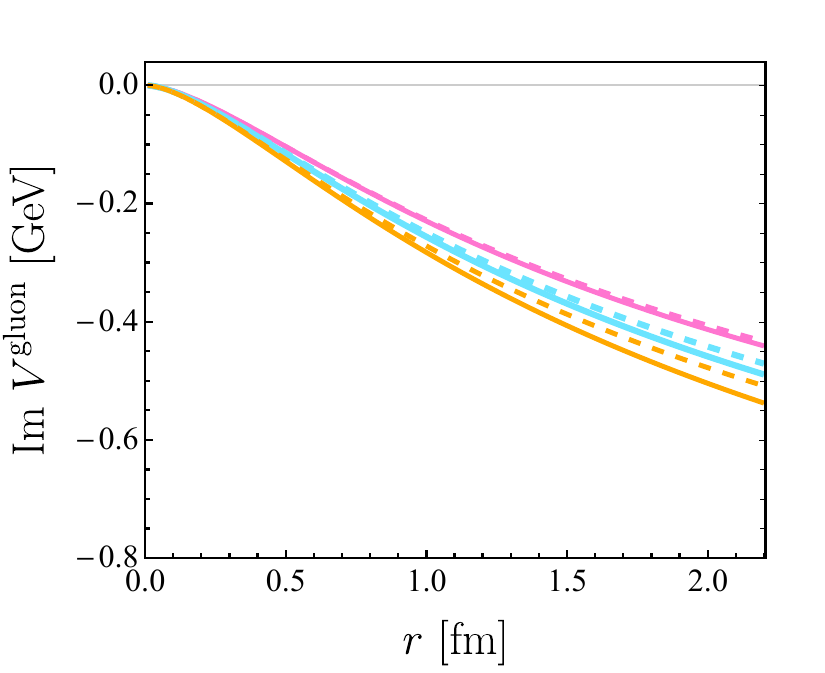}}

\vspace{-5 mm}
\subfigure{\includegraphics[width=0.33\textwidth]{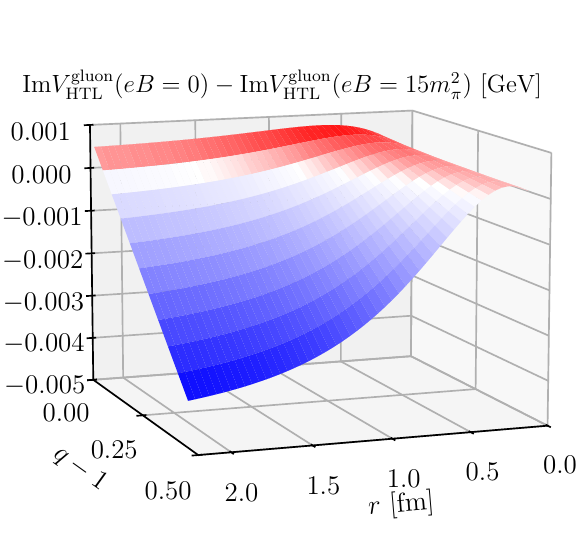}
\includegraphics[width=0.33\textwidth]{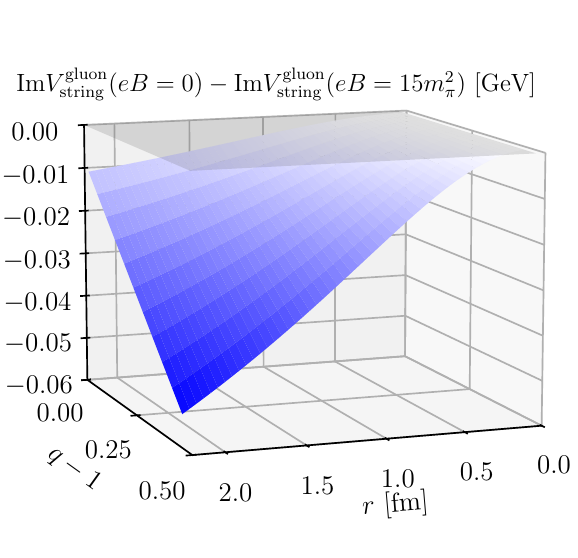}
\includegraphics[width=0.33\textwidth]{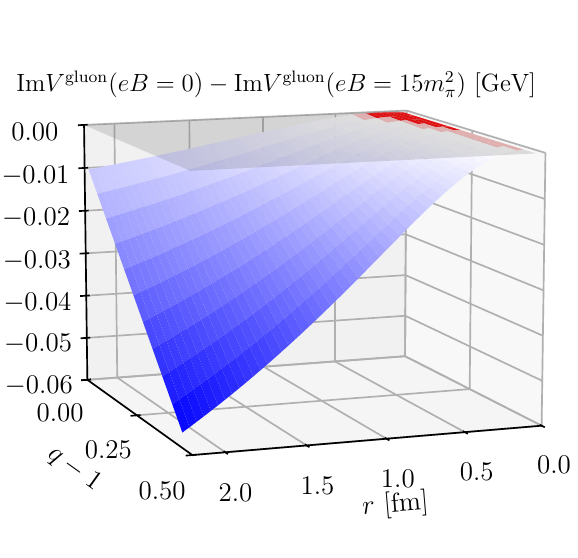}} 
\caption{Same as figure~\ref {Plot_ReV}  but for the various components of the imaginary part of the heavy quark potential originated from the gluonic contributions to the gluon self-energy, $\mathrm{Im}\,V^{\rm gluon}$.}\label{Plots:ImV_gluon}
\end{figure}

In the upper panel of figure~\ref{Plots:ImV_gluon}, we show the dependence of the imaginary part of in-medium heavy quark potential originated from the gluonic contributions to the gluon self-energy, $\mathrm{Im} V^{\rm gluon}$, on the quark-antiquark separation distance $r$ at different values of non-extensive parameter $q$.  Results are presented for $eB=0$ (solid lines) and  $eB=15~m_{\pi}^2$ (dashed lines) at $T=0.3~$GeV. 
It is clear that the magnitudes of both the HTL and string components of $\mathrm{Im}\,V^{\rm gluon}$ exhibit an increasing trend with $r$ and $q$.
In the absence of non-extensivity $(q=1)$, the magnetic field dependence of $\mathrm{Im}\, V^{\rm gluon}$ is also minimal, especially in the range $B=0$ to $eB=15~m_{\pi}^2$. When the non-extensive correction is considered, the presence of a magnetic field reduces the magnitude of  $\mathrm{Im}\, V_{\rm string}^{\rm gluon}$, whereas $\mathrm{Im}\, V_{\rm HTL}^{\rm gluon}$ is still insensitive to the magnetic field. 
This can also be understood from the lower panel of figure~ \ref{Plots:ImV_gluon}. We see that in the large $r$ region, the gradually increasing non-extensive correction raises the absolute value of $\mathrm{Im}\, V_{\rm HTL}^{\rm gluon}$ in the absence of a magnetic field, which is slightly more pronounced than that in the presence of a magnetic field. The absolute value of $\mathrm{Im}\, V_{\rm string}^{\rm gluon}$ is consistently lower in the presence of a magnetic field compared to that without a magnetic field, and this difference becomes more evident with increasing $q$. Since the difference in $\mathrm{Im}\, V^{\rm gluon}$ between the zero magnetic field and the finite magnetic field is dominated by $\mathrm{Im}\, V_{\rm string}^{\rm gluon}$,  the trend of $\mathrm{Im}\, V^{\rm gluon}$ with the variation of $r$ and $q$ is qualitatively consistent with the $\mathrm{Im}\, V_{\rm string}^{\rm gluon}$.

Next, we examine the non-extensive correction to the imaginary part of the heavy quark potential originating from the quark contribution to gluon self-energy $\mathrm{Im}\,V^{\rm quark}$, both in the absence and presence of a magnetic field.  By inserting the sixth term from the right-hand side of eq.~(\ref{eq:epsilon})  into eq.~(\ref{eq:V_general}), $\mathrm{Im}\,V^{\rm quark}$ to order $(q-1)^0$, in the absence of a magnetic field is computed as
 \begin{align}
\mathrm{Im}\,V_{(0)}^{\rm quark}(r,T)&=\mathrm{Im}\,V_{\rm HTL,(0)}^{\rm quark}(r,T)+\mathrm{Im}\,V_{\rm string,(0)}^{\rm quark}(r,T)\\
&=-C_F\alpha_sT\frac{({m}_{D}^{\rm quark})^2}{{m}_{D}^2}\phi_2({m}_{D}r)-\frac{2\sigma T ({m}_{D}^{\rm quark})^2}{{m}_{D}^4}\chi_2({m}_{D} r).\label{eq:ImV_quark}
\end{align}  
Putting the seventh and eighth terms from the right-hand side of eq.~(\ref{eq:epsilon})  into eq.~(\ref{eq:V_general}), the non-extensive correction term  of  $\mathrm{Im}\,V^{\rm quark}$ up
to  order  $(q-1)^1$, arrives at
 \begin{align}
\mathrm{Im}\,V_{(1)}^{\rm quark}(r,q,T)&=\mathrm{Im}\,V_{\rm HTL,(1)}^{\rm quark}(r,q,T)+\mathrm{Im}\,V_{\rm string,(1)}^{\rm quark}(r,q,T)\\
&=-C_F\alpha_sT\bigg[\frac{(m_{D,R,(1)}^{\rm quark})^2}{m_D^2}\bigg(\psi(m_D r)-\phi_2(m_D r)-\phi_3(m_D r)\bigg)\nonumber\\
&+\frac{(m_{D,F,(1)}^{\rm quark})^2}{{m}_{D}^2}\phi_2({m}_{D}r)\bigg]-\frac{2\sigma T}{m_D^2}\bigg[\frac{(m_{D,F,(1)}^{\rm quark})^2}{{m}_{D}^2}\chi_2({m}_{D} r)\nonumber\\
&+\frac{ ({m}_{D,R,(1)}^{\rm quark})^2}{{m}_{D}^2}\bigg(\phi_3(m_D r)-\chi_2(m_D r)-\chi_3(m_D r)\bigg)\bigg].\label{eq:ImV_quark}
\end{align} 

Similarly, inserting the sixth term from the right-hand side of eq.~(\ref{eq:epsilon_B}) into eq.~(\ref{eq:V_general}), $\mathrm{Im}V^{\rm quark}$ to order $(q-1)^0$ in the presence of a magnetic field is computed as follows:
\begin{align}
\mathrm{Im}\,V_{(0)}^{\rm quark}(r,\chi,T,eB)&=\mathrm{Im}\,V_{\rm HTL,(0)}^{\rm quark}(r,\chi,T,eB)+\mathrm{Im}\,V_{\rm string,(0)}^{\rm quark}(r,\chi,T,eB)\\
&=\int\widetilde{q}^2 \sin\theta d\theta d\widetilde{q}\left[J_{0}(\widetilde{q}r\sin\theta\sin\chi)e^{i\widetilde{q}r\cos\theta\cos\chi}-1\right]\nonumber\\
&\times\left[
\sum_{f}\sum_{n=1}^{\infty}\sum_{b=\pm}\frac{ 
 \alpha_s4n|e_fB|^2H^f_{b}(E_{\widetilde{q}_z/2,n}^f)}{({\widetilde{q}}^2+m_{D,B}^2)^2|\widetilde{q}_z|E^f_{\widetilde{q}_z/2,n}}\right]\left(\frac{C_F\alpha_s }{\pi}+\frac{2\sigma}{\widetilde{q}^2\pi}\right).
\end{align}
By inserting the seventh, eighth, and ninth terms from the right-hand side of eq.~(\ref{eq:epsilon_B}) into eq.~(\ref{eq:V_general}), we obtain the non-extensive correction term of  $\mathrm{Im}V^{\rm quark}$ in the presence of the magnetic field, 
to order  $(q-1)^1$, which is written as
\begin{align}
\mathrm{Im}\,V_{(1)}^{\rm quark}(r,\chi, q,T,eB)&=\mathrm{Im}\,V^{\rm quark}_{\rm HTL,(1)}(r,\chi,q,T,eB)+\mathrm{Im}\,V^{\rm quark}_{\rm string,(1)}(r,\chi, q,T,eB)\\
&=\int \widetilde{q}^2\sin\theta d\theta  d\widetilde{q}\left[J_{0}(\widetilde{q}r\sin\theta\sin\chi)e^{i\widetilde{q}r\cos\theta\cos\chi}-1\right]\nonumber\\
&\times\bigg\{
\sum_{f}\sum_{n=1}^{\infty}\sum_{b=\pm}\frac{ 
\alpha_s4n|e_fB|^2}{({\widetilde{q}}^2+m_{D,B}^2)^2|\widetilde{q}_z|E^f_{\widetilde{q}_z/2,n}}\bigg[W^f_{b}(E_{\widetilde{q}_z/2,n}^f)\frac{q-1}{2}\nonumber\\
&-\frac{2H_b^f(E_{\widetilde{q}_z/2,n}^f )m_{D,R,B,(1)}^2}{\widetilde{ q}^2+m_{D,B}^2}\bigg]\bigg\}\left(\frac{C_F\alpha_s }{\pi} +\frac{2\sigma }{\widetilde{q}^2\pi} \right)\nonumber\\
&-\frac{ C_F\alpha_s T}{2\pi} \frac{\sum_{f} \alpha_s|e_fB|
m_{D,R,B,(1)}^2}{m_{D,B}^4}\nonumber\\
&\times\bigg[m_{D,B}^2r^2K_2(m_{D,B}r\sin\chi)(\sin\chi)^2-2\bigg]\nonumber\\
&-\frac{4\sigma T}{\pi}\int d\widetilde{q}\left[J_{0}(\widetilde{q} r \sin\chi)-1\right]\frac{\sum_f\alpha_s|e_fB| m_{D,R,B,(1)}^2}{\widetilde{q}(\widetilde{ q}^2+m_{D,B}^2)^3}.
\end{align}
Here, $E_{\widetilde{q}_z/2,n}^f$ needs to rewrite as $E_{\widetilde{q}_z/2,n}^f=\sqrt{(\widetilde{q}\cos\theta/2)^2+2n|e_fB|}$ and $K_{m}(x)$ is the Bessel function of the second kind.
It is noted that  $\mathrm{Im}\, V^{\rm quark}=\mathrm{Im}\,V_{(0)}^{\rm quark}+\mathrm{Im}\, V_{(1)}^{\rm quark}$ in a magnetic field becomes directional or angular $\chi$ dependent.

\begin{figure}[tb]
\centering
\includegraphics[width=1\textwidth]{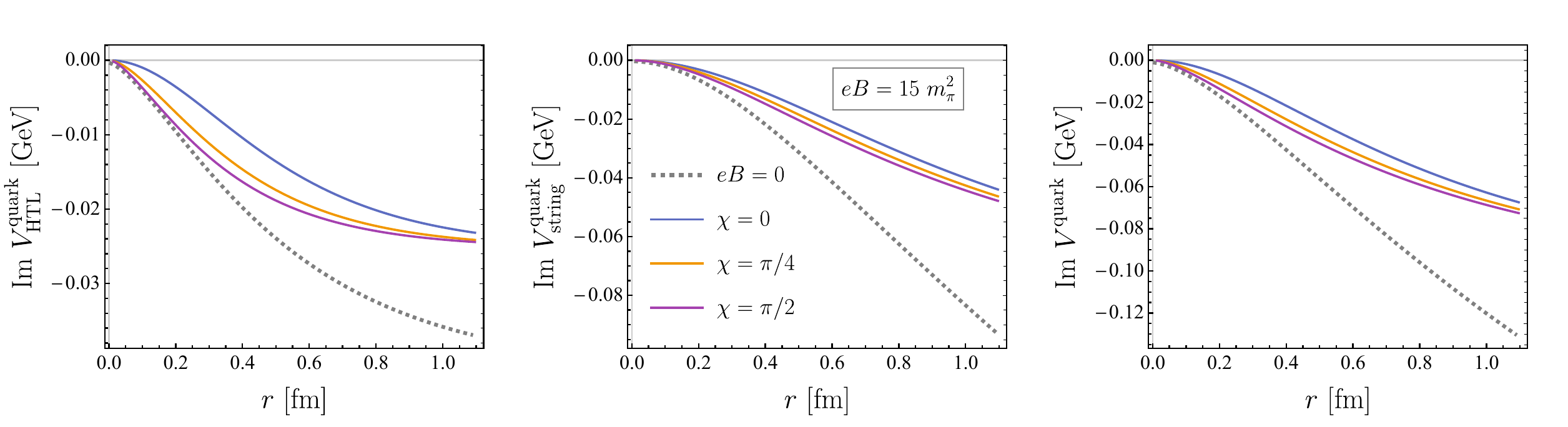}
\caption{The HTL component (left), string component (middle), and total component (right) of the imaginary part of heavy quark potential originating from quark contribution to gluon self-energy, $\mathrm{Im}\,V^{\rm quark}$, as a function of quark-antiquark separation distance $r$ in $eB=0$ (gray dashed line) and $eB=15~m_{\pi}^2$ with $\chi=0, \pi/4, \pi/2$ (solid lines). These results are calculated in the absence of non-extensivity.}\label{Plot:ImV_quark_chi}
\end{figure}

\begin{figure}[tb]
\subfigure{\includegraphics[width=0.33\textwidth]{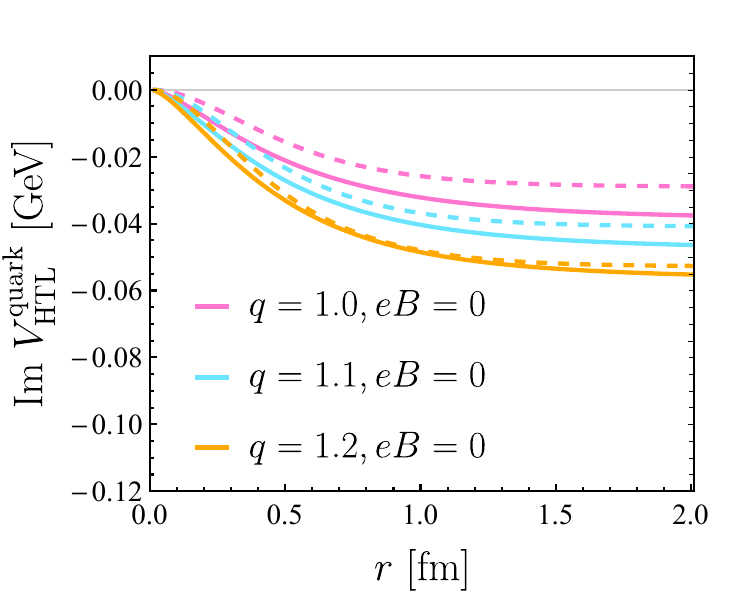}\includegraphics[width=0.33\textwidth]{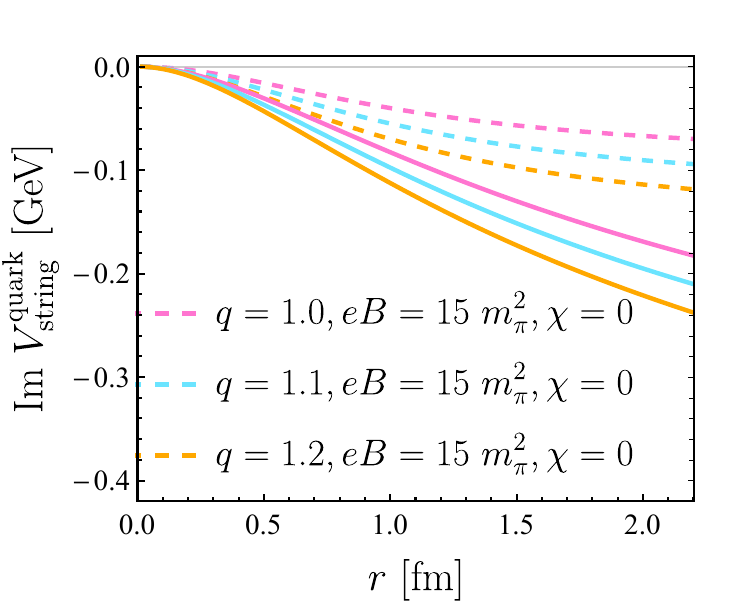}\includegraphics[width=0.33\textwidth]{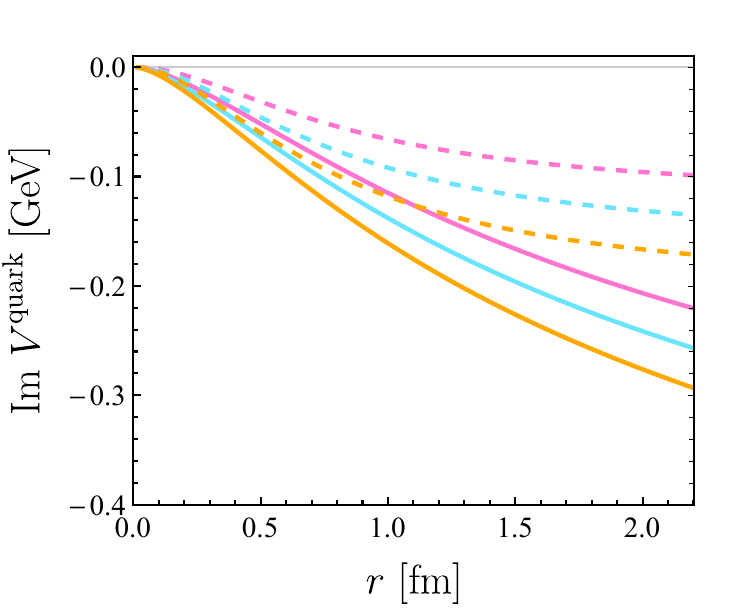}}

\vspace{-5 mm}
\subfigure{\includegraphics[width=0.334\textwidth]{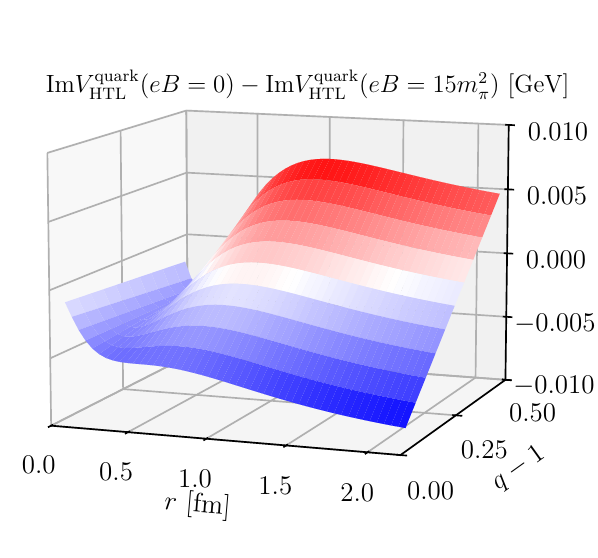}
\includegraphics[width=0.328\textwidth]{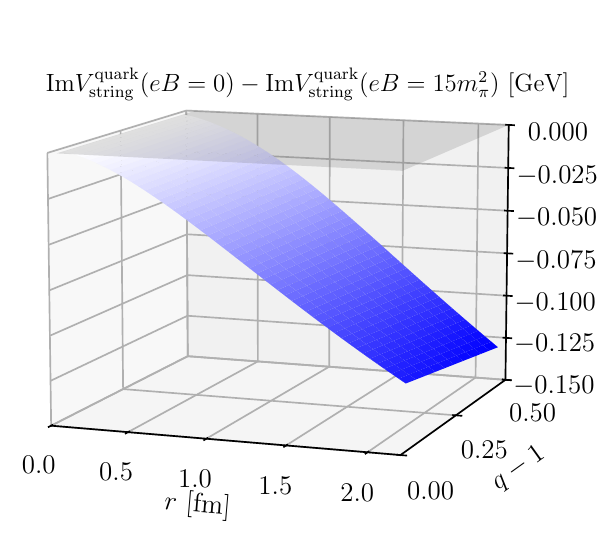}
\includegraphics[width=0.328\textwidth]{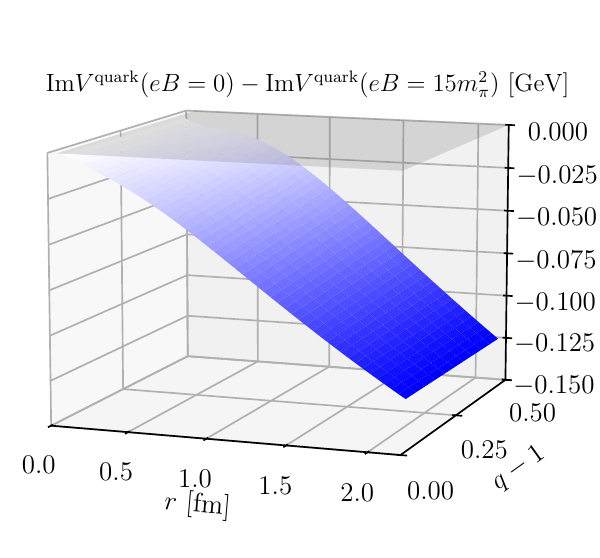}} 
\caption{Same as figure~\ref{Plot_ReV}  but for the various components of $\mathrm{Im}\,V^{\rm quark}$. The solid lines and dashed lines represent the results for the vanishing magnetic field and the finite magnetic field $eB=15~m_{\pi}^2$ with $\chi=0$, respectively. }\label{Plot:ImV_quark}
\end{figure}
\begin{figure}[tb]
	\subfigure{
		\includegraphics[width=0.5\textwidth]{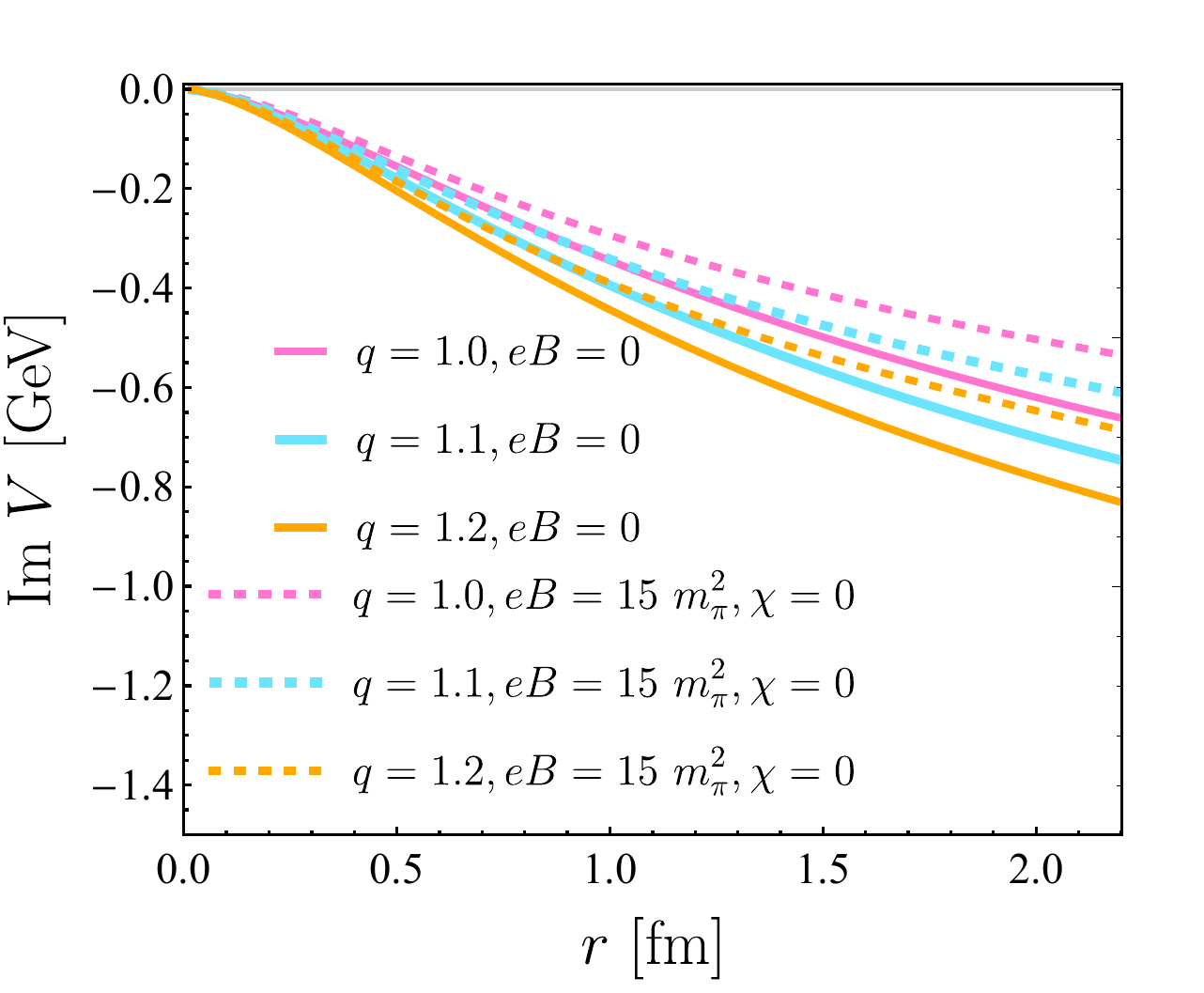}}
\subfigure{\includegraphics[width=0.5\textwidth]{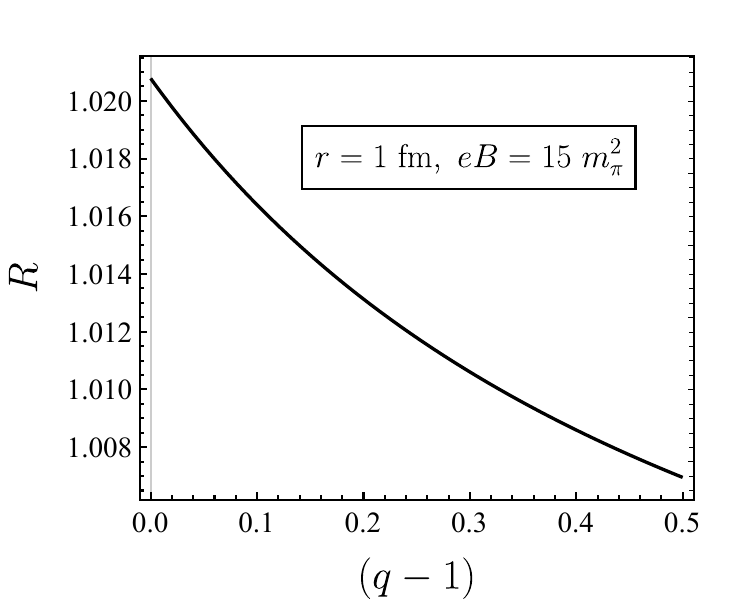}}
	\caption{(Left panel) The variation of the total imaginary part of the heavy quark potential, $\mathrm{Im}\,V$,  with quark-antiquark separation distance $r$  for different values of  $q$ at a fixed temperature $T=0.3$~GeV. The solid lines represent the results for the vanishing magnetic field, while the dashed lines represent the results for a finite magnetic field  $eB=15~m_{\pi}^2$ with $\chi=0$.
		(Right panel) The anisotropy  ratio of  $\mathrm{Im}V$,   defined as $R=\mathrm{Im}\,V(r,\chi=\pi/2,q,T,eB)/\mathrm{Im}\,V(r,\chi=0,q,T,eB)$ is plotted  as a function of $(q-1)$ at $r=1~$fm and $eB=15~m_{\pi}^2$. }
	\label{Plot:ImV}
\end{figure}

In figure~\ref{Plot:ImV_quark_chi}, we show the angular dependence of each component of  $
\mathrm{Im}\,V^{\rm quark}$ in the presence of a  magnetic field $eB=15~m_{\pi}^2$.
Both HTL and string components of  $|\mathrm{Im}\,V^{\rm quark}|$ exhibit a maximum at $\chi=\pi/2$, where the quark-antiquark dipole axis alignment along the direction perpendicular to the magnetic field. Conversely, a minimum is observed at $\chi=0$, where the quark-antiquark dipole axis parallels the magnetic field direction.  This observation confirms the anisotropy of $\mathrm{Im}\,V^{\rm quark}$ within a magnetic field. 
Furthermore, the magnitude of $\mathrm{Im}\,V^{\rm quark}$ is smaller in a finite magnetic field than in a zero magnetic field. 
Quantitatively, the increasing trend in each component of $|\mathrm{Im}\, V^{\rm quark}|$  with increasing $r$ remains consistent, regardless of the presence or absence of a magnetic field.

In the upper panel of figure~\ref{Plot:ImV_quark}, we plot the separation distance $r$ dependence of each component of $\mathrm{Im} \,V^{\rm quark}$ in the absence of a magnetic field, for different values of $q$. We also compare with the corresponding numerical results in a magnetic field of  $eB=15~m_{\pi}^2$ with  $\chi=0$.
In both zero and nonzero magnetic fields, the HTL and string components of $|\mathrm{Im}\,V^{\rm quark}|$  increase as $r$ and $q$ increase, which is naturally reflected in the overall $|\mathrm{Im}\,V^{\rm quark}|$. 
The increasing $q$ results in the absolute value of $\mathrm{Im}\,V^{\rm quark}_{\rm HTL}$ in a finite magnetic field gradually exceeding that in a zero magnetic field.
For $\mathrm{Im}\,V^{\rm quark}_{\rm string}$, in the presence of a magnetic field exhibits a consistently lower magnitude than in the absence of a magnetic field at different values of $q$, as evident from the upper panel of figure~\ref{Plot:ImV_quark}.
Notably, the dependence of  $(q-1)$ dependence of the difference in $\mathrm{Im}\,V^{\rm quark}_{\rm HTL}$ between a zero magnetic field and a finite magnetic field  (specifically, $\mathrm{Im}\,V^{\rm quark}_{\rm HTL}(eB=0)-\mathrm{Im}\,V^{\rm quark}_{\rm HTL}(eB=15m_{\pi}^2)$) is opposite to that of the corresponding difference for $\mathrm{Im}\,V^{\rm quark}_{\rm string}$ (i.e., $\mathrm{Im}\,V^{\rm quark}_{\rm string}(eB=0)-\mathrm{Im}\,V^{\rm quark}_{\rm string}(eB=15m_{\pi}^2)$).
This opposing dependence counterbalances each other, resulting in a situation where the overall difference in $\mathrm{Im}\,V^{\rm quark }$ between the presence and absence of a magnetic field remains nearly unchanged with variations in $(q-1)$ for a fixed separation distance $r$. Instead, this overall difference is primarily dependent on $r$.

In the left panel of figure~\ref{Plot:ImV}, we present the behavior of the total imaginary part of the potential, $\mathrm{Im}V$, with separation distance $r$ in both zero (solid lines) and finite magnetic field for the case of $\chi=0$ (dashed lines).  As $q$ and $r$ increase, the magnitudes of both $\mathrm{Im}\,V^{\rm gluon}$ and $\mathrm{Im}\,V^{\rm quark}$ increase, leading to a continuous increase in  $|\mathrm{Im}V|$. 
In the presence of a magnetic field, the magnitude of $\mathrm{Im} V$ for a fixed $q$ value is smaller compared to that in the purely thermal case. These variations in $\mathrm{Im} V$, influenced by both non-extensivity and the magnetic field, are reflected in the decay widths of heavy quarkonia, subsequently affecting their dissociation process. 
In the right panel of figure~\ref{Plot:ImV}, we present the anisotropy ratio $R$ of  $\mathrm{Im}\, V$, defined as $R=\mathrm{Im}\,V(r,\chi=\pi/2,q,T,eB)/\mathrm{Im}\,V(r,\chi=0,q,T,eB)$, to quantify the anisotropic response of  $\mathrm{Im}\, V$ in a finite magnetic field to non-extensive correction.   After accounting for the isotropic contribution from $\mathrm{Im}V^{\rm gluon}$, the anisotropic feature of  $\mathrm{Im}\,V$ in a finite magnetic field becomes negligible. 
Varying the values of $q$ does not significantly change the anisotropy ratio of $\mathrm{Im}\,V$ in quantitative.

\section{Non-extensive correction to in-medium properties of heavy quarkonia}\label{sec:E&Decay} 
The calculation results in section~\ref{sec:potential} indicate that both the non-extensive correction and the presence of a magnetic field can alter the heavy quark potential. Therefore, 
based on the obtained non-extensive modified heavy quark potential,
in this section, we further investigate the effects of non-extensivity and the magnetic field on in-medium properties of heavy quarkonium states including binding energy ($E_{ bin}$), decay width  ($\Gamma$), as well as melting temperature ($T_{ melt}$).

To obtain the binding energies of heavy quarkonia, one can solve time-independent Schr$\ddot{\mathrm{o}}$dinger equation for the radial 
 quarkonia wave function  $\psi (r)$ with the real part of in-medium heavy quark potential~\cite{Thakur:2020ifi},  
\begin{align}\label{eq:equation}
    -\frac{1}{m_{HQ}}\left(\psi''(r)+\frac{2}{r}\psi'(r)-\frac{l(l+1)}{r^2}\psi (r)\right)+\mathrm{Re}\,V(r,q,T,eB)\psi (r)=\epsilon_{ln}\psi (r),
\end{align} 
where $\epsilon_{nl}$ are eigenvalues of heavy quarkonia with principal quantum number $n$ and orbital quantum number $l$,   and $m_{HQ}$ denotes  heavy quark mass. At small distances,  the real part of the potential reduces to  Coulomb-like potential, given by $\mathrm{Re}\,V\sim -\frac{C_F\alpha_s}{r}$. We are only interested in the ground states  ($n=0,~l=0$) of charmonium $J/\Psi$ and bottomonium $\Upsilon $. We consider 
 the radial quarkonia wave function as the Coulomb wave function, which is given as  $\psi(r)=\frac{1}{\sqrt{\pi a_0^3}}e^{-r/a_0}$, where $a_0=2/(C_F\alpha_{s} m_{HQ})$~\cite{Jamal:2018mog}. By solving eq.~(\ref{eq:equation}) using the real part of heavy quark potential at a small distance limit, we determine the eigenvalue of quarkonium ground state, which is given as $\epsilon_{00}=-(C_F\alpha_s)^2m_{HQ}/4$.

Following  Ref.~\cite{Thakur:2020ifi},  the binding energies of heavy quarkonia at high temperatures are determined by the difference between the asymptotic value of the real part of the potential and the associated eigenvalue,
\begin{align}\label{eq:bindingenergy}
    E_{bin}(q,T, eB)=\mathrm{Re}V(r\to \infty,q,T,eB)-\epsilon_{nl}.
\end{align}
At large distances, the real part of the potential in the zero magnetic field (in the finite magnetic field) can also be simplified into  Coulomb-like potential by identifying with a different coefficient $C_F\alpha_s\to 2\sigma/\widetilde{m}_{D,R}^2$ ($C_F\alpha_s\to 2\sigma/\widetilde{m}_{D,R,B}^2$ ), i.e., $\mathrm{Re}\,V\sim -\frac{2\sigma}{\widetilde{m}_{D,R}^2 r }$ ($\mathrm{Re}\,V\sim -\frac{2\sigma}{\widetilde{m}_{D,R,B}^2 r }$).
 In this situation, the eigenvalues of heavy quarkonium ground states are obtained as $\epsilon_{00}=-\frac{m_{HQ}\sigma^2}{\widetilde{m}_{D,R}^2}$ ($\epsilon_{00}=-\frac{m_{HQ}\sigma^2}{\widetilde{m}_{D,R,B}^2}$). 
Then, applying eq.~(\ref{eq:bindingenergy}), the binding energies of heavy quarkonium ground states at high temperatures are determined as  $E_{\rm bin}=-C_F\alpha_s\widetilde{m}_{D,R}+\frac{2\sigma}{\widetilde{m}_{D,R}}+\frac{m_{HQ}\sigma^2}{\widetilde{m}_{D,R}^2}\simeq \frac{m_{HQ}\sigma^2}{\widetilde{m}_{D,R}^2}$ ($E_{\rm bin}=-C_F\alpha_s\widetilde{m}_{D,R,B}+\frac{2\sigma}{\widetilde{m}_{D,R,B}}+\frac{m_{HQ}\sigma^2}{\widetilde{m}_{D,R,B}^2}\simeq \frac{m_{HQ}\sigma^2}{\widetilde{m}_{D,R,B}^2}$).
In the numerical calculation, the charm and bottom masses are taken as $m_c=1.275~\mathrm{GeV}$ and $m_b=4.66~\mathrm{GeV}$, respectively.

We also provide an estimate for the decay widths ($\Gamma$) of heavy quarkonia.  In a first-order perturbative theory, by using the obtained non-extensive modified imaginary part of the potential and folding with the radial quarkonia wave function, the decay widths of heavy quarkonia are computed as 
\begin{eqnarray}
\Gamma (q, T, eB)&=&-\frac{\int d^3 {\bm r}(\psi(r))^2\mathrm{Im}\,V(r,\chi,q,T,eB)}{\int d^3\bm{r}(\psi(r))^2}\\
&=&-\frac{ \int d\chi dr r^2\sin\chi (\psi(r))^2\mathrm{Im}\,V(r,\chi, q,T,eB) }{ \int d\chi dr r^2\sin\chi (\psi(r))^2},
\end{eqnarray}
where $(\psi (r))^2$ is just the probability density, based on the simple  Coulomb wave function.

\begin{figure}[h]
	\subfigure{
		\hspace{-0mm}	\includegraphics[width=0.5\textwidth]{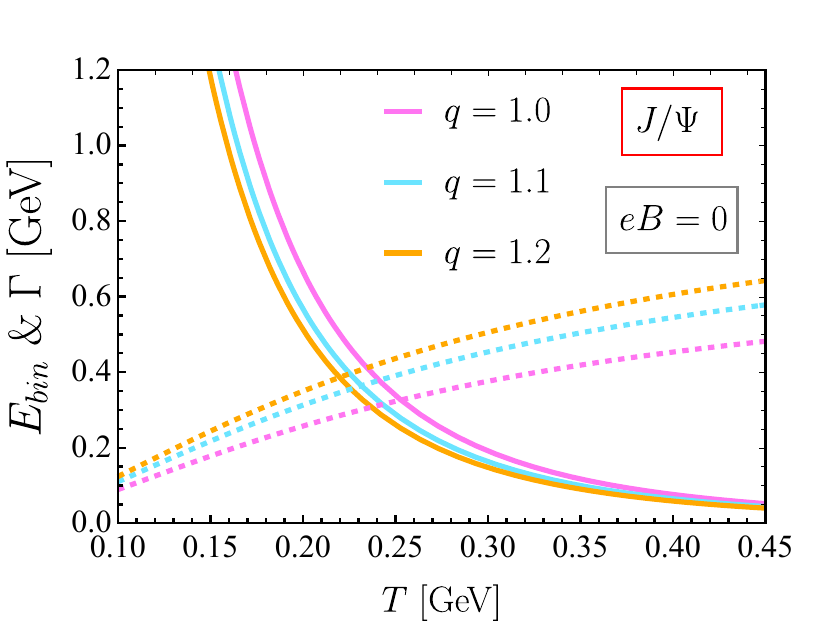}}
  \subfigure{	\includegraphics[width=0.5\textwidth]{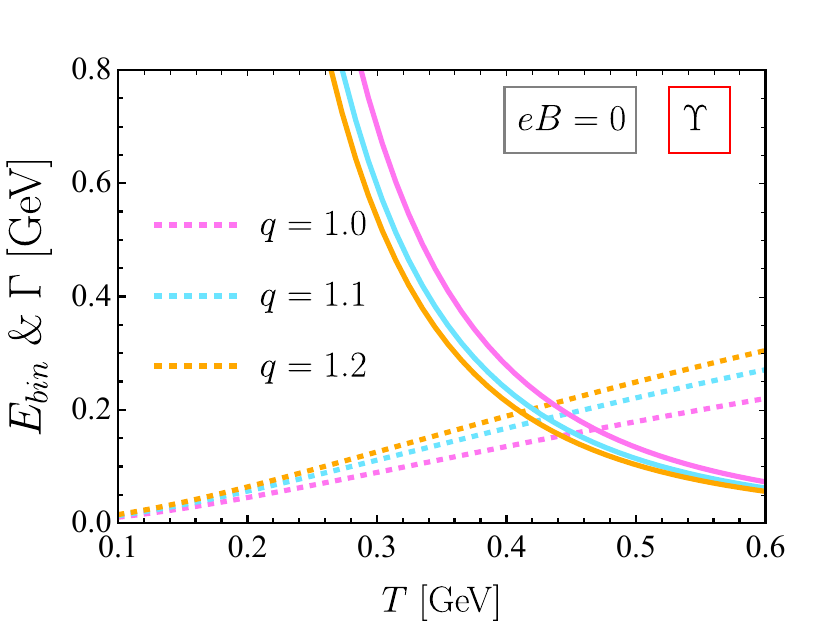}}
  \subfigure{
\includegraphics[width=0.5\textwidth]{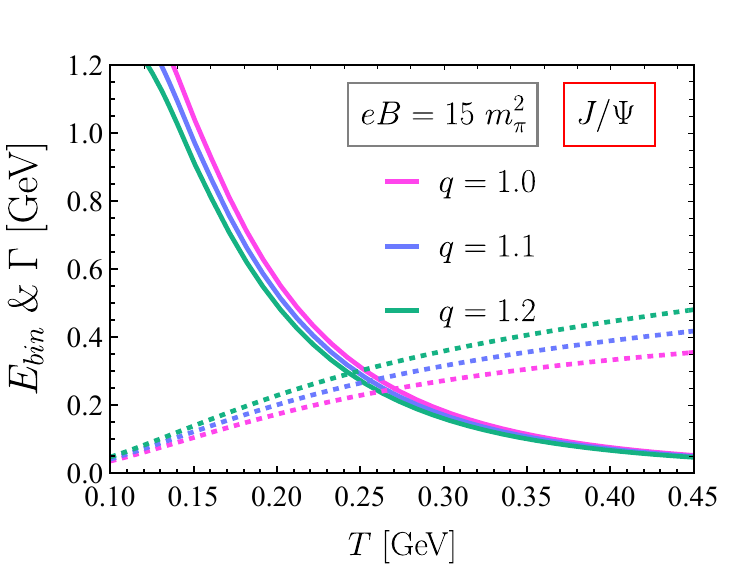}}
  \subfigure{
\includegraphics[width=0.5\textwidth]{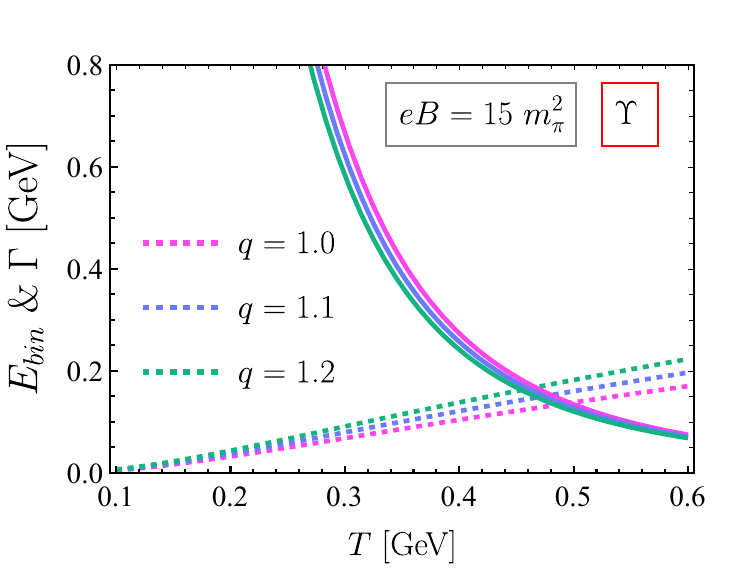}}
	\caption{The binding energies $E_{bin}$ (solid lines) and decay widths $\Gamma$ (dashed lines) for charmonium $J/\Psi$ (left panel) and bottomonium $\Upsilon$ (right panel)
    as a function of temperature at different values of the non-extensive parameter. In all plots, a fixed temperature of $T=0.3$~GeV and zero chemical potential are chosen. The upper and lower panels are the numerical results performed at zero magnetic field and finite magnetic field $eB=15~m_{\pi}^2$, respectively.}\label{Plot:Ebind&Width}
\end{figure}

Based on the computations of the binding energies and decay widths for heavy quarkonia, we can estimate the melting temperatures $(T_{melt})$ of heavy quarkonia, which are determined by a common criterion that the binding energy coincides with the decay width for a quarkonium state, i.e., $\Gamma(T_{ melt})=E_{bin}(T_{ melt})$~\cite{Mocsy:2007jz}.

Next, we will explore the influence of non-extensivity on the binding energies and decay widths of heavy quarkonia.
In figure~\ref{Plot:Ebind&Width}, we depict the binding energies (solid lines) and decay widths (dashed lines) for charmonium $J/\Psi$ (left panel) and bottomonium $\Upsilon$ (right panel) as functions of temperature, both in the absence (upper panel) and presence (lower panel) of a magnetic field. 
We observe that as the temperature increases, the binding energies of heavy quarkonia rapidly decrease, signifying that the binding between the quark and antiquark becomes increasingly incompact. 
Concurrently, the decay widths of heavy quarkonia gradually increase with the rising temperature.
We note that the introduction of non-extensive correction has a dual effect: it suppresses the binding energies of heavy quarkonia and elevates their decay widths. This shift results in the intersection point of these quantities (intersection points of solid and dashed lines of the same color in figure~\ref{Plot:Ebind&Width}) moving towards lower temperatures, implying that non-extensive correction lowers the melting temperature of heavy quarkonia, leading to an earlier quarkonium dissociation.

\begin{table}[h]  
\centering 
\begin{tabular}{|l|c|c|c|c|c|} 
\hline  
Magnetic field & State & $T_{ melt}^{q=1}$ & $T_{ melt}^{q=1.1}$  & $T_{ melt}^{q=1.2}$ & lattice QCD\\ \hline  \hline
 $eB=0$ & $J/\Psi$ & 0.254 & 0.232   & 0.219 & 0.267\\ \hline  
 $eB=15m_{\pi}^2$ & $J/\Psi$  & 0.270 &  0.255 & 0.243  & \\ \hline 
$eB=0$ & $\Upsilon$ & 0.468 & 0.431 &  0.411 & 0.440\\ \hline   
$eB=15m_{\pi}^2$ & $\Upsilon$ & 0.496 & 0.472 & 0.453& \\ \hline   
\end{tabular}  
 \caption{Melting temperatures of charmonium ($J/\Psi$) and bottomonium ($\Upsilon$) both in the absence and presence of a magnetic field for different values of $q$. The results at $q=1$ are compared with a recent work based on lattice QCD~\cite{Lafferty:2019jpr}.} 
 \label{Table}
\end{table}

The melting temperatures of both $J/\Psi$ and  $\Upsilon$ for different values of $q$ are summarized in table \ref{Table}.  In the absence of non-extensivity, our computed melting temperatures align reasonably well with lattice data.  In contrast, as shown in  figure~\ref{Plot:Ebind&Width},  the presence of a magnetic field has an opposite effect on the binding energies and decay widths of heavy quarkonia compared to non-extensive correction. Specifically, it increases the melting temperatures of heavy quarkonia, as presented in table \ref{Table}. This, in turn, impedes the heavy quarkonium dissociation. 

\section{Summary}\label{sec:summary}

In this paper, we incorporated the concept of non-extensivity into the HTL resummed perturbative theory,  utilizing the non-extensive statistical mechanics parameterized by a non-extensive parameter $q$.
Utilizing the real-time formalism, we systematically investigated the non-extensive correction to the retarded, advanced, and symmetric (time-ordered) HTL gluon self-energies, as well as the resulting resummed gluon propagators, both in the absence and presence of a magnetic field.
In the massless limit and without a magnetic field, we found that the introduction of non-extensivity leads to distinct shifts in the retarded/advanced and symmetric Debye masses. Specifically, in the leading order of $(q-1)$, the retarded/advanced Debye mass at zero chemical potential is modified from
$ \frac{g^2T^2}{6}(2N_c+N_f)$  to: 
\begin{equation}
    \frac{g^2T^2}{6}2N_c\left[1+\left(q-1\right)\left(\frac{18\zeta(3)}{\pi^2}-2\right)\right]+\frac{g^2T^2}{6}N_f\left[1+\left(q-1\right)\left(\frac{27\zeta(3)}{\pi^2}-2\right)\right],\nonumber
\end{equation} while the symmetric Debye mass is modified to:
\begin{equation}
    \frac{g^2T^2}{6}2N_c\left[1+\left(q-1\right)\left(\frac{36\zeta(3)}{\pi^2}-3\right)\right]+\frac{g^2T^2}{6}N_f\left[1+\left(q-1\right)\left(\frac{54\zeta(3)}{\pi^2}-3\right)\right].\nonumber\\
\end{equation} 
When a magnetic field is considered,  the Landau quantization of light quark motions leads to distinct forms of the Debye masses in the one-loop contributions from quarks and gluons to the retarded/advanced HTL gluon self-energies.
We observed that the retarded/advanced Debye mass becomes larger in a finite magnetic field than in a zero magnetic field. 
When non-extensivity of the system is taken into account, the slope of retarded/advanced Debye mass with respect to $q$ for a zero magnetic field becomes steeper than that for a finite magnetic field. Consequently, the retarded/advanced Debye mass enhances more significantly as $q$ increases in the absence of a magnetic field, eventually exceeding the value observed in the presence of a magnetic field.

Applying non-extensive modified resummed gluon propagators, we derived the dielectric permittivity of the QGP medium,  which is then used to compute the heavy quark potential through convolution with a Cornell potential. Our numerical result revealed that the real part of the potential at $q=1$ is more screened in the presence of a magnetic field than in the absence of a magnetic field.
When accounting for non-extensivity, the real part of the potential becomes flatter due to the enhanced screening effect. Furthermore, as both $q$ and quark-antiquark separation distance $r$ increase, the difference in the real part of the potential without and with magnetic field scenarios changes from positive to negative. 
On the other hand, increasing $r$ and introducing non-extensivity enhance the magnitude of the imaginary part of the potential, whereas the presence of a magnetic field reduces it.   
Notably, the imaginary part of the potential arising from the quark contributions to the gluon self-energy exhibits significant anisotropy in a magnetic field, although this anisotropy is largely mitigated by the isotropic contribution from $\mathrm{Im}\, V^{\rm gluon}$.

Using the obtained real part of heavy quark potential, we solved the time-independent Schr$\ddot{\rm o}$dinger equation for the radial quarkonium wave function, which allows us to determine the binding energies of heavy quarkonia,  specifically $J/\Psi$ and $\Upsilon$. On the other hand, after performing the coordinate space integration by folding with the probability density and imaginary part of the potential, we calculated the decay widths of $J/\Psi$ and $\Upsilon$. 
Inheriting traits from heavy quark potential, the binding energies of $J/\Psi$ and $\Upsilon$ decrease, while their decay widths increase with rising temperature and $q$.
Subsequently, we estimated the melting temperatures of $J/\Psi$ and $\Upsilon$. Our findings indicated that the melting temperatures decrease as $q$ increases, leading to earlier heavy quarkonium dissociation. Conversely, the presence of a magnetic field results in an increase in the melting temperatures, thereby delaying heavy quarkonium dissociation.


\section*{Acknowledgments}
This work is supported by the National Natural Science Foundation of China under Grants Nos. 11935007.
H. Zhang is supported by the Talent Scientific Start-up Foundation of Quanzhou Normal University. 


~

\appendix

\section{Non-extensive  gluon self-energy in the  zero magnetic field} \label{sec:appendixA}

\begin{figure}[h]
\centering
\subfigure{\includegraphics[width=14cm]{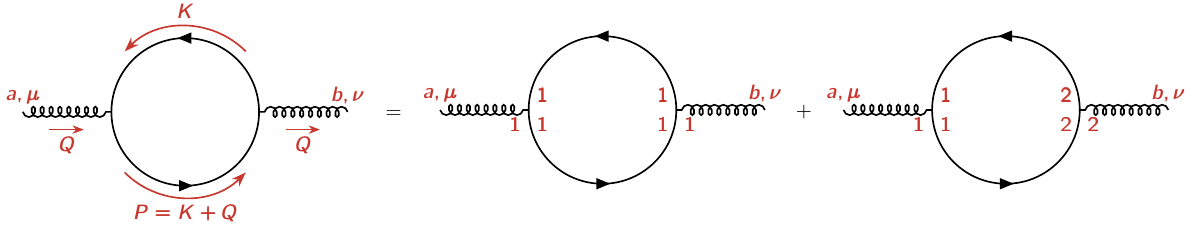}}
	\caption{  One-loop diagram for quark contributions to retarded gluon polarization tensor.
}\label{Plot:gluon_selfenergy}
\end{figure}

We derive the gluon self-energy, mainly focusing on the one-loop contribution from quarks, within the HTL perturbative theory and non-extensive statistics, using real-time formalism.
 Utilizing  eq.~(\ref{eq:Matrix_S}) and   eq.~(\ref{eq:Pirelation1}), as well as applying the Feynman rule, the one-loop contribution from quarks to the retarded gluon polarization tensor depicted in figure~\ref{Plot:gluon_selfenergy} can read  as
\begin{align}
	\Pi^{\mu\nu,\rm quark}_{R}(Q)
 =&-iN_fg^2
	\int\frac{d^4K}{(2\pi)^4} \big\{\mathrm{Tr}[t_b\gamma^\mu S_{11}(K)t_a\gamma^\nu S_{11}(P)]\nonumber\\
 -&\mathrm{Tr}[t_b\gamma^\mu S_{21}(K)t_a\gamma^\nu S_{12}(P)]\big\}\label{eq:Pimunu_p1}\\
 =&-iN_f\frac{g}{4}^2
	\int\frac{d^4K}{(2\pi)^4} 4(K^\mu P^\nu+K^\nu P^\mu-g^{\mu\nu}K\cdot P)[\Delta_F(K)\Delta_R(P)\nonumber\\
+&\Delta_A(K)\Delta_F(P)+\Delta_R(K)\Delta_R(P)+\Delta_A(K)\Delta_A(P)],\label{eq:Pimunu_p2}
	\end{align}
where $P=K+Q$, $t_{a,b}$ are the generators of color group, and  $\displaystyle{\not}  K\Delta_{R/A/F}(K)=S_{R/A/F}(K)$. 
The minus sign in front of the second square bracket of eq.~(\ref{eq:Pimunu_p1}) appears due to the vertex of type-2 field~\cite{Landsman:1986uw}. 
In deriving eq.~(\ref{eq:Pimunu_p2}) from eq.~(\ref{eq:Pimunu_p1}), we utilized the eq.~(\ref{eq:SRA}) and eq.~(\ref{eq:SF}), and performed the trace over the gamma matrices as well as suppressed the color indices.
It should be noted that  $\Delta_R(K)\Delta_R(P)$ and $\Delta_A(K)\Delta_A(P)$ in the integrand of eq.~(\ref{eq:Pimunu_p2}) are zero upon integration over $k_0$. 
Given our focus on the temporal component of the gluon self-energy, we set $\mu=\nu=0$  directly to obtain  $(K^\mu P^\nu+K^\nu P^\mu-g^{\mu\nu}K\cdot P)=(k_0p_0+\bm{k}\cdot\bm{p})$.  Consequently, the temporal component of  $\Pi_{R}^{\mu\nu,\rm quark}(Q)$, denoted as $\Pi_{R}^{\rm quark}(Q)$  is expressed as 
 \begin{align}
 \Pi_{R}^{\rm quark}(Q)
 =&-iN_fg^2
	\int\frac{d^4K}{(2\pi)^4}
(k_0p_0+\bm{k}\cdot\bm{p})\left[{\Delta}_F(K){\Delta}_R(P)+{\Delta}_A(K){\Delta}_F(P)\right]\\
 =&-N_fg^2\int\frac{d^4K}{(2\pi)^3}(k_0p_0+\bm{k}\cdot\bm{p})\Bigg[ \frac{1-2\Theta(k_0)f_{q,F}^{+}(k_0)-2\Theta(-k_0)f_{q,F}^{-}(-k_0)}{P^2+i\ \mathrm{sgn}(p_0)\epsilon}\nonumber\\
     \times &\delta(K^2)+\frac{1-2\Theta(p_0)f_{q,F}^{+}(p_0)-2\Theta(-p_0)f_{q,F}^{-}(-p_0)}{K^2-i\ \mathrm{sgn}(k_0)\epsilon} \delta(P^2)\Bigg]\label{eq:PiR_t1}\\
   =&2N_fg^2 \int\frac{d^4K}{(2\pi)^3}\bigg\{\Big[(\Theta(k_0)\left(f_{q,F}^{+}(k_0)+f_{q,F}^{-}(k_0)\right)\nonumber\\
   +&\Theta(-k_0)\left(f_{q,F}^{+}(-k_0)+f_{q,F}^{-}(-k_0)\right)\Big] \delta (k_0^2-\bm{k}^2)\nonumber\\
     \times & \frac{k_0(k_0+\omega)+\bm{k}\cdot(\bm{k}+\widetilde{\bm{q}})}{(k_0+\omega)^2-(\bm{k}+\widetilde{\bm{q}})^2+i\mathrm{sgn}(k_0+\omega)\epsilon}\bigg\}\label{eq:PiR_t2}\\
     =&N_fg^2\int\frac{dk k^2d \Omega_{k}}{(2\pi)^3}\frac{1}{k}\left(f_{q,F}^{+}(k)+f_{q,F}^{-}(k)\right)\nonumber\\
     \times &\bigg[\frac{k(k+\omega)+\bm{k}\cdot(\bm{k}+\widetilde{\bm{q}})}{(k+\omega)^2-(\bm{k}+\widetilde{\bm{q}})^2+i\epsilon}+\frac{-k(-k+\omega)+\bm{k}\cdot(\bm{k}+\widetilde{\bm{q}})}{(-k+\omega)^2-(\bm{k}+\widetilde{\bm{q}})^2-i\epsilon}\bigg].\label{eq:PiR_t3}
 \end{align}
 From eq.~(\ref{eq:PiR_t1}) to eq.~(\ref{eq:PiR_t3}), we used the momentum substitution $K\to -P$ and $d\Omega_{k}=d\phi \sin\theta d\theta$. Here, the vacuum part has been neglected because it is suppressed compared to the medium part in the HTL approximation. 
Considering that  $\omega/k$ and $\widetilde{q}/k$  in  the HTL approximation are small terms,  the square bracket in eq.~(\ref{eq:PiR_t3}) can be further  expanded as follows:
 \begin{align}
\frac{k(k+\omega)+\bm{k}\cdot(\bm{k}+\widetilde{\bm{q}})}{(k+\omega)^2-(\bm{k}+\widetilde{\bm{q}})^2+i\epsilon}&=\frac{2k^2+k\omega+kq\cos\theta}{2k\omega+Q^2-2k\widetilde{q}\cos\theta +i\epsilon}\\
     &\simeq\frac{k}{\omega-\widetilde{q}\cos\theta+i\epsilon}-\frac{1}{2}\frac{Q^2}{(\omega-q\cos\theta +i\epsilon)^2}\nonumber\\
     &+\frac{1}{2}\frac{\omega+\widetilde{q}\cos\theta}{\omega-\widetilde{q}\cos\theta+i\epsilon},\label{eq:bracket_p1}\\
     \frac{-k(-k+\omega)+\bm{k}\cdot(\bm{k}+\widetilde{\bm{q}})}{(-k+\omega)^2-(\bm{k}+\widetilde{\bm{q}})^2-i\epsilon}&=\frac{2k^2-k\omega+k\widetilde{q}\cos\theta}{-2k\omega+Q^2-2k\widetilde{q}\cos\theta -i\epsilon}\\
     &\simeq-\frac{k}{\omega+\widetilde{q}\cos\theta+i\epsilon}-\frac{1}{2}\frac{Q^2}{(\omega+\widetilde{q}\cos\theta +i\epsilon)^2}\nonumber\\
      &+\frac{1}{2}\frac{\omega-\widetilde{q}\cos\theta}{\omega+\widetilde{q}\cos\theta+i\epsilon},\label{eq:bracket_p2}
 \end{align}
 where $x=\bm{k}\cdot \widetilde{\bm{q}}/(k\widetilde{q})=\cos\theta$. 
 Since the first terms in both eq.~(\ref{eq:bracket_p1}) and eq.~(\ref{eq:bracket_p2}) are odd functions of $x$, they integrate to zero in the range from $-1$ to 1. Therefore, eq.~(\ref{eq:PiR_t3}) is simplified to the following result:
\begin{align}
   \Pi_{R}^{\rm quark}(Q) &=N_fg^2\int\frac{ kdk d \Omega_k}{2(2\pi)^3}\left(f_{q,F}^{+}(k)+f_{q,F}^{-}(k)\right)\nonumber\\
   &\times\left[\frac{1-x^2}{[x+(\omega+i\epsilon)/\widetilde{q}]^2}+\frac{1-x^2}{[-x+(\omega+i\epsilon)/\widetilde{q}]^2}\right].
\end{align}

 By utilizing  eq.~(\ref{eq:Matrix_S}) and eq.~(\ref{eq:Pirelation3}), the one-loop contribution from quarks  to the symmetric gluon self-energy tensor is expressed as
 \begin{align}\label{eq:Pimunu}
	\Pi^{\mu\nu,\rm quark}_{F}(Q)
 =&-iN_fg^2
	\int\frac{d^4K}{(2\pi)^4} \big\{\mathrm{Tr}[t_b\gamma^\mu S_{11}(K)t_a\gamma^\nu S_{11}(P)]\nonumber\\
 +&\mathrm{Tr}[t_b \gamma^\mu S_{22}(K)t_a\gamma^\nu S_{22}(P)]\big\}\\
 =&-iN_f g^2
	\int\frac{d^4K}{(2\pi)^4} (K^\mu P^\nu+K^\nu P^\mu-g^{\mu\nu}K\cdot P)\nonumber\\
\times& [\Delta_F(K)\Delta_F(P)-(\Delta_R(K)-\Delta_A(K))(\Delta_R(P)-\Delta_A(P))].
	\end{align}
Applying the relation  $\Delta_R(K)-\Delta_A(K)=-i 2\pi\delta(K^2)\mathrm{sgn}(k_0)$, 
 and focusing specifically on the temporal component of $\Pi^{\mu\nu,\rm quark}_{F}(Q)$ where $\mu=\nu=0$, we obtain 
 \begin{align}
     \Pi_{F}^{\rm quark}(Q)&=iN_fg^2
	\int\frac{d^4K}{(2\pi)^2} (k_0 p_0+\bm{k}\cdot \bm{p}) \bigg\{\left[1-2\Theta(k_0)f_{q,F}^{+}(k_0)-2\Theta(-k_0)f_{q,F}^{-}(-k_0)\right]\nonumber\\
 &\times\left[1-2\Theta(p_0)f_{q,F}^{+}(p_0)-2\Theta(-p_0)f_{q,F}^{-}(-p_0)\right]-\mathrm{sgn}(k_0)\mathrm{sgn}(p_0)\bigg\}\nonumber\\
 &\times \delta(K^2)\delta(P^2)\\
&=iN_fg^2
	\int\frac{k^2dk  d\Omega_k}{(2\pi)^2} \left(k (k+\omega)+\bm{k}\cdot (\bm{k}+\widetilde{\bm{q}})\right)\frac{1}{2k}\delta(Q^2+2k\omega-2\bm{k}\cdot\widetilde{\bm{q}})\nonumber\\
&\times  \left[-2f_{q,F}^{+}(k)- 2f_{q,F}^{+}(k+\omega)+4f_{q,F}^{+}(k)f_{q,F}^{+}(k+\omega)\right]\nonumber\\
&+iN_fg^2
	\int\frac{dk k^2 d\Omega_k}{(2\pi)^2} (-k (-k+\omega)+\bm{k}\cdot (\bm{k}+\widetilde{\bm{q}}))\frac{1}{2k}\delta(Q^2-2k\omega-2\bm{k}\cdot\widetilde{\bm{q}})\nonumber\\
&\times \left[-2f_{q,F}^{-}(k)- 2f_{q,F}^{-}(k-\omega)+4f_{q,F}^{-}(k)f_{q,F}^{-}(k-\omega)\right].\label{eq:PiF_p2}
 \end{align}
In the HTL approximation within  soft momentum limit $\omega/T\ll 1$,  $ {f}_{q,F}^{\pm}(k\pm  \omega)$ can be expanded to the leading order of $\omega/T$, yields
 \begin{align}\label{eq:expansion_f}
     {f}_{q,F}^{\pm}(k\pm  \omega)\approx   {f}_{q,F}^{\pm}(k)\mp \frac{\omega q}{T}\exp_q\left(\frac{k\mp \mu}{T}\right)\left[{f}_{q,F}^{\pm}(k)\right]^2.
 \end{align}
In eq.~(\ref{eq:PiF_p2}), the Delta function $\delta (Q^2\pm 2k\omega -2\bm{k}\cdot \widetilde{\bm{q}})$ can be rewritten as $\frac{1}{2k\widetilde{q}}\delta(Q^2/(2k\widetilde{q})\pm \omega/\widetilde{q}-x)$, subsequently,  the term $Q^2/(2k\widetilde{q})$  in the HTL approximation with $k\gg \omega$ becomes irrelevant and can be reasonably removed.  Then, eq.~(\ref{eq:PiF_p2}) is rewritten as
 \begin{equation}\label{eq:PiF_quark}
  \Pi_{F}^{\rm quark}(Q)
 =-iN_fg^2
	\int\frac{dk k^2}{(2\pi)}    \left[f_{q,F}^{+}(k)( 1-f_{q,F}^{+}(k))+f_{q,F}^{-}(k)( 1-f_{q,F}^{-}(k))\right] \frac{2}{\widetilde{q}} \Theta (\widetilde{q}^2-\omega^2),
 \end{equation}
where we have used the integral over the angle, yielding 
\begin{align}
    \int d\Omega_k \delta(\omega/\widetilde{q}\pm x)=2\pi \Theta (\widetilde{q}^2-\omega^2).
\end{align}
Since we are interested in the space-like region within the static limit  ($\omega\to 0$), which results in $ \Theta (\widetilde{q}^2-\omega^2)=1$.

The detailed derivation of the one-loop contribution from gluons (gauge fields) to the gluon self-energy can be found in Refs.~\cite{Wang:2022dcw,Gorda:2023zwy, Bellac:2011kqa}.
By repeating the computation procedure in Ref.~\cite{Bellac:2011kqa} and applying the non-extensive bare gluon  propagators listed in eqs.~(\ref{eq:SRA_gluon}) and (\ref{eq:SF_gluon}), the gluonic contributions (including the gluon-loop, ghost-loop, and tadpole contributions) to the retarded gluon self-energy tensor is given as 
\begin{align}
   \Pi_{R}^{\mu\nu,\rm gluon}(Q)&=
   -i2N_cg^2\frac{d^4K}{(2\pi)^4}(K^{\mu}P^{\nu}+K^{\nu}P^{\mu}-g^{\mu\nu}K\cdot P)\nonumber\\
&\times[\tilde{\Delta}_F(K)\tilde{\Delta}_R(P)+\tilde{\Delta}_A(K)\tilde{\Delta}_F(P)].
\end{align}
Here,   $ \tilde{\Delta}_{R/A/F}(K)\equiv G_{R/A/F}(K)$. 
Using the HTL approximation  and concerning the temporal component, a straightforward calculation leads to
\begin{align}
    \Pi_{R}^{\rm gluon }(Q)
&=2N_cg^2\int\frac{k^2 dk d \Omega_k}{(2\pi)^3}\frac{f_{q,B}(k) }{2k}\bigg[\frac{1-x^2}{[x+(\omega+i\epsilon)/\widetilde{q}]^2}+\frac{1-x^2}{[-x+(\omega+i\epsilon)/\widetilde{q}]^2}\bigg].
\end{align}

For the one-loop contribution from gluons to the symmetric gluon self-energy tensor, it can be expressed as  \begin{align}\Pi^{\mu\nu,\rm gluon}_{F}(Q)
 =&-i2N_cg^2	\int\frac{d^4K}{(2\pi)^4} (K^\mu P^\nu+K^\nu P^\mu-g^{\mu\nu}K\cdot P)\nonumber\\&\times \left[\tilde{\Delta}_F(K)\tilde{\Delta}_F(P)-\left(\tilde{\Delta}_R(K)-\tilde{\Delta}_A(K)\right)\left(\tilde{\Delta}_R(P)-\tilde{\Delta}_A(P)\right)\right].
	\end{align}
In the HTL approximation within  the small $\omega$ limit, the temporal component of the above equation yields,
 \begin{align}\label{eq:PiF_gluon}
 \Pi^{\rm gluon}_{F}(Q)
=-i2N_cg^2
	\int\frac{dk k^2}{2\pi}    f_{q,B}(k)( 1+f_{q,B}(k))\frac{2}{\widetilde{q}}\Theta (\widetilde{q}^2-\omega^2),
	\end{align}
 which has the same structure as eq.~(\ref{eq:PiF_quark}) but with $f_{q,F}(k)$ replaced by a non-extensive deformed Bose-Einstein distribution function $f_{q,B}(k)$.

 \section{Non-extensive gluon self-energy in the finite magnetic field} \label{sec:appendixB}
Similar to the calculation procedure in  a zero magnetic field, the one-loop contribution from quarks to the retarded gluon self-energy tensor in a finite magnetic field is expressed as
 \begin{equation}\label{eq:PI_R_quark_B}
     \Pi^{\mu\nu,\rm quark}_{R,B}(Q)
     =-i\sum_{f}\sum_{n,l=0}\frac{g^2}{4}\int\frac{d^2K_{\|}}{(2\pi)^2} L^{\mu\nu}
\left[\Delta^{n,f}_F(K)\Delta^{l,f}_R(P)+\Delta^{n,f}_A(K)\Delta^{l,f}_F(P)\right],
 \end{equation} 
 where $\exp\left(-\frac{\bm{k}_{\perp}^2}{|e_f B|}\right)\sum_{n=0}^{\infty}(-1)^nD_{n}^f(K)\Delta_{R/A/F}^{n,f}(K)=S_{R/A/F}^{n,f}(K)$ and the expression of tensor $L^{\mu\nu}$ is given as
 \begin{equation}
 L^{\mu\nu}=\int\frac{d^2\bm{k}_{\perp}}{(2\pi)^2}(-1)^{n+l}\exp\left(\frac{-\bm{k}_{\perp}^2-\bm{p}_{\perp}^2}{|e_f B|}\right)\mathrm{Tr}[\gamma^\mu D_{n}^f(K)\gamma^\nu D_{l}^f(P)]
 \end{equation}
 In the HTL approximation with the hierarchy of scale $\widetilde{q}^2\ll eB\sim T^2$, the tensor $L^{\mu\nu}$ can be further simplified to
    $L^{\mu\nu}=(-1)^{n+l}
    \frac{|e_f B|}{\pi}
	\big[4|e_fB|n\delta_{l-1}^{n-1} g_{\|}^{\mu\nu}+(\delta^n_{l}+\delta^{n-1}_{l-1})(K_{\|}^\mu P_{\|}^\nu+K_{\|}^\nu P_{\|}^\mu-g_{\|}^{\mu\nu}(K_{\|}\cdot P_{\|}))-(\delta_{l}^{n-1}+\delta^n_{l-1})(K_{\|}\cdot P_{\|})g_{\perp}^{\mu\nu}\big]$~\cite{Zhang:2023ked}. 
    The metric is defined as $g^{\mu\nu}=g^{\mu\nu}_{\|}+g_{\perp}^{\mu\nu}$, where $g^{\mu\nu}_{\|}=\mathrm{diag}(1,0,0,-1)$ and $\mathrm{diag}(0,-1,-1,0)$.
In the soft momentum transfer limit,
the perturbative QCD interaction cannot induce the Landau level transition for light (anti-)quarks, because they do not possess enough energy to jump across the energy gap separating the Landau levels, which is proportional to $ \sqrt {eB}$. Consequently, by taking  $n=l$, the tensor $L^{\mu\nu}$ resonantly simplifies to: $L^{\mu\nu}=
\frac{e_fB}{\pi}\alpha_{0n}[2n|e_fB|g_{\|}^{\mu\nu}+(K_{\|}^\mu P_{\|}^\nu+K_{\|}^\nu P_{\|}^\mu)-g_{\|}^{\mu\nu}(K_{\|}\cdot P_{\|})]$, where $\alpha_{0n}=(2-\delta_{0,n})$ is the Landau level-dependent spin degeneracy. 
Focusing solely on the temporal component of  $  \Pi^{\mu\nu,\rm quark}_{R,B}(Q)$,  we arrive at the following result:
 \begin{align}\label{eq:PI_R_quark_B}
     \Pi^{\rm quark}_{R,B}(Q)
    =&-\sum_f\sum_{n=0}\frac{g^2}{4} \int\frac{d^2K_{\|}}{(2\pi)}\bigg\{\bigg[(1-2\Theta(k_0)f_{q,F}^{+}(k_0)-2\Theta(-k_0)f_{q,F}^{-}(-k_0))\nonumber\\
     &\times \frac{ L^{00}(k_0)\delta(K_{\|}^2-2n|e_fB|)}{P_\|^2-2n|e_fB|+i\,\mathrm{sgn}(p_0)\epsilon}\bigg]+\bigg[(1-2\Theta(p_0)f_{q,F}^{+}(p_0)-2\Theta(-p_0)f^{-}_{q,F}(-p_0))\nonumber\\
     &\times \frac{L^{00}(k_0)\delta(P_\|^2-2n|e_fB|)}{K_\|^2-2n|e_fB|-i\,\mathrm{sgn}(k_0)\epsilon}\bigg]\bigg\}.
 \end{align} 
 Here, $L^{00}(k_0)=\alpha_{0n}(k_0p_0+k_zp_z+2n|e_fB|)\frac{|e_f B|}{\pi}$. In the presence of a magnetic field, it is essential to isolate the vacuum part and medium part in $\Pi^{\rm quark}_{R,B }(Q)$. 
In a magnetic field,  the vacuum part in $\Pi^{\rm quark}_{R,B }(Q)$, denoted as $\Pi_{R,B,\rm vac}^{\rm quark}(Q)$,  becomes significant, which has been extensively computed in the literature (see Refs.~\cite{Fukushima:2015wck, Hattori:2012ny, Fukushima:2011nu}, for details). It is given as:
\begin{equation}\label{eq:Pi_quark_vac}
    \Pi_{R,B,\rm vac}^{\rm quark}(Q)=\frac{\widetilde{q}_z^2}{Q_{\|}^2}\sum_f\frac{\alpha_s|e_fB|}{\pi}-i\frac{\omega}{2}\sum_f\alpha_s|e_fB|
    \left[\delta(\omega-\widetilde{q}_z)+\delta(\omega+\widetilde{q})\right],
\end{equation}
where the form factor $\exp(-\frac{\bm{\widetilde{q}}_{\perp}^2}{2|e_fB|})$ has been removed in the HTL approximation with soft momentum limit $\widetilde{q}^2\ll T^2\sim eB$.

We proceed to the computation of medium part of $\Pi^{\rm quark}_{R,B }(Q)$, denoted as $\Pi^{\mathrm{quark}}_{R,B,\rm med}(Q)$, its real part can be computed as follows:
   \begin{align}\label{eq:RePi}
   \mathrm{Re}\,\Pi^{\rm quark}_{R,B,\rm med}(Q)
     =&-\sum_f\sum_{n=0}\frac{g^2}{4} \int\frac{dk_z}{2\pi}\bigg\{\bigg[-\frac{f^{+}_{q,F}(E_{k_z,n}^f)}{E_{k_z,n}^f}\frac{L^{00}(E_{k_z,n}^f)}{(\omega+E_{k_z,n}^f)^2-(E_{p_z,n}^f)^2}\nonumber\\
     -&\frac{f^{-}_{q,F}(E_{k_z,n}^f)}{E_{k_z,n}^f}\frac{L^{00}(-E_{k_z,n}^f)}{(\omega-E_{k_z,n}^f)^2-(E_{p_z,n}^f)^2}-\frac{f^{+}_{q,F}(E_{p_z,n}^f)}{E_{p_z,n}^f}\frac{L^{00}(-\omega+E_{p_z,n}^f)}{(\omega+E_{p_z,n}^f)^2-(E_{k_z,n}^f)^2}\nonumber\\
     -&\frac{f^{-}_{q,F}(E_{p_z,n}^f)}{E_{p_z,n}^f}\frac{L^{00}(-\omega-E_{p_z,n}^f)}{(\omega+E_{p_z,n}^f)^2-(E_{k_z,n}^f)^2}\bigg]\bigg\},
\end{align}
 where $L^{00}(\pm E^f_{k_z,n})=2(E^f_{k_z,n})^2\pm \omega E^f_{k_z,n}+\widetilde{q}_zk_z$,  and $L^{00}(-\omega\pm E^f_{p_z,n})=(E^f_{p_z,n})^2\mp \omega E^f_{p_z,n}+(E^f_{k_z,n})^2+k_z\widetilde{q}_z$. 
By taking the  static limit ($\omega\to 0$), the above equation can be simplified as   
  \begin{align}\label{eq:RePi}
  \underset{\omega\to 0}{\mathrm{lim}} \mathrm{Re}\,\Pi^{\rm quark}_{R,B,\rm med}(Q)
     =&-\sum_f\sum_{n=1}\frac{g^2|e_fB|}{4\pi}
     \int\frac{dk_z }{(2\pi)}\bigg\{-\frac{2(E^f_{k_z,n})^2+\widetilde{q}_zk_z}{E_{k_z,n}^f(E_{k_z,n}^f)^2-(E_{p_z,n}^f)^2}\nonumber\\
    \times & \left[f^{+}_{q,F}(E_{k_z,n}^f)+f^{-}_{q,F}(E_{k_z,n}^f)\right]-\frac{(E^f_{p_z,n})^2+(E^f_{k_z,n})^2+\widetilde{q}_zk_z}{E_{p_z,n}^f(E_{p_z,n}^f)^2-(E_{k_z,n}^f)^2}\nonumber\\
     \times & \left[f^{+}_{q,F}(E_{p_z,n}^f)+f^{-}_{q,F}(E_{p_z,n}^f)\right]\bigg\}.
\end{align}

Extracting the imaginary part from the medium part of eq.~(\ref{eq:PI_R_quark_B}), i.e., the imaginary part of $\Pi^{\rm quark}_{R,B,\rm med }(Q)$, and performing a momentum substitution $P\to -K$  in the first square bracket of eq.~(\ref{eq:PI_R_quark_B}), we obtain
 \begin{align}
    \mathrm{Im} \, \Pi^{\rm quark}_{R,B,\rm med}(Q)
 =&-\sum_{f}\sum_{n=0}^{\infty}\frac{g^2}{8}\int d^2K_{\|}L^{00}(k_0) \delta (K_{\|}^2-2n|e_fB|)\delta(P_{\|}^2-2n|e_fB|)\nonumber\\
   \times & \bigg\{ \mathrm{sgn}(-k_0)\left[\Theta(-p_0) f^{+}_{q,F}(-p_0)+\Theta(p_0)f^-_{q,F}(p_0)\right]\nonumber\\
  -&\mathrm{sgn}(k_0)\left[\Theta(p_0) f^+_{q,F}(p_0)+\Theta(-p_0)f^-_{q,F}(-p_0)\right]\bigg\}\\
  =&-\sum_{f}\sum_{n=0}^{\infty}\frac{g^2}{8}\int dk_z\frac{L^{00}(E_{k_z,n}^f)}{4E_{k_z,n}^fE_{p_z,n}^f}\left[ -f^-_{q,F}(\omega+E_{k_z,n}^f)-f^+_{q,F}(\omega+E_{k_z,n}^f)\right]\nonumber\\
  \times&\left[\delta (\omega+E_{k_z,n}^f-E_{p_z,n}^f)+\delta (\omega+E_{k_z,n}^f+E_{p_z,n}^f)\right]\nonumber\\
-&\sum_{f}\sum_{n=0}^{\infty}\frac{g^2}{8}\int dk_z\frac{L^{00}(-E_{k_z,n}^f)}{4E_{k_z,n}^fE_{p_z,n}^f}\left[ f^{+}_{q,F}(E_{k_z,n}^f-\omega)+f^-_{q,F}(E_{k_z,n}^f-\omega)\right]\nonumber\\
  \times&\left[\delta (\omega+E_{k_z,n}^f-E_{p_z,n}^f)+\delta (\omega+E_{k_z,n}^f+E_{p_z,n}^f)\right].\label{eq:Pi_quark_med}
 \end{align} 
In eq.~(\ref{eq:Pi_quark_med}), only the delta functions related to the decay processes  ($\delta (\omega+E^f_{k_z,n}-E^f_{p_z,n})$ and $\delta (\omega-E^f_{k_z,n}+E^f_{p_z,n})$) contribute to our work. 
Within the static limit ($\omega\to 0 $),  these delta functions of our interest can be explicitly worked out as
\begin{equation}
\delta(E_{k_z,n}^f-E_{p_z,n}^f)=(E_{\widetilde{q}_z/2,n}^f/|\widetilde{q}_z|)\delta(k_z+\widetilde{q}_z/2).
\end{equation}
By utilizing eq.~(\ref{eq:expansion}) and performing the integration over $k_z$,  eq.~(\ref{eq:Pi_quark_med}) in the static limit ($\omega\to 0$), finally  is computed as
 \begin{align}
 \underset{\omega\to 0}{\mathrm{lim}}    \frac{\mathrm{Im}\, \Pi^{\rm quark}_{R,B,\rm med}(Q)}{\omega}=&-\sum_{f}\sum_{n=0}^{\infty}\frac{g^2}{4\pi}\frac{\alpha_{0n}2n|e_fB|^2}{E_{\widetilde{q}_z/2,n}^f|\widetilde{q}_z|}\frac{ q}{2T}\bigg[\exp_q\left(\frac{E_{\widetilde{q}_z/2,n}^f+\mu}{T}\right)\left(f_{q,F}^{-}(E_{\widetilde{q}_z/2,n}^f)\right)^2\nonumber\\
+&\exp_q\left(\frac{E_{\widetilde{q}_z/2,n}^f- \mu}{T}\right)\left(f_{q,F}^{+}(E_{\widetilde{q}_z/2,n}^f)\right)^2\bigg].
 \end{align} 
Since we consider the massless QCD limit, when light quarks occupy the lowest Landau level,   $\mathrm{Im}\,\Pi_{R,B, \rm med}^{\rm quark}(Q)$ becomes zero.

The one-loop contribution from quarks to the symmetric gluon self-energy tensor in a magnetic field, denoted as $  \Pi^{\mu\nu,\rm quark}_{F,B}(Q)$, is purely imaginary and can be expressed as
 \begin{align}\label{eq:Pimunu_F_quark_B}
	\Pi^{\mu\nu,\rm quark}_{F,B}(Q)
 =&-i\sum_f \sum_{n,l}^{\infty}\frac{g^2}{4}\int\frac{d^2K_{\|}}{(2\pi)^4}L^{\mu\nu} 
\bigg[\Delta_F^{n,f}(K)\Delta_F^{l,f}(P)\nonumber\\
-&\left(\Delta_R^{n,f}(K)-\Delta_A^{n,f}(K)\right)\left(\Delta_R^{l,f}(P)-\Delta_A^{l,f}(P)\right)\bigg].
	\end{align}
By inserting the definite expressions of  $\Delta_{R/A}^{n,f}(K)$ and $\Delta_{F}^{n,f}(K)$ into the above  equation,  and  following a similar calculation procedure as for $\Pi^{\mu\nu,\rm quark}_{R,B}$,
 the temporal component of eq.~(\ref{eq:Pimunu_F_quark_B}) in the HTL approximation is computed  as:
\begin{align}\label{eq:appendix_Pi_quark_F1}
     \Pi^{\rm quark}_{F,B}(Q)=&i\sum_f\sum_{n=0}^{\infty}\frac{g^2}{4}
	\int d^2K_{\|}L^{00}(k_0) \bigg\{\left[1-2\Theta(k_0)f_{q,F}^{+}(k_0)-2\Theta(-k_0)f_{q,F}^{-}(-k_0)\right]\nonumber\\
\times&\left[1-2\Theta(p_0)f_{q,F}^{+}(p_0)-2\Theta(-p_0)f_{q,F}^{-}(-p_0)\right]-\mathrm{sgn}(p_0)\mathrm{sgn}(k_0)\bigg\}\nonumber\\
\times& \delta(K_\|^2-2n|e_fB|)\delta(P_\|^2-2n|e_fB|).
\end{align}
By performing integration over $k_0$ and using the identities for Dirac delta functions, the above equation is further computed as 
\begin{align}\label{eq:appendix_Pi_quark_F2}
\Pi^{\rm quark}_{F,B}(Q)
&=-i\sum_f\sum_{n=0}^{\infty}\frac{g^2}{4}\int\frac{ dk_z}{4E_{k_z,n}^fE_{p_z,n}^f} \bigg\{L^{00}(E_{k_z,n}^f)\bigg[2f_{q,F}^{+}(E_{k_z,n}^f)
+2f_{q,F}^{+}(\omega+E_{k_z,n}^f)\nonumber\\
&-4f_{q,F}^{+}(E_{k_z,n}^f)f_{q,F}^{+}(\omega+E_{k_z,n}^f)\bigg]\nonumber\\
&+L^{00}(-E_{k_z,n}^f)\bigg[2f_{q,F}^{-}(E_{k_z,n}^f)+2f_{q,F}^{-}(-\omega+E_{k_z,n}^f)\nonumber\\
&-4f_{q,F}^{-}(E_{k_z,n}^f)f_{q,F}^{-}(-\omega+E_{k_z,n}^f)\bigg]\bigg\}\nonumber\\
&\times\left[\delta(\omega+E_{k_z,n}^f-E_{p_z,n}^f)+\delta(\omega+E_{k_z,n}^f+E_{p_z,n}^f)\right].
 \end{align}
By utilizing eq.~(\ref{eq:expansion}) again and considering the decay processes, the above equation  in the static limit ($\omega\to 0$) 
  finally is obtained as
\begin{align}\label{eq:appendix_Pi_quark_F3}
      \underset{\omega\to 0}{\mathrm{lim}} \Pi^{\rm quark}_{F,B}(Q)
=&-i\sum_f\sum_{n=0}^{\infty}\frac{g^2}{4\pi}\frac{\alpha_{n0}4n|e_fB|^2}{E_{\widetilde{q}_z/2,n}^f|\widetilde{q}_z|} \bigg\{\bigg[f_{q,F}^{+}(E_{\widetilde{q}_z/2,n}^f)\left(1-f_{q,F}^{+}(E_{\widetilde{q}_z/2,n}^f)\right)\nonumber\\
+&f_{q,F}^{-}(E_{\widetilde{q}_z/2,n}^f)
\left(1-f_{q,F}^{-}(E_{\widetilde{q}_z/2,n}^f)\right)\bigg]\bigg\}.
 \end{align}

Since thermal gluons are not directly coupled to the magnetic field, the gluonic contributions to the symmetric HTL gluon self-energy in a magnetic field is consistent with that in a zero magnetic field, as presented in eq.~(\ref{eq:PiF_gluon}).


\bibliographystyle{JHEP}
\bibliography{biblio.bib}






\end{document}